%% file: jcp_ltpic_arxiv-v2.tex
\documentclass[preprint,1p,times]{elsarticle}       

\usepackage{amssymb}
\usepackage{amsthm}
\usepackage{epsfig}

\usepackage{amsmath,amssymb}
\usepackage{bbm}

\def\bs{\boldsymbol}

\def\rmd{\, {\rm d}} 
\def\Ds{{\Delta s}}

\def\bse{{\bs e}}

\def\NGpieces{{N_G}}
\def\NGpoints{{N'_G}}

\def\Dtrec{\Dt_{\rm r}}
\def\Dsrec{\Ds_{\rm r}}

\def\ex{{\rm ex}}

\def\vp{{\varphi}}
\def\ve{{\varepsilon}}
\def\Dt{{\Delta t}}
\def\Ds{{\Delta s}}

\def\jrho0{{j_0^{\, \rho}}}

\def\cB{\mathcal{B}}

\def\cF{\mathcal{F}}

\def\RR{\mathbb{R}}
\def\ZZ{\mathbb{Z}}

\DeclareMathOperator{\diam}{diam}

\providecommand{\abs}[1]{\lvert#1\rvert}

\providecommand{\norm}[1]{\lVert#1\rVert}

\DeclareMathOperator{\supp}{supp}

\theoremstyle{remark}

\newtheorem{algo}{Algorithm}[section]

\journal{J. Comp. Phys.}

\begin{document}

\begin{frontmatter}



\title{Noiseless Vlasov-Poisson simulations with linearly transformed particles}

\author[lbnl,ljll-cnrs,ljll-upmc]{M. Campos Pinto\corref{cor1}}
\ead{campos@ann.jussieu.fr}
\author[irma,calvi]{E. Sonnen{\-}dr\"ucker}
\ead{sonnen@math.unistra.fr}
\author[llnl,lbnl]{A. Friedman}
\ead{af@llnl.gov}
\author[llnl,lbnl]{D. Grote}
\ead{grote1@llnl.gov}
\author[llnl,lbnl]{S. Lund}
\ead{smlund@llnl.gov}

\cortext[cor1]{Corresponding author. LJLL, Universit\'e Pierre et Marie Curie, Bo\^ite courrier 187,
75252 Paris Cedex 05, France. Tel.: +33 1 44 27 91 54; fax: +33 1 44 27 72 00.}
\address[lbnl]{Lawrence Berkeley National Laboratory, Berkeley, CA 94720, USA}
\address[ljll-cnrs]{CNRS, UMR 7598, Laboratoire Jacques-Louis Lions, F-75005, Paris, France}
\address[ljll-upmc]{UPMC Univ Paris 06, UMR 7598, Laboratoire Jacques-Louis Lions, F-75005, Paris, France}
\address[irma]{IRMA, UMR 7501, Universit\'e de Strasbourg \& CNRS, 7 rue Ren\'e Descartes, F-67084 Strasbourg Cedex, France}
\address[calvi]{Project-team CALVI, INRIA Nancy Grand Est, 7 rue Ren\'e Descartes, F-67084 Strasbourg Cedex, France}
\address[llnl]{Lawrence Livermore National Laboratory, Livermore, CA 94550, USA}

\begin{abstract}

We introduce a deterministic discrete-particle simulation approach, the Linearly-Transformed Particle-In-Cell (LTPIC) method, 
that employs linear deformations of the particles to reduce the noise traditionally associated with particle schemes. 
Formally, transforming the particles is justified by local first order expansions of the characteristic flow in phase space. 
In practice the method amounts to using deformation matrices within the particle shape functions; these matrices are updated via 
local evaluations of the forward numerical flow. 
Because it is necessary to periodically remap the particles on a regular grid to avoid excessively deforming their shapes, 
the method can be seen as a development of Denavit's Forward Semi-Lagrangian (FSL) scheme 
[J.~Denavit, {\em J. Comp. Physics} {\bf 9}, 75 (1972)]. However, it has recently been established 
[M.~Campos Pinto, ``Smooth particle methods without smoothing'', arXiv:1112.1859 (2012)] 
that the underlying Linearly-Transformed Particle scheme converges for abstract transport problems, with no need to remap the particles; 
deforming the particles can thus be seen as a way to significantly lower the remapping frequency needed in the FSL schemes, 
and hence the associated numerical diffusion. To couple the method with electrostatic field solvers, 
two specific charge deposition schemes are examined, and their performance compared with that of the standard deposition method. 
Finally, numerical 1d1v simulations involving benchmark test cases and halo formation in an initially mismatched thermal sheet beam 
demonstrate some advantages of our LTPIC scheme over the classical PIC and FSL methods. 
Benchmarked test cases also indicate that, for numerical choices involving similar computational effort, 
the LTPIC method is capable of accuracy comparable to or exceeding that of state-of-the-art, high-resolution Vlasov schemes.

\end{abstract}


\end{frontmatter}

\date{\today}



\section{Introduction}

Although considered very efficient in many practical cases, particle-in-cell (PIC) simulations sometimes present 
levels of noise that make fine plasma phenomena very numerically expensive to resolve. The fact that particles are usually initialized with
random procedures explains part of the statistical noise, yet there is another reason for the birth of strong oscillations
in the numerical solutions. 
Indeed, it is known from the mathematical analysis of deterministic particle methods \cite{Beale.Majda.1982b.mcomp, Raviart.1985.lnm} 
that a typical requirement for smooth convergence is that the radius $\ve$ of the 
particles tend to 0 at a much slower rate than the average grid spacing $h$ used for their initialization, a property that is 
expensive to satisfy in practice.
Here by ``smooth convergence'' we mean the pointwise convergence of the density function carried by the macro-particles,
towards the exact solution $f$ of the Vlasov equation. If the latter is a continuous function of the phase-space coordinates and
if the convergence is pointwise, numerical solutions are indeed free of spurious oscillations, at least asymptotically.

Specifically, smooth convergence requires $\ve \sim h^q$ with $q <1$, which can be interpreted as an 
{\em extended overlapping} condition: as the initialization grid gets finer, more and more particles must overlap.
In PIC schemes (and more precisely, weighted PIC schemes with uniform Poisson-solver grids) the particle size is implicitly dictated 
by the $d$-dimensional mesh used in the field solver through its number of cells $N_c  \sim \ve^{-d}$, 
whereas the (average, if random) initial spacing can be derived from the number of particles
$N_{\rm p} \sim h^{-d-d'}$, with $d'$ denoting the dimension of the velocity variable.
Therefore, to guarantee the smooth convergence of the numerical density one should increase the number of particles per cell
consistent with the number of cells, i.e.,
$$
\frac{N_{\rm p}}{N_{\rm c}} \sim  N_{\rm c}^{\frac{d+d'}{dq}-1}.
$$ 
Here the exponent is always positive, and when $d = d'$ it is greater than unity, e.g., 1.5 for $q = 0.8$. 
Hence for smooth particle simulations, the number of particles per cell
should increase {\em significantly faster} than the number of cells. In practice such a condition is usually not met. 

On the mathematical level, particle methods that do not meet the extended overlapping condition may still converge 
towards a smooth $f$ but only in a weak sense, i.e., in the sense that the local integrals of the particle density function
tend to the same integrals of $f$. 
This case typically corresponds to simulations with strongly oscillating density functions, where 
accurate results can be obtained for certain integral quantities such as the electric field, or for the density itself, 
through appropriate smoothing procedures. And since the accuracy of the electric field is what matters most for the (electrostatic) dynamics 
of the system, strongly oscillating simulations can very well give satisfactory results on the longer scale sizes of physical interest.

However, in cases where the physics of interest is in a region of low plasma density, smooth convergence seems to be necessary
for precise measurements. For a variety of practical problems indeed (including backward Raman scattering \cite{Ghizzo.al.1992.nfus},
plasma-wall transitions \cite{Shoucri.al.2004.epjd, Valsaque.Manfredi.2001.jnucmat}, halo formation in beams \cite{Sonnendrucker.al.2001.nima}
and development of electron holes in the presence of a guide field \cite{Goldman.Newman.Pritchett.2008.grl}) physicists often need to
resort to (grid based) Vlasov or Semi-Lagrangian solvers in order to obtain sufficient accuracy. Unfortunately these methods are known
to be numerically expensive to run and challenging to implement, as they require the mesh to cover the whole phase space and can suffer
from diffusive effects.

To reduce noise, Denavit \cite{Denavit.1972.jcp} proposed a particle method later revisited as a Forward Semi-Lagrangian (FSL) scheme 
\cite{Nair.Scroggs.Semazzi.2003.jcp, Crouseilles.Respaud.Sonnendrucker.2009.cpc}, where the distribution function carried by the particles
is periodically remapped to the nodes of a phase-space grid. This has a smoothing effect which in practice eliminates the need for
extended overlaping. However, frequent remappings can introduce unwanted numerical diffusion which in many cases contradicts 
the benefit of using low-diffusion particle schemes. Resulting numerical diffusion from the remappings can be reduced by use of 
high order adaptive schemes; see, e.g., \cite{Bergdorf.Koumoutsakos.2006.MMS,Wang.Miller.Colella.2011.jsc}.
Other methods to reduce the noise have also been studied, such as wavelet-based denoising techniques; see e.g., 
\cite{Chehab.al.2005.irma, Terzic.Pogorelov.2005.nyac, Gassama.al.2007.esaim}.

In this article, we present a new particle scheme where in addition to pushing the particle centers along their trajectories, one
updates the particle shapes through the use of local linear transformations to better follow the local shear and rotation flows in phase space.
As in the FSL scheme, the method is purely deterministic, and to prevent particles from being arbitrarily stretched, the particles need to be remapped 
periodically. However, significantly lower remapping frequencies are needed in practice, which results in higher accuracy and less numerical diffusion. 
On a theoretical level this advantage is supported by the fact that for transport problems with prescribed characteristic flow, 
the linearly transformed particle solutions are shown to converge in the uniform norm as $h$ tends to 0, {\em without} any remappings; 
see \cite{Campos-Pinto.2012.sub}.

Deforming the particles is not a new idea. For instance, our scheme can be viewed as a variation on Hou's formal vortex method 
\cite{Hou.1990.sinum} where the particles are deformed through a global mapping. In our method, each particle is transported by the linearized 
flow around its trajectory. Still on a formal level this approach coincides with a method presented by Cohen 
and Perthame \cite{Cohen.Perthame.2000.sima} who established its first-order convergence, but 
they did not provide a numerical scheme to compute the deformation matrices. In the context of plasma simulation, 
an important class of deformed particle methods is offered by the Complex Particle Kinetic (CPK) schemes introduced by Bateson and Hewett 
\cite{Bateson.Hewett.1998.jcp, Hewett.2003.jcp}. In the CPK method, in addition to having the Gaussian shape 
of the particles transformed by the local shearing of the flow, the particles can also be fragmented to probe for emerging features, 
and merged where fine particles are no longer needed. 
Another exciting method is the Cloud in Mesh (CM) scheme of Alard and Colombi \cite{Alard.Colombi.2005.ras} 
that has been brought to our attention after the writing of this article. 
CM particles have Gaussian shapes as in the CPK method, and they are deformed by local linearizations of the force field,
in a manner similar to ours. Moreover, in \cite{Alard.Colombi.2005.ras} the authors also describe locally refined algorithms for charge deposition
and phase-space sampling, based on adaptive refinement trees. When mature, our Linearly-Transformed PIC (LTPIC) scheme will incorporate some 
of the multilevel refinement features presented in \cite{Campos-Pinto.2012.sub}, and the resulting adaptive scheme should be compared
with the CPK and CM methods to determine which classes of problems best fit each method.

The outline is as follows. In Section~\ref{sec. method} we describe the LTPIC scheme for 1d1v electrostatic plasmas:
in Section~\ref{sec. numsols} we introduce the main notations and present the general form of the numerical solutions. 
Although a wide range of shape-functions is supported by our approach, to illustrate the method we review in Section~\ref{sec. Ah} 
one deterministic algorithm for initialization and remapping of B-spline particles, and we provide a correction scheme to make 
the remappings conservative. In Section~\ref{sec. transport} we then
define a particle transport scheme that transforms the shapes in phase space, and is solely based on pointwise evaluations
of the (forward) numerical flow. 
Two specific charge deposition schemes are then presented in Section~\ref{sec. field-solver}, and a leap-frog
time advance scheme implementing the method is described in Section~\ref{sec. march}. 
Numerical results involving several standard benchmark tests and a halo problem associated with an initially mismatched thermal distribution are 
presented in Section~\ref{sec. results}. The results obtained demonstrate some advantages of our method compared to the classical PIC or
FSL approaches. They also indicate that for numerical choices involving similar computational effort, 
the LTPIC method is capable of accuracy comparable to or greater than state-of-the-art high-resolution Vlasov schemes.

\section{The numerical method}
\label{sec. method}

To describe our method we may consider the normalized 1d1v Vlasov-Poisson equation 
\begin{equation}
\left\{ \partial_t + v \partial_x + E(x,t) \partial_v \right\} f(t,z) = 0
\quad \text{ with } ~~ t \ge 0, ~~  z = (x,v) \in \RR^2, 
\label{norm-Vlasov}
\end{equation}
which models the evolution of simplified plasmas and sheet beams; see for instance Ref. \cite{Lund.2011.prstab}. 
Here, $x$ and $v$ are dimensionless positions and velocities and $E$ is a dimensionless electric field satisfying
\begin{equation}
\label{E-abs}
\partial_x E(t,x) = \int_\RR  f(t,x,v) \rmd v - n_e,
\end{equation}
where $n_e$ is the density of a uniform neutralizing background cloud.

\subsection{Structure of the numerical solutions}
\label{sec. numsols}

\begin{figure}[ht!]
\medskip
\hspace{1.5cm} \includegraphics[width=0.45\textwidth]{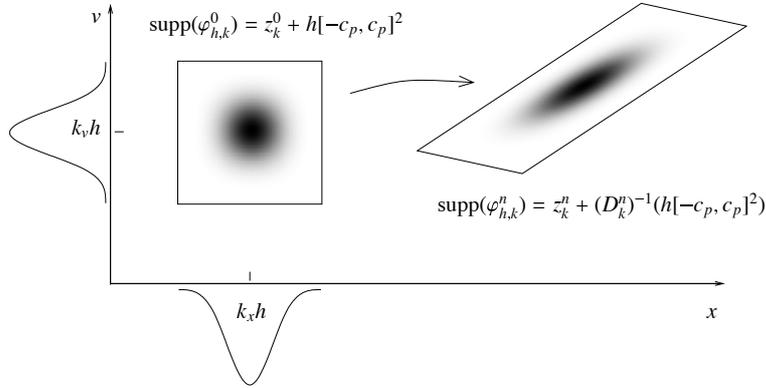}
\caption{Structured particles (left) are defined at initialization and remapping steps, as tensor-product B-splines centered on regular nodes $z^0_k = hk$,
    $k \in \ZZ^2$. Unstructured particles (right) are obtained by pushing the particle centers along their trajectories $z^n_k$ and transforming their 
    shapes with a matrix $D^n_k$ representing the local Jacobian of the characteristic flow.}
\label{fig. particles}
\end{figure}

As in standard particle methods, we represent the phase-space density $f$ with weighted collections of finite-size particles 
(index $k$) which are pushed along their trajectories $z^n_k$ corresponding to the discrete times $t_n = n\Dt$, $n = 0, 1 \ldots, N_t$.
However, in our method the particles also have their shape transformed to better represent the local shear and rotation 
flows in phase space, as illustrated in Figure~\ref{fig. particles}. The particles can either be structured or unstructured. 
The first case corresponds to the initialization and remapping steps, where particles are defined as 
tensor-product B-splines and centered on regular nodes 
\begin{equation}
\label{regnodes}
z^0_k = (x^0_k, v^0_k) \equiv hk \qquad \text{ with } \qquad  k \in \ZZ^2 = \{\ldots, -1, 0, 1, \ldots \}^2.
\end{equation}
Specifically, the univariate (i.e., one-dimensional) 
centered B-spline $\cB_p$ is recursively defined as the piecewise polynomial of degree $p$ satisfying
\begin{equation}
\label{Bp}
\cB_0(x) \equiv \begin{cases} 1, & -\tfrac 12 \le x \le \tfrac 12 \\ 0 & \text{otherwise} \end{cases} \quad \text{ and } \quad
\cB_p(x) \equiv \int_{x-\tfrac 12}^{x+\tfrac 12} \cB_{p-1}(\tilde x)\rmd \tilde x \quad \text{ for } \quad p \ge 1.
\end{equation}
Thus $\cB_1(x) = \max\{1-\abs{x},0\}$ is the traditional ``hat-function'', $\cB_3$ is the well-known cubic B-spline supported on
$[-2,2]$, and so on, see e.g. Ref.~\cite{deBoor.B.1978}. The fundamental shape function is then defined on the two-dimensional phase space as a tensor product
\begin{equation}
\label{vp}
\vp(z) \equiv \cB_p(x)\cB_p(v) \quad \text{ with support } ~~ \supp(\vp) = [-c_p,c_p]^2, \quad
c_p \equiv \tfrac{p+1}{2},
\end{equation}
from which we derive a normalized ($\int \vp_h = 1$), grid-scaled shape function
$
\vp_h(z) \equiv h^{-2} \vp(h^{-1}z). 
$ 
Structured particles are then defined as translated versions of the latter,
\begin{equation}
\label{struct-part}
\vp^0_{h,k}(z) \equiv \vp_h(z - z^0_k) = h^{-2} \vp(h^{-1}z - k), \qquad k \in \ZZ^2.
\end{equation}
When transported by our method, particles become unstructured in the sense that their centers $z^n_k$ leave the
nodes of the structured phase-space grid and their shapes are linearly transformed. 
That is, the positions of different parts of the ``cloud'' associated with a single particle advance with their own peculiar velocities, 
the velocities advance with their own peculiar accelerations, and the cloud distorts, but the distortion is constrained to be linear.
Generic particles are then characterized by the $2 \times 2$ deformation matrices $D^n_k$ 
(initialized with $D^0_k \equiv \big(\begin{smallmatrix} 1 & 0 \\ 0 & 1 \end{smallmatrix}\big)$) 
which determine the linear transformation of their shape, and numerical solutions take the form
\begin{equation}
\label{f-parts}
f_h^n(z) = \sum_{k \in \ZZ^2} w_k^n \vp_{h,k}^n(z)
\quad \text{ with } \quad
\vp_{h,k}^n(z) \equiv \vp_{h}(D^n_k(z-z_k^n)).
\end{equation}
In Section~\ref{sec. Ah} we shall describe a structured particle approximation operator 
$$
A_h : ~~ f( z) ~\mapsto~ \sum_{k \in \ZZ^2} w_k(f) \vp_{h}( z -  z^0_k)
$$ 
acting on a generic density $f$, and in Section~\ref{sec. transport} and \ref{sec. march} we will construct a time-dependent transport operator
$$
T_h^n : ~~ \sum_{k \in \ZZ^2} w^n_k \vp_{h}(D^n_k(z-z_k^n)) ~\mapsto~ \sum_{k \in \ZZ^2} w^{n+1}_k \vp_{h}(D^{n+1}_k(z-z_k^{n+1})).
$$ 
The deformation matrices will be transformed with an area preserving scheme ($\det(D^{n+1}_k) = \det(D^n_k) = 1$), 
so that the charges carried by the particles read (up to a constant factor)
$$
\int w^n_k \vp^n_{h,k}(z) \rmd z = 
\int w^n_k \vp_{h}(D^n_k(z-z_k^n))\rmd z = 
w^n_k \int \vp_{h}(\tilde {z}) \rmd \tilde {z} = w^n_k.
$$ 
In particular, the particle weights will not be modified by our transport operator.

To prevent the deformed particles from being arbitrarily stretched in one direction, we have chosen to periodically remap them 
onto the regular grid \eqref{regnodes}. Note that, as in Semi-Lagrangian methods, the remappings are likely to introduce 
unwanted numerical diffusion. 
However, since our particle method is mathematically proven to converge without remappings \cite{Campos-Pinto.2012.sub},
we expect the optimal remapping frequencies to be significantly lower than with Semi-Lagrangian schemes.
This point will be numerically demonstrated in Section~\ref{sec. results}.

The global structure of the scheme is then formulated as follows. First, a collection of weighted particles is initialized with
$$
f_h^{0} \equiv A_h f(t=0),
$$
then for $n = 0, \ldots, N_t-1$, we compute
\begin{equation}
\label{struct-scheme}
f_h^{n+1} \equiv T_h^n f_h^{n,0} 
\qquad \text{ where } \qquad
f_h^{n,0} \equiv 
\begin{cases}
    ~ A_h f_h^n ~  & \text{if }~ n > 0 ~\text{ and } \mspace{-10mu} \mod(n\Dt,\Dtrec) = 0,
    \\
    ~ f_h^n   & \text{otherwise}
\end{cases}
\end{equation}
to evolve the distribution with a given remapping period $\Dtrec$.

\subsection{Initialization and remappings}
\label{sec. Ah}

Since the fundamental shape function $\vp_h$ is a B-spline with the same scale as the spacing $h$ of the regular grid \eqref{regnodes},
arbitrary polynomials with coordinate degree less or equal to $p$ can be obtained by 
linear combinations of structured particles \eqref{struct-part} derived by shifting $\vp_h$ on the grid,  see, e.g., \cite{deBoor.B.1978}.
Therefore, to initialize and remap the particle densities we can use existing high order approximation schemes that rely on that property.
One attractive method is given by the quasi-interpolation schemes described in Refs.~\cite{Chui.Diamond.1990.nm} and \cite{Unser.Daubechies.1997.ieee_tsp}.
Such schemes pass through data points when they are described by polynomial target functions $f$ of a certain degree,
and they have the advantage of computing high order B-spline approximants from local evaluations of the target function, 
unlike standard spline interpolation which requires solving a global system.
Thus, in the univariate case the approximation $A^{(1d)}_h$ takes the form
\begin{equation}
\label{Ah-1d}
A^{(1d)}_h : f(x) \mapsto \sum_{k \in \ZZ} w_{k}(f) \vp_{h}(x-hk)
\quad \text { with weights } ~~ 
w_{k}(f) \equiv h \sum_{\abs{l} \le m_p} a_{l} \, f(h(k+l)).
\end{equation}
Here we have denoted $\vp_h(x) \equiv h^{-1} \cB_p(h^{-1}x)$, 
and the $a_l=a_{-l}$ are symmetric coefficients defined in such a way that $A^{(1d)}_h f = f$ for any 
$f(x) = \alpha_0 +\cdots +\alpha_p x^p$. 
Specifically, they can be computed with the algorithm from Ref.~\cite[Section 6]{Chui.Diamond.1990.nm}. For the first orders we find
\begin{itemize}
\item
$m_p = 0$ and $a_0 = 1$ for $p=1$,
\item
$m_p = 1$ and $(a_0,a_1) = (\frac 86, -\frac 16)$ for $p=3$,
\item
$m_p = 4$ and $(a_0,a_1,a_2,a_3,a_4) = (\frac{503}{288}, -\frac{1469}{3600},\frac{7}{225},\frac{13}{3600},\frac{1}{14400})$ for $p=5$.
\end{itemize}
In the bivariate case we can ``tensorize'' the above scheme, as it is easily checked that the operator
\begin{equation}
\label{Ah}
A_h : f(z) \mapsto \sum_{k \in \ZZ^2} w_{k}(f) \vp_{h}(z - z^0_k)
~~  \text { with } ~~ 
w_{k}(f) \equiv h^2 \sum_{\norm{l}_\infty \le m_p} a_{l} \, f(z^0_{k+l}) 
, \quad 
a_{l} \equiv a_{l_x} a_{l_v}
\end{equation} 
reproducts any polynomial of coordinate degree less than or equal to $p$ (here, $\norm{l}_\infty \equiv \max\{\abs{l_x},\abs{l_v}\}$).
With standard arguments one can then show that the resulting approximation error converges as $h^{p+1}$ for smooth functions $f$, 
see e.g., Ref.~\cite{Campos-Pinto.2012.sub}.

When the above scheme is used to remap a generic particle density $f^n_h$ of the form \eqref{f-parts}, the mass of the resulting approximation is 
(setting $a_l \equiv 0$ for $\norm{l}_\infty > m_p$ and using $\sum_{l \in \ZZ^2} a_{l} = 1$)
\begin{align*}
\int A_h f^n_h(z) \rmd z 
&= \sum_{k \in \ZZ^2} w_k(f^n_h)  \int \vp_{h}(z-z^0_k) \rmd z  = \sum_{k \in \ZZ^2} w_k(f^n_h) 
= h^2 \!\! \sum_{k,l,k' \in \ZZ^2} a_{l} w^n_{k'} \vp^n_{h,k'}(z^0_{k+l})
= \sum_{k' \in \ZZ^2} \hat w^n_{k'}
\end{align*}
where $\hat w^n_{k'} =  h^2 w^n_{k'} \sum_{i \in \ZZ^2} \vp^n_{h,k'}(z^0_{i})$ represents the charge deposited by the particle 
$k'$ in the remapping process. Due to the shape transformations, this quantity generally differs from the original $w^n_{k'}$ 
(indeed it vanishes if the support of $\vp^n_{h,k'}$ misses the grid $h\ZZ^2$). This shows that the quasi-interpolation 
is not conservative, but a locally conservative correction is easily implemented by depositing
the local error $w^n_{k'}-\hat w^n_{k'}$ with a PIC-like method, which results in defining
\begin{equation}
\label{cons-correct}
w^{n,0}_k = w_k(f^n_h) + \sum_{k'\in\ZZ^2} (w^n_{k'}-\hat w^n_{k'}) h^2 \vp_h(z^0_{k}-z^n_{k'}).
\end{equation}
Note that in practice the deposited fractions $\hat w^n_{k'}$ can be evaluated by summing over $i \in \ZZ^2$ the values 
$\vp^n_{h,k'}(z^0_{i})$ involved in the quasi-interpolation scheme.

\subsection{Particle transport with linear transformations}
\label{sec. transport}

The LTPIC scheme is based on a LTP (linearly-transformed particle) transport operator $T_h[\cF]$ that transforms the particles through
local linearizations of a given characteristic flow $\cF $. In this section we present the LTP transport operator in this general setting,
and reserve for the following sections the description of the numerical flow. Schematically, one could indeed decompose the transport operator 
$T_h^n$ appearing in \eqref{struct-scheme} as follows.

\begin{itemize}
\item[1.]
From the density carried by the particles one computes a numerical flow
$$
f^n_h \mapsto \cF^n_h
$$
that approximates the (exact) characteristic flow of the Vlasov equation~\eqref{norm-Vlasov} over one time step $[t_n,t_{n+1}]$. 
Namely, the mapping $\cF^n_\ex: (x,v) \mapsto (X,V)(t_{n+1})$ that associates any phase-space point to 
the advanced-time point of its corresponding trajectory defined by
\begin{equation}
\label{ode-n}
\left\{
    \begin{array}{l}
    X'(t) = V(t)
    \\ \noalign{\medskip}
    V'(t) = E(t,X(t))
    \end{array}
\right.
\quad \text{with initial data } \quad (X,V)(t_n) = (x,v).
\end{equation}
\item[2.] The particles are transported by the associated LTP transport operator,
$$
T_h^n = T_h[\cF^n_h].
$$ 
\end{itemize}

In Section~\ref{sec. march} we will derive a leap-frog version of this approach, involving two intermediate flows $\cF^{n,0}_h$ and $\cF^{n,1}_h$
such that $\cF^{n,1}_h \circ \cF^{n,0}_h \approx \cF^n_\ex$. 
The LTP operator $T_h[\cdot]$ will then be applied twice per time step, i.e., we will define $T_h^n \equiv T_h[\cF^{n,1}_h]T_h[\cF^{n,0}_h]$.
However, to simplify our presentation we shall consider in the remainder of this section that we are given a single flow 
$\cF^n_h \approx \cF^n_\ex$.

Applied to a generic particle $\vp^n_{h,k}$ with deformation matrix $D^n_k$, the LTP transport operator is 
\begin{equation}
\label{ltp}
T_h[\cF_h^n]: ~~ \vp^n_{h,k} \equiv \vp_h(D^n_k(\cdot-z^n_k)) ~~ \mapsto ~~ \vp^{n+1}_{h,k}\equiv \vp_h( D^{n+1}_k(\cdot-z^{n+1}_k) )
~~~ \text{ with } ~~ 
\begin{cases} 
~ z^{n+1}_{k} \equiv \cF_h^n(z^n_{k}) 
\\
~ D^{n+1}_{k} \equiv D^n_{k}(J^n_{k})^{-1}
\end{cases}
\end{equation}
where $J^n_k$ is a matrix representing the Jacobian of the flow at $z^n_k$, defined as follows. 
An approximated Jacobian matrix $\tilde J^{n}_{k}$ is first defined with a centered 
finite difference scheme,
\begin{equation}
\label{fd}
(\tilde J^{n}_{k})_{i,j} \equiv (2 h)^{-1}\big[(\cF_h^n)_i(z^n_{k}+h \bse_j)-(\cF_h^n)_i(z^n_{k}-h \bse_j)\big]
\approx \partial_j (\cF_\ex^n)_i(z^n_{k})
\quad \text{ for } 
~~~  1 \le i,j \le 2,
\end{equation}
where we have denoted $\bse_j = (\delta_{i,j})_{1 \le i \le 2}$. Here $h$ is the grid spacing of the remapping grid, but a different
spacing could be used as well. Next we observe that, while the exact flow has a Jacobian with uniform determinant equal to 1, there 
is no reason why this should be true for the finite difference approximation \eqref{fd}. To obtain a conservative transport scheme 
(in the sense that $\int \vp^{n+1}_{h,k}(z) \rmd z = \int \vp^n_{h,k}(z) \rmd z $), we then define $J^n_k$ as 
\begin{equation}
\label{consJacob}
J^{n}_{k} \equiv \det(\tilde J^n_{k})^{-\frac 12} {\tilde J}^{n}_{k}.
\end{equation}

To justify the above approximations inherent in Eqs.~\eqref{ltp}-\eqref{consJacob} we temporarily assume that we 
can apply the exact flow $\cF^n_\ex$. Pushing a fixed-shape particle as in standard PIC schemes gives
\begin{equation}
\label{translate}
T_{\rm PIC}[\cF^n_\ex]: \vp_{h}(z-z^n_k) \mapsto \vp_{h}(z-z^{n+1}_k) \qquad \text{ with } \qquad z^{n+1}_k \equiv \cF_\ex^n(z^n_k).
\end{equation}
Now, since \eqref{ode-n} is reversible, the exact transport of 
an arbitrary phase-space density $f(t_n,z)$ over the time step is $f(t_{n+1},z) = f(t_n,(\cF_\ex^n)^{-1}(z))$. In particular, 
for the particle $\vp_{h}(z-z^n_k)$ we have
$$
T_\ex[\cF^n_\ex]: \vp_{h}(z-z^n_k) \mapsto \vp_{h}((\cF_\ex^n)^{-1}(z)-z^n_k),
$$ 
to which \eqref{translate} can be seen as the lowest order approximation. Enhanced accuracy is obtained by a first order expansion around $z^{n}_k$: 
writing $J_{\cF}(z) = \big(\partial_j \cF_i(z)\big)_{1\le i,j\le 2}$ the Jacobian matrix 
associated with an arbitrary flow $\cF: \RR^2 \to \RR^2$, we let
$$
\cF^n_{\ex, z^n_k}(z) \equiv \cF_\ex^n(z^{n}_k) + J_{\cF_\ex^n}(z^{n}_k)(z - z^{n}_k)
$$
denote the linearized flow around $z^n_k$. We then define a ``formal'' LTP transport operator 
as the {\em exact} tranport corresponding to this linearized flow, namely
\begin{equation}
\label{ltp-def}
T_{\rm LTP}[\cF^n_\ex] \equiv T_{\rm ex}[\cF^n_{\ex, z^n_k}] \qquad \text{ for the particle associated to $z^n_k$.}
\end{equation}
Applied to a structured particle, we observe that it reads
\begin{equation}
\label{ltp-cont}
T_{\rm LTP}[\cF^n_\ex]: \vp_{h}(z-z^n_k) \mapsto \vp_{h}((J^n_k)^{-1}(z-z^{n+1}_k)) \quad \text{ with } \quad z^{n+1}_k \equiv \cF_\ex^n(z^n_k),
\quad J^n_k \equiv J_{\cF_\ex^n}(z^{n}_k).
\end{equation}
Now, replacing the flow $\cF^n_\ex$ by its numerical approximation $\cF^n_h$ and using a finite difference scheme for the forward Jacobian leads then
to the practical LTP transport operator as defined by Eqs.~\eqref{ltp}-\eqref{consJacob}.
A rigorous error analysis of this procedure is presented in \cite{Campos-Pinto.2012.sub}. This error analysis demonstrates the global convergence of both 
the discrete \eqref{ltp} and continuous \eqref{ltp-cont} schemes (without remappings) in the uniform norm, as $h$ tends to 0, provided that the exact flow 
is approximated with sufficient accuracy.

\subsection{Conservative charge deposition schemes for linearly transformed particles}
\label{sec. field-solver}

To complete the LTPIC scheme we now describe how to compute the field from the linearly transported particles.
For this purpose we equip the (1d) physical space with regular nodes
$$
x_i = ih' \qquad \text{ with } \qquad i \in \ZZ,
$$ 
and as in Section~\ref{sec. Ah} we let
$\vp_{h'}(x) \equiv \tfrac{1}{h'}\cB_p(\tfrac{x}{h'})$ denote the
scaled B-spline of mass 1 in the physical variable.
Following the standard approach \cite{Hockney.Eastwood.1988.tf} we represent the charge density on the grid with
\begin{equation}
\label{rho-num}
\rho^n_{h'}(x) \equiv \sum_{i \in \ZZ} \rho^n_i \vp_{h'}(x-x_i),
\end{equation}
and solve the Poisson equation on the same grid. Specifically we represent the electric field with
\begin{equation}
\label{E-num}
E^n_{h'}(x) \equiv \sum_{i \in \ZZ} h' E^n_i \vp_{h'}(x-x_i)
\end{equation}
with coefficients computed by a centered finite difference scheme, such as
$$
E_i^n = -\frac{\phi^n_{i+1}-\phi_{i-1}}{2h'},
\qquad
-\frac{\phi^n_{i+1}-2\phi_{i}+\phi_{i-1}}{(h')^2} = \rho^n_h(x_i).
$$
Here the different normalization in \eqref{rho-num} and \eqref{E-num} allow the coefficients $E^n_i$ to be on the order of the point values $E^n_{h'}(x_i)$,
whereas the coefficients $\rho^n_i$ are on the order of the local charges 
$h'\rho^n_{h'}(x_i)$,
consistent with standard notations, see \cite[Chapter 5]{Hockney.Eastwood.1988.tf}.
To compute the local charges $\rho^n_i$, several methods can be considered.
\begin{enumerate}
\item[1.]
    In the simplest approach the particles are seen as point particles and they deposit their charges in 
    a way similar to PIC schemes, i.e.,
    \begin{equation}
    \label{pic-deposition}
    \rho^n_i \equiv  h'\sum_{k \in \ZZ^2} w^n_k \vp_{h'}(x^n_k-x_i).
    \end{equation}
    Note that in this case the numerical scheme still differs from a standard PIC method, because the particles are 
    periodically remapped on the regular grid with a smoothing effect.
\item[2.]
    To take into account the shape of the particles, specific deposition schemes can be used instead. 
    They rely on an intermediate charge density defined as the exact integral of $f^n_h$ along the velocity variable,
    $$
    \tilde \rho^n_h(x) = \sum_{k \in \ZZ^2} \tilde \rho^n_{h,k}(x) 
    \equiv \sum_{k \in \ZZ^2} w^n_k \int_\RR\vp^n_{h,k}(x,v) \rmd v = \int_\RR f^n_h(x,v) \rmd v, 
    $$
    and on the use of univariate quasi-interpolation \eqref{Ah-1d} to compute the local charges,
    \begin{equation}
    \label{qi-deposition}
    \rho^n_{h'} \equiv A^{(1d)}_{h'} \tilde \rho^n_{h}
    \qquad \text{ i.e., } \qquad
    \rho^{n}_i \equiv h' \sum_{\abs{l}\le m_{p}} a_{l} \, \tilde \rho^n_{h}(x_{i+l})
    = h' \, \sum_{k \in \ZZ^2} \sum_{\abs{l}\le m_{p}} a_{l} \, \tilde \rho^n_{h,k}(x_{i+l}).
    \end{equation}
    Note that a correction similar to \eqref{cons-correct} can be used here to make the deposition conservative,
    in the sense that $\int \rho^n_h(x) \rmd x= \int f^n_h(z) \rmd z$.
    In the above deposition formula \eqref{qi-deposition}, we observe that the evaluation of the ``integrated particles'' 
    $\tilde \rho^n_{h,k}$ is not straightforward, due to the linear transformation of their shape. 
    To compute them we have considered two methods.
    \begin{enumerate}
    \item[2.a.]
        In the first method (see Algorithm~\ref{algo. gauss-dep} and Figure~\ref{fig. gauss-dep}) 
        we use a Gaussian quadrature $\hat \rho^n_{h,k}(x_{i+l})$ to evaluate each velocity integral $\tilde \rho^n_{h,k}(x_{i+l})$. 
        In order to be accurate this approximation requires a few quadrature intervals fitted to the 
        particle support (projected along the velocity variable) and a few Gauss points per interval. This makes it hard to apply in higher dimensions.
    \item[2.b.]
        In the second method (see Algorithm~\ref{algo. moment-dep} and Figure~\ref{fig. moment-dep})
        we simply replace each integrated particle by a univariate weighted B-spline sharing the same first 3 moments. 
        Specifically, we approximate $\tilde \rho^n_{h,k}$ as
        \begin{equation}
        \label{approx_B}
        \tilde \rho^n_{h,k}(x) \approx \hat \rho^n_{h,k}(x) \equiv
        \frac {w^n_{h,k}}{\lambda_{h,k}^n}\cB_p\Big(\frac {x-x^n_k}{\lambda_{h,k}^n}\Big)
        ~~~ \text{ with } ~~~
        \lambda_{h,k}^n 
        \equiv h\sqrt{((D^n_k)_{2,2})^2 + ((D^n_k)_{1,2})^2},
        \end{equation}
        so that $\int_\RR x^m \hat \rho^n_{h,k}(x) \rmd x = \int_\RR x^m \tilde \rho^n_{h,k}(x) \rmd x$ holds for $m = 0, 1, 2$.
        The resulting implementation is much simpler, and the extension to higher dimensions is straightforward.
    \end{enumerate}        
\end{enumerate}

\begin{algo}[Charge deposition with Gaussian quadrature]  \label{algo. gauss-dep}
Let $\NGpieces$ and $\NGpoints$ denote the prescribed number of quadrature intervals and Gauss points per interval in the $v$ dimension, per particle.
\begin{itemize}
\item[1.] Loop over every active particle $\vp^n_{h,k}$, i.e., over $k \in \ZZ^2$ such that $w^n_{k} \neq 0$. For conciseness we denote the 
deformation matrix by 
$D \equiv D^n_{k}$, and we observe that the particle support is
\begin{equation}
\label{supp-part}
\supp(\vp^n_{h,k}) = (x^n_k, v^n_k) + D^{-1}(h[-c_p,c_p]^2) = \big\{ (x,v) : \norm{ D(x-x^n_k,v-v^n_k) }_\infty \le h c_p \big\}.
\end{equation}
\item[2.] Determine the $x$-projection $[x^n_k - h_x, x^n_k + h_x]$ of the support \eqref{supp-part} of $\vp^n_{h,k}$, i.e., set
\begin{equation} \label{hx-gauss}
h_x = h_x(h,k,n) 
 \equiv \tfrac 12 \diam\big(\{x^n_k + (D^{-1}r)_x : \norm{r}_\infty \le h c_p \}\big) 
 =  h c_p (\abs{D_{2,2}} + \abs{D_{1,2}}),
\end{equation}
where we have used $D^{-1} = \left(\begin{smallmatrix} D_{2,2} & -D_{1,2} \\ -D_{2,1} & D_{1,1} \end{smallmatrix}\right)$ since $\det(D) = 1$.
\item[3.]
\label{chdep-update}
Loop over the non-vanishing point values $\tilde \rho^n_{h,k}(x_{l'})$, namely over $l' \in \ZZ $ such that
\begin{equation}
\label{loop-l'}
\abs{x^n_k - x_{l'}} <  h_x, 
~~ \text{ i.e.,  } ~~
l' \in \left\{ \, \Big\lfloor \tfrac{1}{h'}(x^n_k- h_x)\Big\rfloor + 1, ~ \cdots ~ ,  
\Big\lceil \tfrac{1}{h'}(x^n_k+h_x)\Big\rceil - 1 \, \right\}.
\end{equation}
Then define $\hat \rho^n_{h,k}(x_{l'}) \approx \tilde \rho^n_{h,k}(x_{l'})$ with $\NGpieces$ Gauss quadrature formulas using $\NGpoints$ points
in the $v$ dimension: from \eqref{supp-part} the interval 
$
[v^-(l'),v^+(l')] \equiv \supp(v\mapsto\vp^n_{h,k}(x_{l'},v))
$ 
is given by
$$
\begin{cases}
~ v^-(l') \equiv v^n_k + \max_{i =1,2}\{(D_{i,2})^{-1}\big(D_{i,1}(x^n_k-x_{l'})-h c_p\big)\}
\\
~ v^+(l') \equiv v^n_k + \min_{i =1,2}\{(D_{i,2})^{-1}\big(D_{i,1}(x^n_k-x_{l'})+h c_p\big)\},
\end{cases}
$$ 
so that we may compute 
\begin{equation}
\label{quad-rho}
\begin{split}
\hat \rho^n_{h,k}(x_{l'}) 
& \equiv  w^n_{k} \Delta v \sum_{m'=0}^{\NGpieces -1} \sum_{m=1}^{\NGpoints} \lambda^G_m \vp^n_{h,k}\big(x_{l'},v^-(l') + (m'+\nu^G_m) \Delta v\big)
\\
&\approx  w^n_{k}\int_{v^-(l')}^{v^+(l')}  \vp^n_{h,k}(x_{l'},v) \rmd v = \tilde \rho^n_{h,k}(x_{l'}).
\end{split}
\end{equation}
Here, $\Delta v \equiv (v^+(l')-v^-(l'))/\NGpieces $, and $\lambda_i^G, \nu^G_i$ are the Gauss weights and nodes corresponding to the interval $[0,1]$, e.g., 
\begin{align*}
\text{ for } \NGpoints = 1: \quad   & \nu^G_1 = \tfrac 12, \quad \lambda_1^G = 1,
\\
\text{ for } \NGpoints = 2: \quad   & \nu^G_i = \tfrac 12 (1 \pm \tfrac{1}{\sqrt 3}), \quad \lambda_i^G = \tfrac 12, 
\quad i = 1,2,
\\
\text{ for } \NGpoints = 3: \quad   & 
\begin{cases}
~ \nu^G_2 = \tfrac 12,  \quad 
\lambda_2^G = \tfrac 49  &
\\
~ \nu^G_i = \tfrac 12 (1 \pm \tfrac{\sqrt {15}}{5}),  \quad 
\lambda_i^G = \tfrac {5}{18}, \quad &i = 1,3.
\end{cases}
\end{align*}
Finally update the appropriate weights (initialized to 0) consistent with \eqref{qi-deposition}, by setting
\begin{equation}
\label{update-wr}
\rho^{n}_{i}  \equiv \rho^{n}_{i}
    + h' a_{l'-i} \hat \rho^n_{h,k}(x_{l'}) \qquad
\text{ for } \quad i = l'-l = l'-m_{p}, ~ \ldots, ~l'+m_{p}~.
\end{equation}
\end{itemize}
\end{algo}

\begin{algo}[Charge deposition with a moment method]  \label{algo. moment-dep}
$~$
\begin{itemize}
\item[1.] Loop over the active particles $\vp^n_{h,k}$, i.e., over $k \in \ZZ^2$ such that $w^n_{k} \neq 0$.
\item[2.] Determine the support $[x^n_k - h_x, x^n_k + h_x]$ of $\hat \rho^n_{h,k}$, i.e., set
$h_x = h_x(h,k,n) \equiv \lambda_{h,k}^n c_p$
with $\lambda_{h,k}^n$ defined as in \eqref{approx_B}.

\item[3.] Deposit the approximated charge contributions as above: loop over $l' \in \ZZ$ satisfying \eqref{loop-l'}, and update 
the weights as in \eqref{update-wr}, with the explicit expression \eqref{approx_B} for $\hat \rho^n_{h,k}$.
\end{itemize}
\end{algo}

\begin{figure}[ht!]
\begin{center}
\input{gauss-dep.pstex_t}
\caption{In the Gauss deposition scheme described in Algorithm~\ref{algo. gauss-dep}, the charge density 
$\tilde \rho^n_{h,k}(x) \equiv w^n_{k}\int_\RR\vp^n_{h,k}(x,v) \rmd v$ associated to a linearly-transformed particle 
is deposited with a quasi-interpolation scheme where the required point values are estimated with Gaussian quadrature along $v$ slices.}
\label{fig. gauss-dep}
\end{center}
\end{figure}
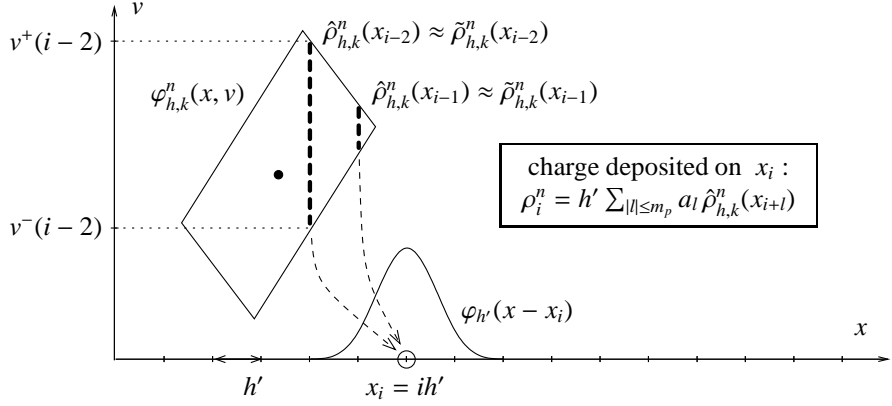

\begin{figure}[ht!]
\begin{center}
\input{moment-dep.pstex_t}
\caption{In the moment deposition scheme described in Algorithm~\ref{algo. moment-dep}, no numerical integration is needed.
Instead, the charge density 
$\tilde \rho^n_{h,k}(x) \equiv w^n_{k}\int_\RR\vp^n_{h,k}(x,v) \rmd v$ is replaced by a B-spline 
$\hat \rho^n_{h,k}(x)$ that shares its first 3 moments, and the latter is deposited with the quasi-interpolation scheme
(compare with Figure~\ref{fig. gauss-dep}).}
\label{fig. moment-dep}
\end{center}
\end{figure}
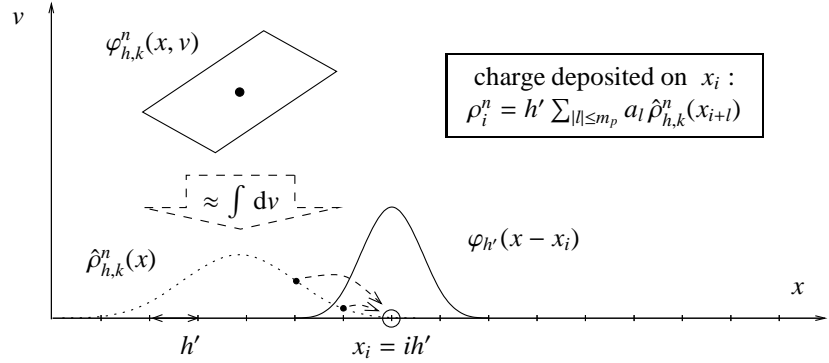

The good news is that in practice it does not seem necessary to resort to accurate piecewise Gauss quadratures.
Indeed, in most of the numerical tests presented in Section~\ref{sec. results} the results obtained with the simple moment 
deposition scheme (displayed) were compared with simulations using a Gauss deposition scheme with $\NGpieces = 4$ 
quadrature intervals and $\NGpoints = 3$ Gauss points per intervals, and the differences were hardly visible.
Maybe more surprisingly, we also compared these results with simulations using the much simpler, PIC-like point deposition scheme,
and again the differences were hardly visible. This suggests that in many cases the oscillatory representation of the density 
(as seen by the field solver) still yields an electric field that is able to drive accurate dynamics. On one hand, we should not be overly surprised by such a fact,
indeed it is routinely observed in PIC (and FSL) simulations. On the other hand, we could expect that in some cases an
accurate resolution of the field requires a smooth representation of the density as seen by the solver, especially
when high order solvers are used. This is an important question that is familiar in the finite element community,
and that shall be addressed in future research.

\subsection{Time marching scheme}
\label{sec. march}

Equipped with the LTP transport operator $T_h[\cF]$ defined in \eqref{ltp} for a generic flow $\cF$, and with the field solver
described in Section~\ref{sec. field-solver}, we are now in position to specify the numerical transport involved in the scheme \eqref{struct-scheme}, namely
the operator
$$
T_h^n: f^{n,0}_h \mapsto f^{n+1}_h.
$$ 
To this end we consider a standard leap-frog time discretization.
We first transport
$$
f^{n,1}_h \equiv T_h[\cF_h^{n,0}]f^{n,0}_h 
\quad 
\text{ where } \quad \cF_h^{n,0}(x,v) \equiv \big(x + \tfrac {\Dt}{2} v, v \big),
$$
then compute an intermediate electric field 
$$
f^{n,1}_h \mapsto E^{n,1}_{h'}
$$
using finite differences as described in Section~\ref{sec. field-solver}, and finally complete the time step with
$$
f^{n+1}_h \equiv T_h[\cF_h^{n,1}]f^{n,1}_h 
\quad 
\text{ where } \quad \cF_h^{n,1}(x,v) \equiv \big(x + \tfrac {\Dt}{2} \tilde v, 
                                            \tilde v \equiv v + \Dt E^{n,1}_{h'}(x) \big).
$$

\section{Numerical simulations}
\label{sec. results}


In this section we apply our LTPIC scheme~\eqref{struct-scheme} on a series of 1d1v test cases and compare the resulting solutions with classical 
PIC or FSL runs, the latter being obtained by freezing the particle shapes in our code -- that is, by setting
$D^n_k \equiv \big(\begin{smallmatrix} 1 & 0 \\ 0 & 1 \end{smallmatrix}\big)$ for all $k$ and $n$.
To facilitate the comparison with LTPIC and FSL, we often indicate the number of particles used in a PIC run
as a product (e.g., 128 $\times$ 128). This may correspond to a uniform grid where (weighted) particles are initialized,
but in most cases the displayed PIC runs use unweighted particles initialized with a standard quiet start method.

In Sections~\ref{sec. LD} and \ref{sec. TSI} we first consider standard benchmark problems for which our results can 
be compared to the existing literature, see e.g. 
Refs.~\cite{Nakamura.Yabe.1999.cpc, Filbet.Sonnendrucker.2003.cpc, Qiu.Christlieb.2010.jcp, Banks.Hittinger.2010.ieee_tps, Crouseilles.Mehrenberger.Vecil.2011.esaim, Rossmanith.Seal.2011.jcp}.
Next, in Section~\ref{sec. MTB} we study a more applied test case consisting of a mismatched beam in a constant (continous) 
focusing channel, derived from the 1d sheet beam model developed in Ref.~\cite{Lund.2011.prstab}.

For the benchmark test cases in Sections~\ref{sec. LD} and \ref{sec. TSI} we classically consider the normalized 
Vlasov equation~\eqref{norm-Vlasov} in $x$-periodic phase space $[0,L]\times \RR$ coupled with a periodic Poisson equation
\begin{equation}
\label{neutral-Poisson}
\partial_x E(t,x) = \int_{\RR} f(t,x,v)\rmd v - n_e, \qquad t > 0, \quad x \in [0,L].
\end{equation}
Here $n_e$ is the uniform (and constant) density of a neutralizing background cloud, 
and we complete \eqref{neutral-Poisson} with the standard condition $\int_{0}^{L} E(t,x)\rmd x = 0$.

\subsection{Weak and Strong Landau damping}
\label{sec. LD}

We first consider the normalized Vlasov-Poisson system described above with perturbed initial distribution
\begin{equation}
\label{landau-fin}
f(t=0,x,v) \equiv \frac{1}{\sqrt{2\pi}} \exp\left(-\frac{v^2}{2}\right)\Big( 1 + A \cos(k x) \Big).
\end{equation}
Consistent with classical benchmarks \cite{Nakamura.Yabe.1999.cpc, Filbet.Sonnendrucker.2003.cpc, Rossmanith.Seal.2011.jcp} we take 
$k \equiv 0.5$ and set the perturbation amplitude $A \equiv 0.01$ for the weak Landau damping test case, 
or $A \equiv 0.5$ for the strong Landau damping test case
(actually, for such a perturbation the field is only damped for times $t \lessapprox 10$). 
In this section we use a cutoff velocity $v_{\rm max} \equiv 6.5$ and periodic boundary conditions at $x=0$ and $x=L\equiv 2\pi/k$.

\begin{figure} [hbtp] 
\begin{center}
\begin{tabular}{cc}
\includegraphics[height=0.4\textwidth]{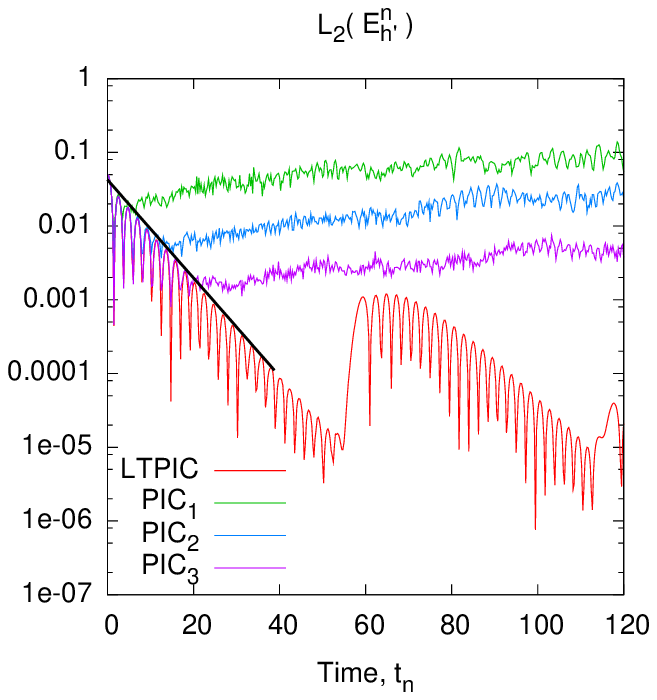}
& \hspace{15pt} 
\includegraphics[height=0.4\textwidth]{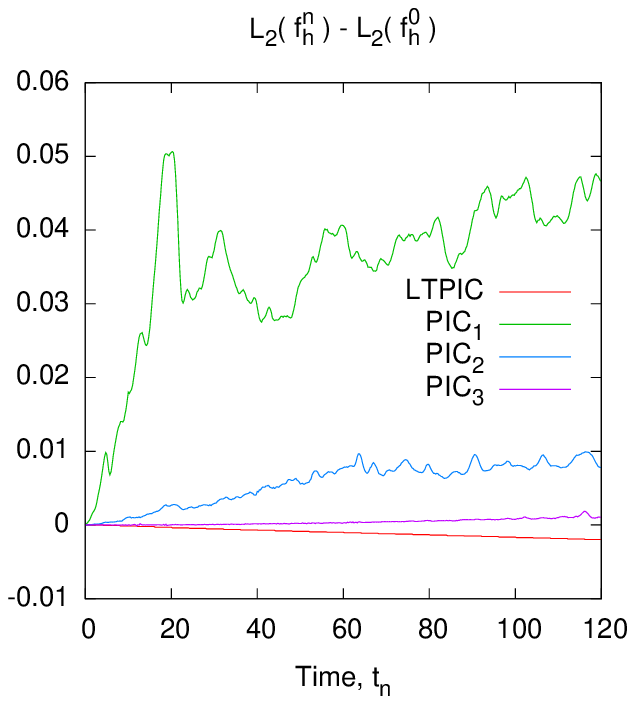}
\end{tabular}
  \caption{Weak Landau damping.
  $L^2$ norms of the electric field $E^n_{h'}$ (left) 
  and of the particle density $f^n_h$ (right) are plotted vs. $t_n = n\Dt$ for LTPIC and PIC simulations.
  All the runs use a time step of $\Dt =1/8$ and 64 cells for the Poisson solver. The PIC runs labelled as ${\rm PIC_1}$, ${\rm PIC_2}$ and ${\rm PIC_3}$ use increasingly
  high numbers of particles, namely $64 \times 64$, $128 \times 128$ and $256 \times 256$.
  The LTPIC run uses $64 \times 64$ particles and a remapping period $\Dtrec = 4$.
  The approximate cpu times for these runs are 40~s (${\rm PIC_1}$), 90~s (${\rm PIC_2}$), 330~s (${\rm PIC_3}$) and 45~s (LTPIC).
  On the left panel the plotted slope ($\gamma = -0.1533$) matches the theoretical damping rate, see Refs.~\cite{Filbet.Sonnendrucker.2003.cpc, Rossmanith.Seal.2011.jcp}.
  The quasi-periodic relaxation in the LTPIC curve is known as a Poincar\'e recurrence, see text for details.
  }
  \label{fig. WLD-energy}
 \end{center}
\end{figure}

In Figure~\ref{fig. WLD-energy} we show the $L^2$ norms of the electric field (left panel, semi-log scale) and of the phase-space density (right panel) 
obtained with PIC and LTPIC simulations of the weak Landau damping. Results clearly show the noiseless aspect of the LTPIC method, 
as the theoretical damping rate ($\gamma = -0.1533$) is matched with a low-resolution run using 64 Poisson cells and 64 particles per cell. We also observe 
the classical recurrent relaxation 
occuring with period $T_R \approx 60$, in good agreement with the theoretical period $L/\Delta v \approx 62$, see e.g., Ref.~\cite{Nakamura.Yabe.1999.cpc}. 
In contrast, a PIC run using the same number of cells and particles (labelled as ${\rm PIC}_1$) is unable to predict the correct damping rate beyond $t\approx 5$.
And even with 1024 particles per cell (and significantly greater cpu time), the ${\rm PIC}_3$ run only predicts the correct rate until $t = 20$.
We also see that the low-resolution LTPIC run does a significantly better job at preserving the $L^2$ norm of the density (a Vlasov invariant),
compared to the low- and moderate-resolution PIC runs. For such a test case, only the high-resolution PIC run performs better with regard to the $L^2$ measure.

\begin{figure} [hbtp] 
\begin{center}
\begin{tabular}{cc}
\includegraphics[height=0.4\textwidth]{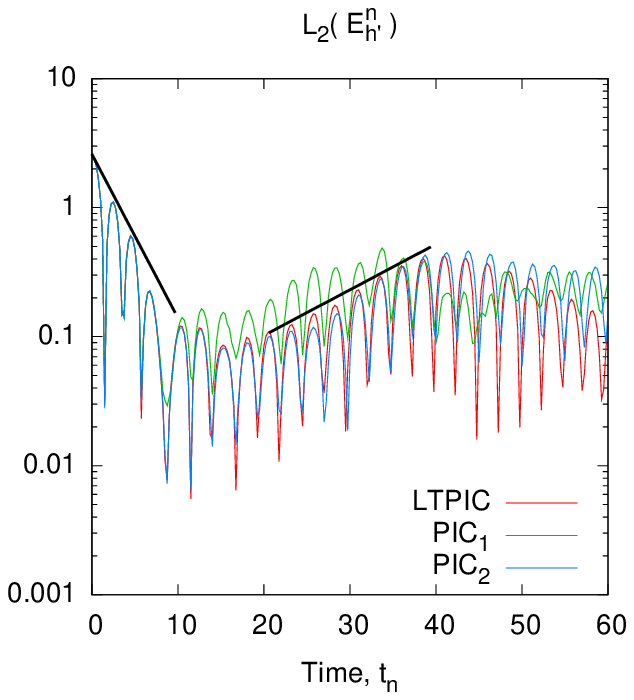}
& \hspace{15pt} 
\includegraphics[height=0.4\textwidth]{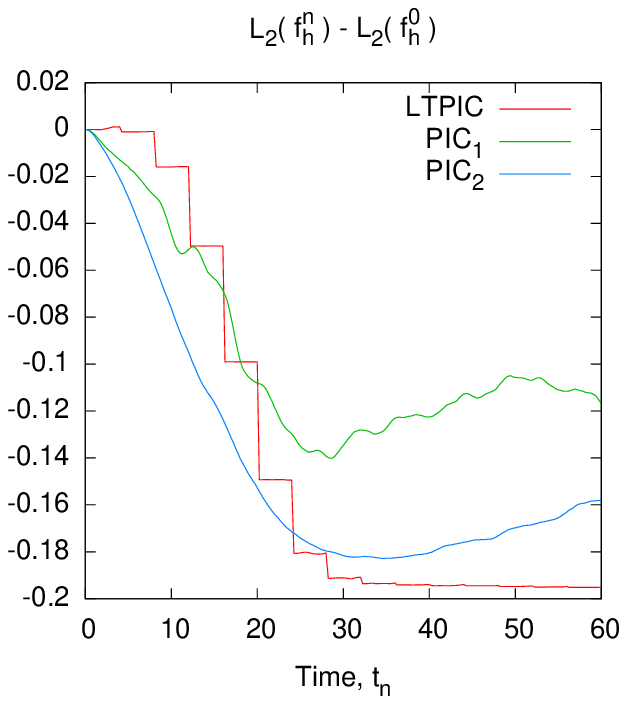}
\end{tabular}
  \caption{Strong Landau damping. Plotted quantities and numerical parameters for the PIC and LTPIC runs are the same 
  than in Figure~\ref{fig. WLD-energy}.
  The approximate cpu times for these runs are 20~s (${\rm PIC_1}$), 45~s (${\rm PIC_2}$) and 24~s (LTPIC).
  The plotted slopes ($\gamma_1 = -0.2920$ for the initial damping and $\gamma_2 = 0.0815$ for the growth between times $t = 20$ and $t = 40$) 
  match benchmarked exponential rates, see e.g. Refs.~\cite{Filbet.Sonnendrucker.2003.cpc, Rossmanith.Seal.2011.jcp}.}
  \label{fig. SLD-energy}
 \end{center}
\end{figure}

In Figure~\ref{fig. SLD-energy} the same quantities are shown for the strong Landau ``damping''. Again, the low-resolution LTPIC run predicts
the benchmarked rates for the initial damping and subsequent exponential growth. The low-resolution ${\rm PIC}_1$ run (with 64 cells and $64 \times 64$ particles)
only predicts the initial damping. The moderate-resolution ${\rm PIC}_2$ run predicts correctly both rates (although with less accuracy for the growth period), 
but at a significantly higher cost in terms of memory and cpu time.
As for the preservation of the $L^2$ norm of the density we observe that LTPIC does not perform better than PIC here, essentially due to the remappings.

Finally, we found that the low-resolution FSL runs (using 64 cells and $64 \times 64$ particles) give energy curves very 
similar to the LTPIC ones, in both the weak and strong damping cases. These curves were omitted for readability.

\begin{figure} [ht!]
\begin{center}
\begin{tabular}{ccc}
\includegraphics[height=0.29\textwidth]{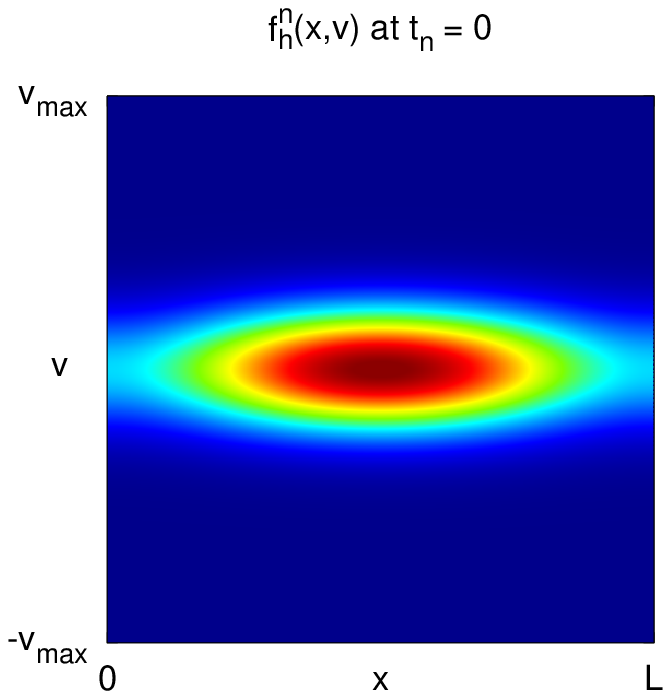}
& 
\includegraphics[height=0.29\textwidth]{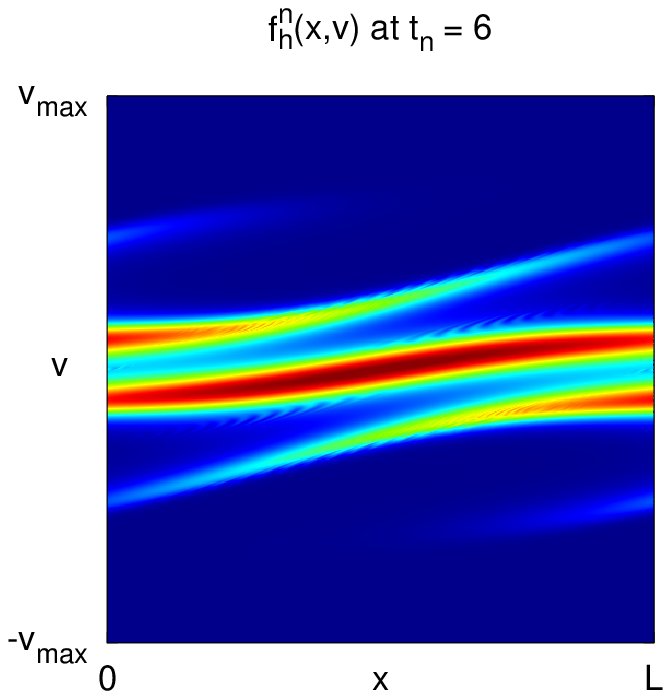}
& 
\includegraphics[height=0.29\textwidth]{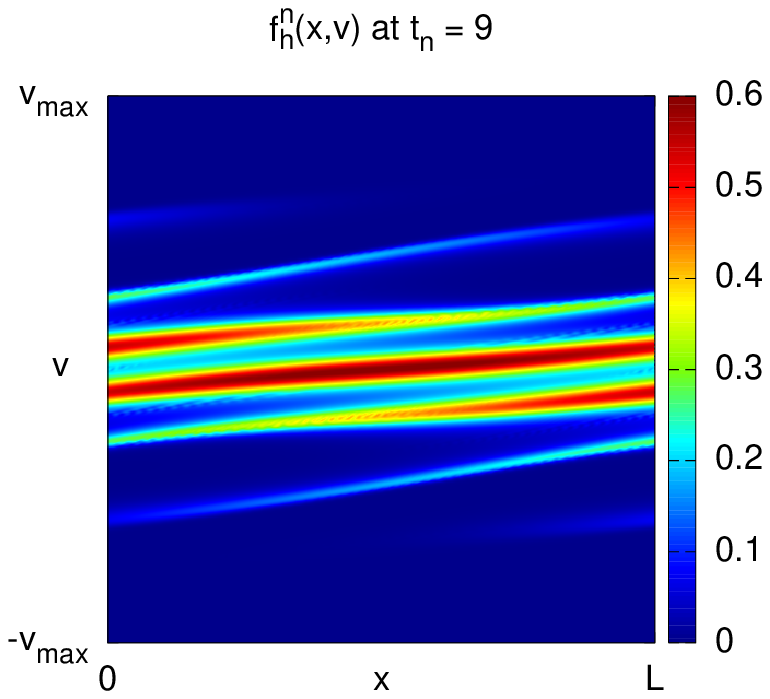}
\vspace{5pt} \\
\includegraphics[height=0.29\textwidth]{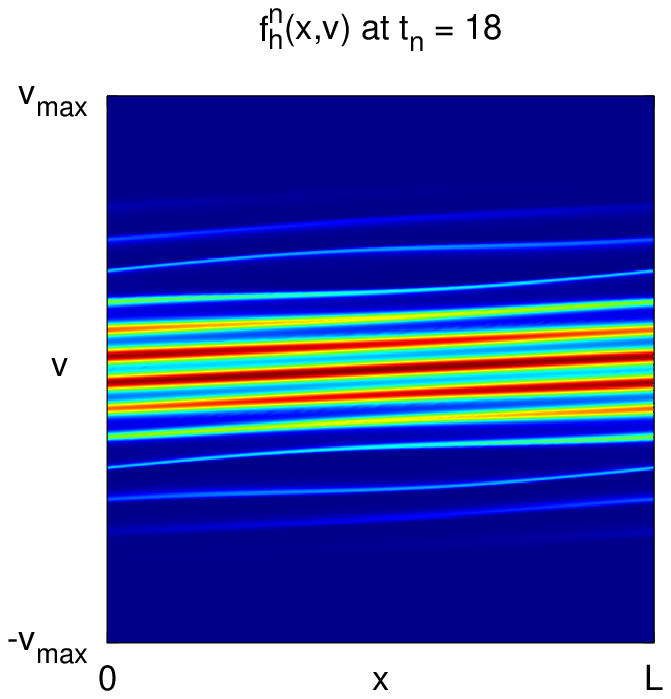}
& 
\includegraphics[height=0.29\textwidth]{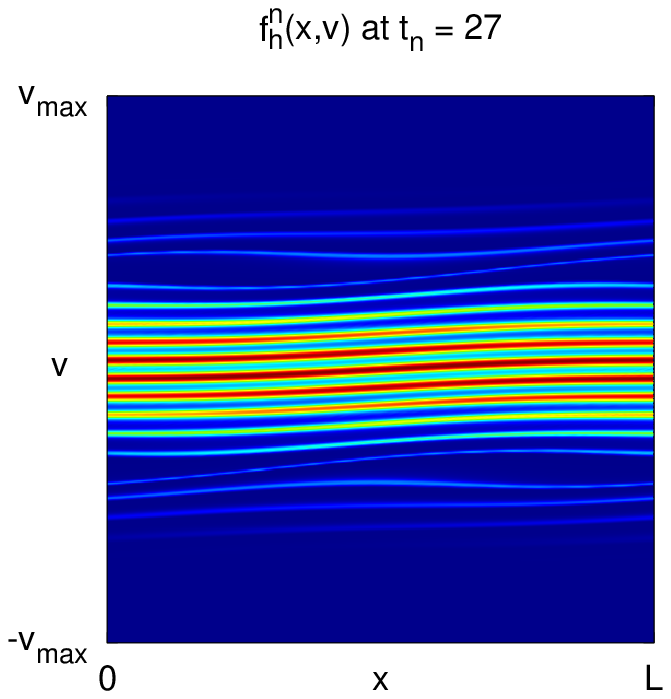}
& 
\includegraphics[height=0.29\textwidth]{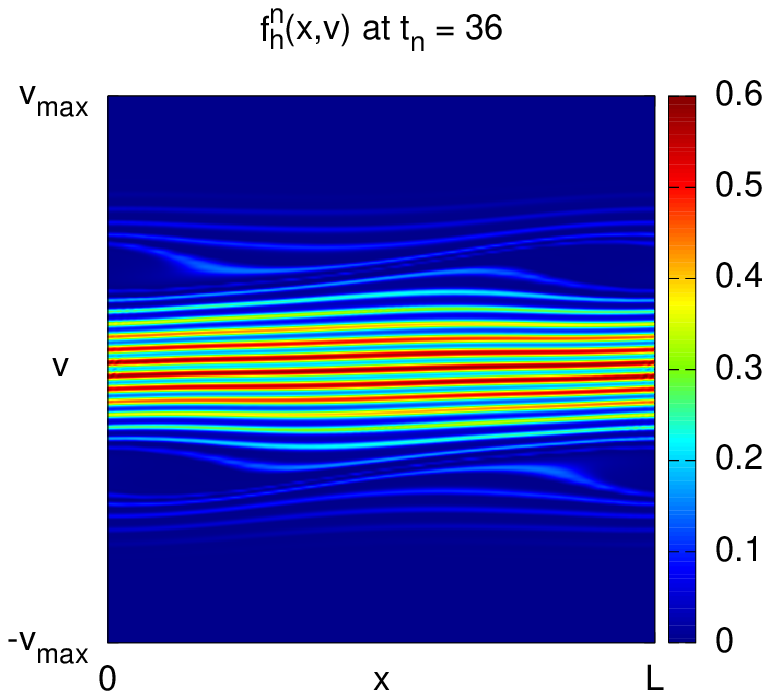}
\vspace{5pt} \\ 
\includegraphics[height=0.29\textwidth]{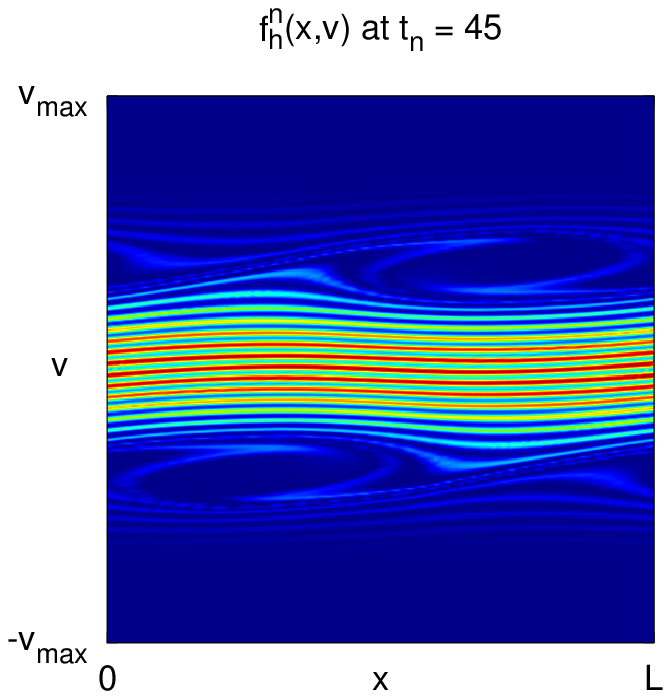}
& 
\includegraphics[height=0.29\textwidth]{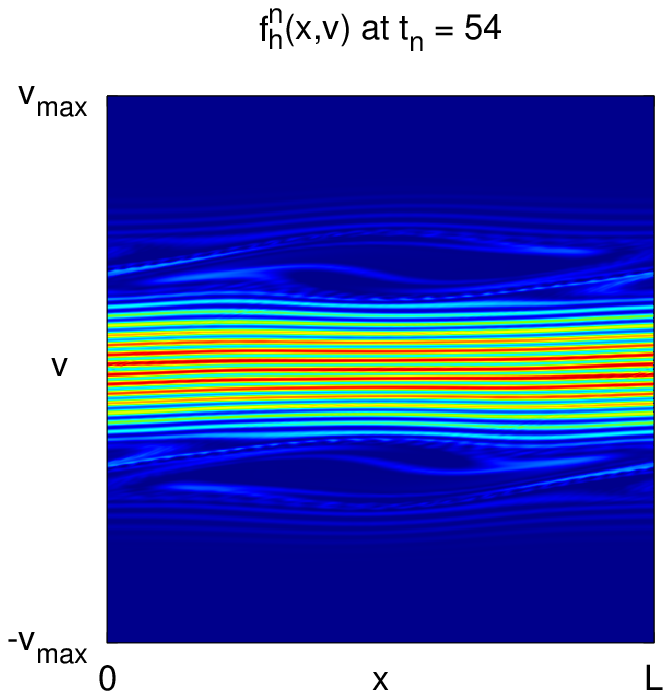}
& 
\includegraphics[height=0.29\textwidth]{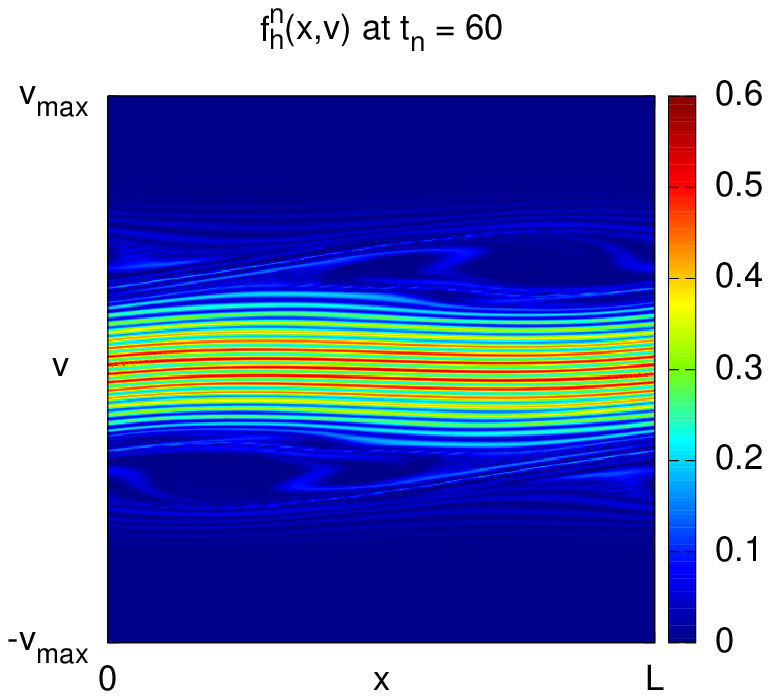}
\end{tabular}
  \caption{Strong Landau damping. Time evolution of the phase-space particle density $f^n_h(x,v)$ obtained 
  with an LTPIC simulation. This run uses a time step $\Dt=1/8$, a remapping period $\Dtrec = 4$, a Poisson solver with 256 cells 
  and $256 \times 256$ particles.
  On the $t_n = 6$ snapshot some oscillations are visible but they almost vanish in the subsequent plots, 
  due to the particle remappings and the strong shearing of the flow. The approximate cpu times for this runs is 660~s.
  }
  \label{fig. SLD-evol}
 \end{center}
\end{figure}

To better assess the noiseless aspect of our method, we also show in Figure~\ref{fig. SLD-evol} 
the phase-space density $f^n_h(x,v)$ as it evolves in the time range $t_n \in [0,60]$, 
obtained with an LTPIC run with periodic ($\Dtrec = 4$) remappings on a grid of $256 \times 256$ particles. 
Here the strong phase-space filamentation is accurately resolved. In particular it agrees very well with similar phase-space 
plots shown on Figure~10 in Ref.~\cite{Rossmanith.Seal.2011.jcp}, obtained with a high order Backward Semi-Lagrangian 
Discontinuous Galerkin (BSL-DG) scheme (our color scale is chosen in order to fit theirs).
In that scheme the phase-space density is computed using a cartesian mesh of lower resolution in the $x$ dimension (128 points), 
but in the $v$ dimension where the filaments are most difficult to resolve the resolution is the same (256 points). 
Since each fifth-order DG cell contains 15 basis functions, we observe that this run involves about the same number 
of degrees of freedom as the LTPIC run, where each particle carries 6 floating numbers
(for the weight, phase space coordinates and normalized deformation matrix).
Close examination reveals slightly better resolution of fine structures with the BSL-DG scheme.
However, due to the CFL constraint we note that in the BSL run the time steps are significantly smaller 
(namely $\Dt \approx 0.03$) than with the present scheme where we have set $\Dt = 1/8$.
Comparison with another fifth-order BSL-DG simulation \cite[Fig. 7, bottom row]{Crouseilles.Mehrenberger.Vecil.2011.esaim}
shows a better resolution for the LTPIC scheme, however in this case the BSL-DG run uses far less degrees of freedom than
the LTPIC one.

\begin{figure} [ht!]
\begin{center}
\begin{tabular}{ccc}
\includegraphics[height=0.29\textwidth]{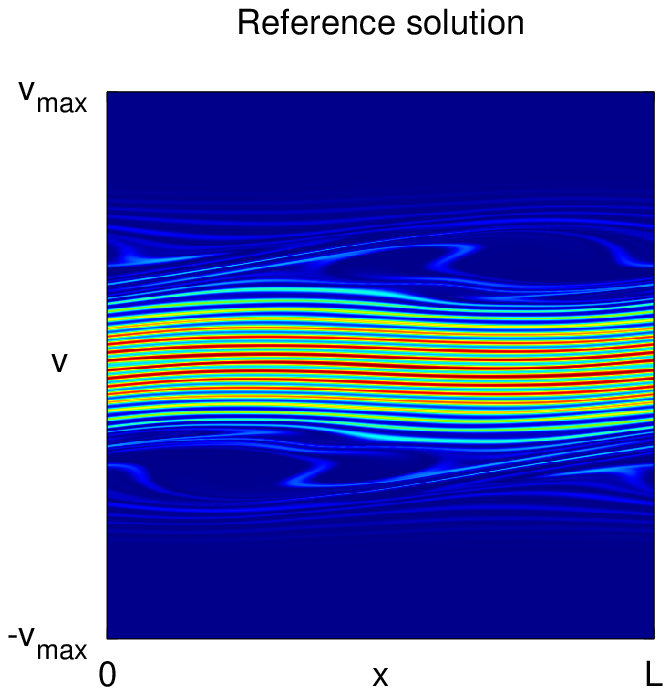} 
&
\includegraphics[height=0.29\textwidth]{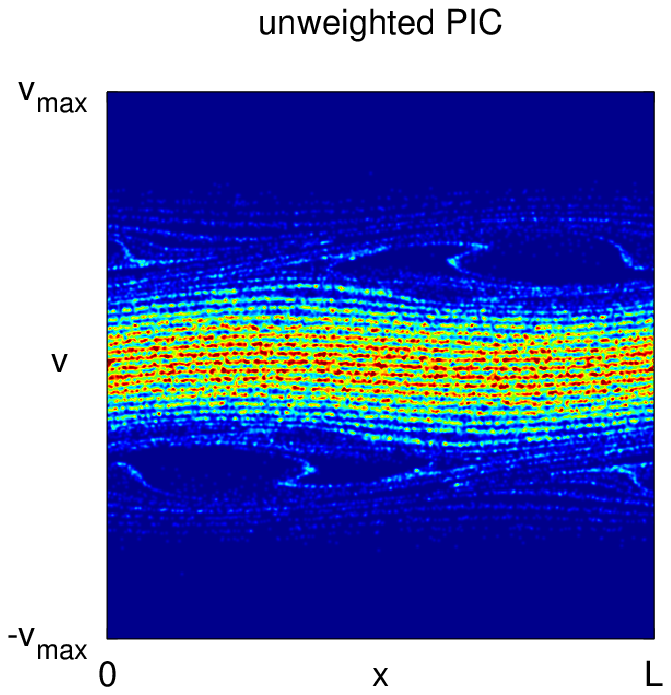} 
& 
\includegraphics[height=0.29\textwidth]{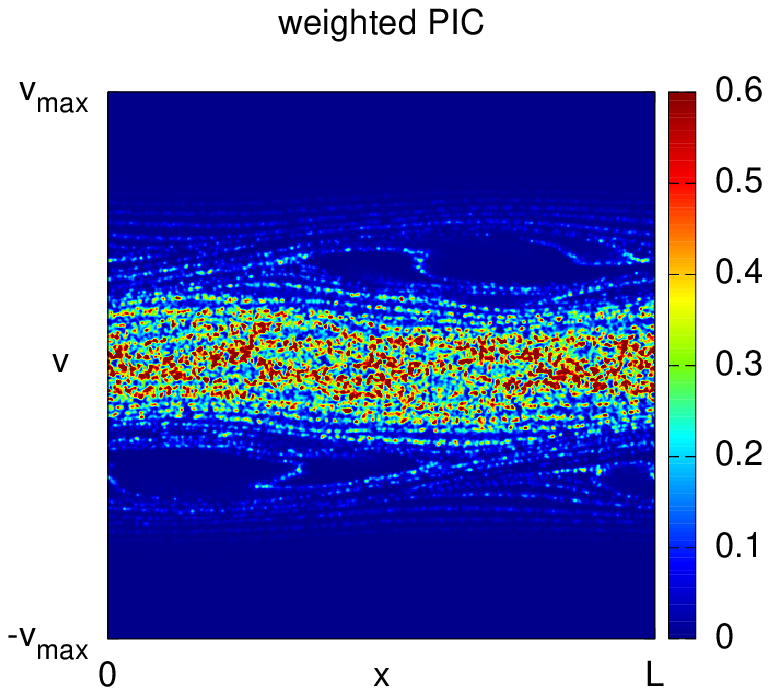} 
\vspace{5pt} 
\\ 
\includegraphics[height=0.29\textwidth]{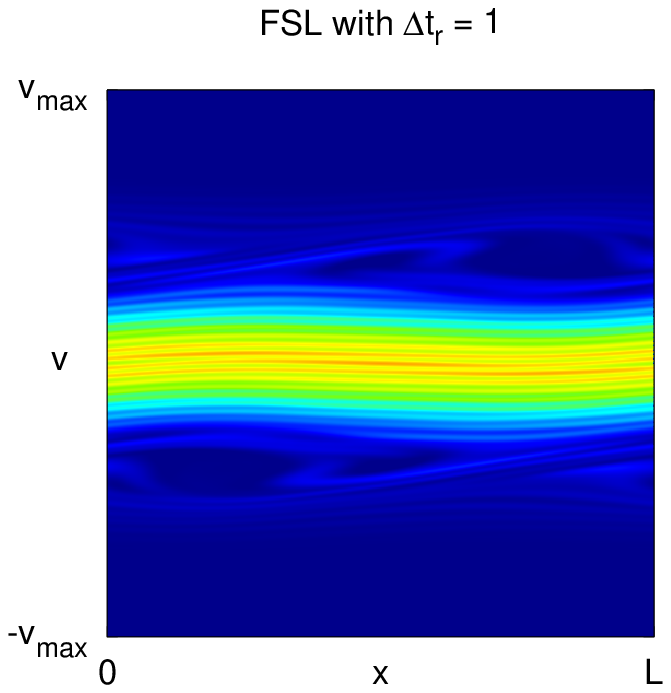} 
&
\includegraphics[height=0.29\textwidth]{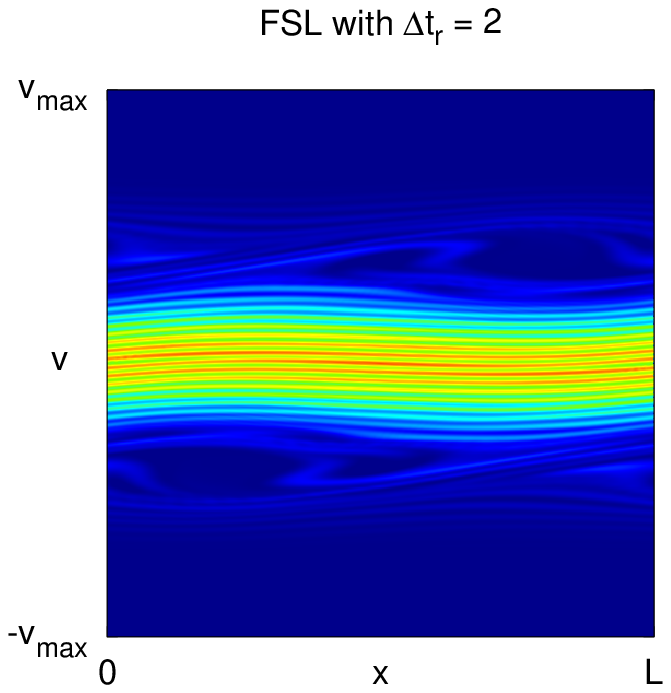} 
&
\includegraphics[height=0.29\textwidth]{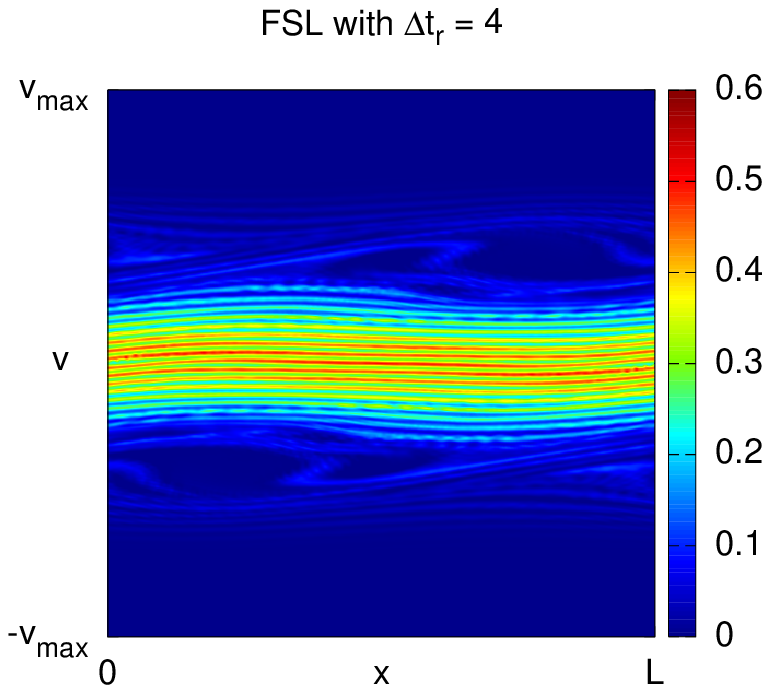} 
\vspace{5pt} 
\\
\includegraphics[height=0.29\textwidth]{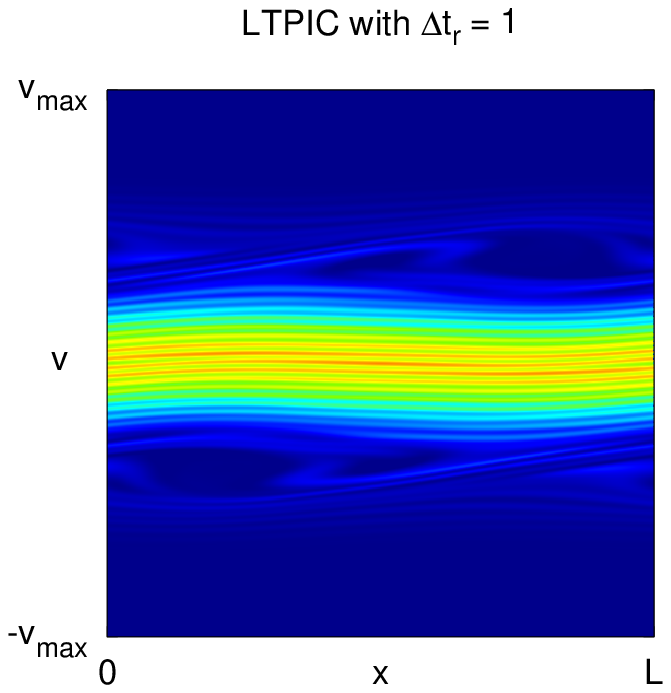} 
&
\includegraphics[height=0.29\textwidth]{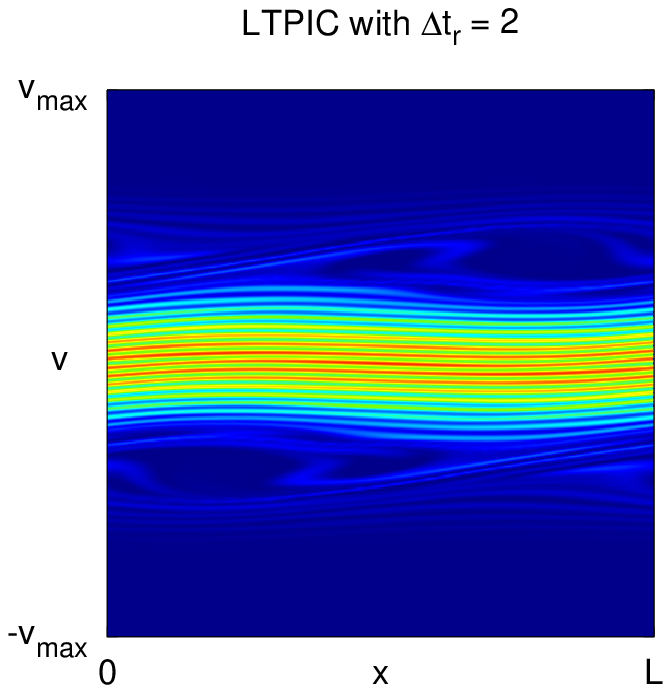} 
&
\includegraphics[height=0.29\textwidth]{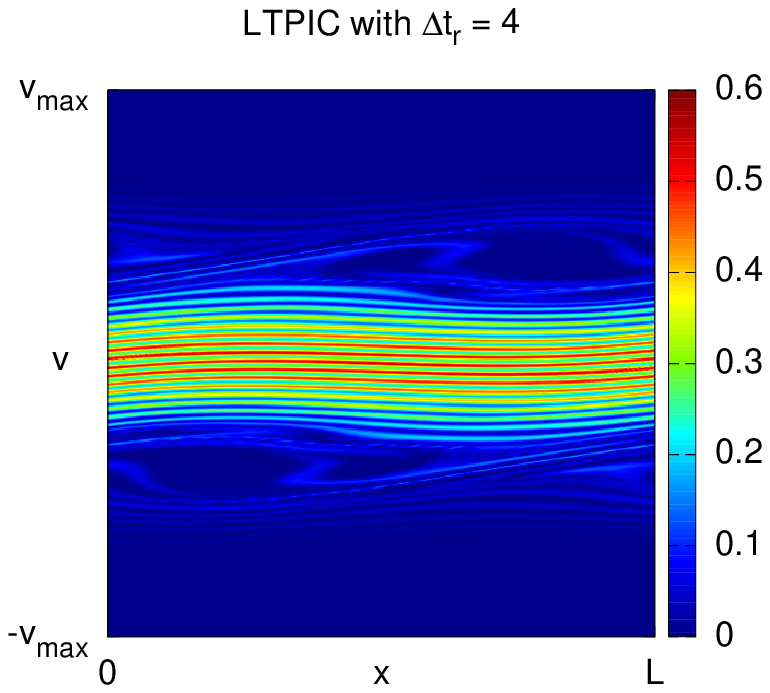} 
\end{tabular}
  \caption{Strong Landau damping. 
  Comparisons of phase-space densities obtained at $t_n=60$ with different methods.
  All the runs use a time step $\Dt = 1/8$, a Poisson solver with 256 cells and $256\times 256$ particles, except for the reference simulation,
  an LTPIC run with 512 cells and $512\times 512$ particles.
  In the FSL and LTPIC runs the remapping period varies as indicated (in the reference run it is $\Dtrec = 4$).
  The approximate cpu times for these runs are 4900~s (reference LTPIC), 625~s (unweighted PIC), 650~s (weighted PIC), 690 to 720~s (FSL runs) and 655 to 665~s (LTPIC runs). 
  }
  \label{fig. SLD-t60}
 \end{center}
\end{figure}

\begin{figure} [h]
\begin{center}
\begin{tabular}{cccc}
\includegraphics[width=0.22\textwidth]{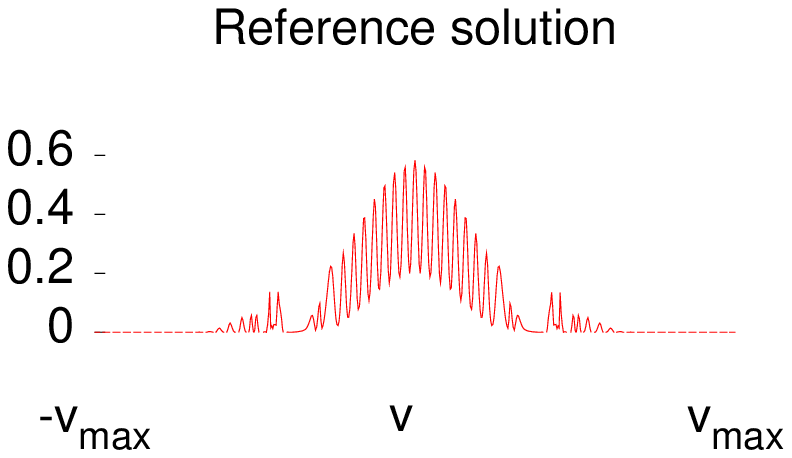}  
&
\includegraphics[width=0.22\textwidth]{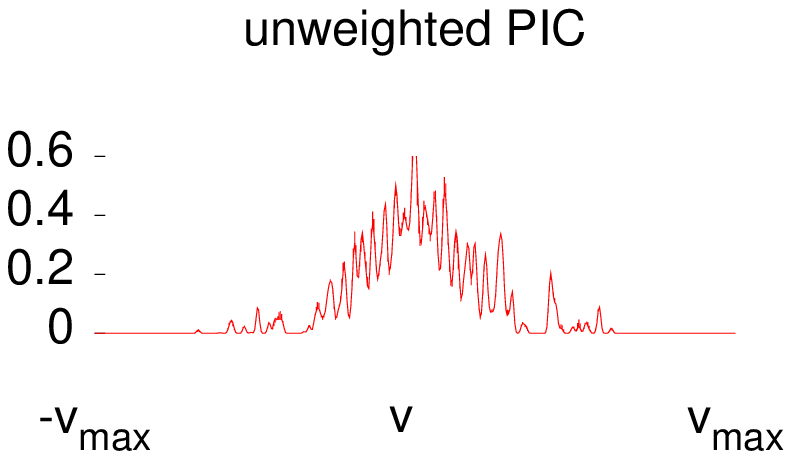}  
& 
\includegraphics[width=0.22\textwidth]{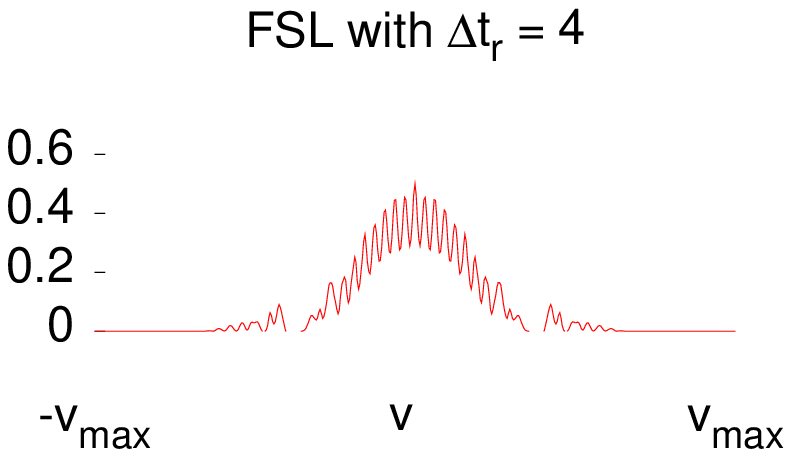}  
& 
\includegraphics[width=0.22\textwidth]{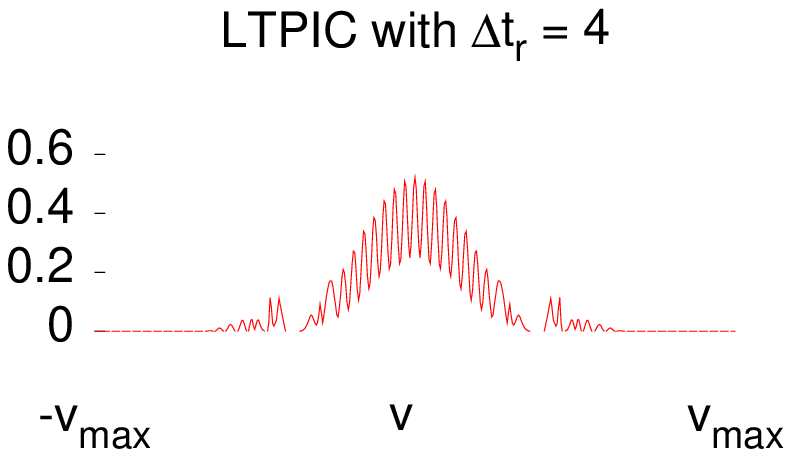}  
\end{tabular}
  \caption{Velocity profiles $f^n_h(x=L/2,v)$ at $t_n=60$ associated with some of the numerical solutions shown on 
  Figure~\ref{fig. SLD-t60}.}
  \label{fig. SLD-v}
 \end{center}
\end{figure}

In Figures~\ref{fig. SLD-t60} and \ref{fig. SLD-v} we then compare how well different particle methods (namely PIC, LTPIC and FSL 
with various numerical parameters indicated in the figure caption) resolve the filaments in the strong Landau ``damping'' test case
at $t=60$. Phase-space densities obtained at $t=60$ with different schemes are shown on Figure~\ref{fig. SLD-t60},
together with a reference solution obtained with an LTPIC run using improved numerical parameters relative 
to Figure~\ref{fig. SLD-evol}.
Again, the results clearly show that LTPIC and FSL are able to remove the noise. With PIC the localization of global patterns 
such as filaments and holes may be accurate, however the noise level is significant (and it remains so with finer simulations, 
not shown here). Results also show the effect of varying the remapping period $\Dtrec$ in the FSL and LTPIC runs.
For low remapping periods both methods give similar results, which is expected since particles do not have time to deform much.
For high remapping periods however the LTPIC performs significantly better: it introduces less diffusion than FSL, and does not present the
unphysical oscillations that start to appear in the filaments computed with the FSL method. This is also expected from the convergence 
analysis of the LTP transport operator \eqref{ltp-def}, which does not require remappings for asymptotic convergence. The good news is that
this improved performance does not come at an expensive price: the measured cpu times are indeed similar for FSL and LTPIC runs,
which indicates that the additional work of updating the deformation matrices does not represent a significant portion of the overall time.

Finally, in Figure~\ref{fig. SLD-v} we show $v$-slices of the distribution at $x=L/2$ and $t_n=60$. Again, results show that the LTPIC 
scheme gives the best results: compared to the PIC method the noise has been removed, and compared to the FSL scheme the numerical 
diffusion is significantly reduced.

\subsection{Two-stream instabilities}
\label{sec. TSI}

Here we consider again the periodic Vlasov-Poisson system and set the initial distribution
function as follows.
\begin{enumerate}
\item 
    (Weak instability.)
    First, to compare our results with Refs.~\cite{Filbet.Sonnendrucker.2003.cpc, Qiu.Christlieb.2010.jcp} we set
    $$
    f(t=0,x,v) \equiv \frac{2(1+5 v^2) }{7\sqrt{2\pi}}e^{-\frac{v^2}{2}} \left( 1 + A \left(\frac{\cos(2 kx) + \cos(3kx)}{1.2} + \cos(kx)\right) \right) 
    $$
    with $k \equiv \frac 12$ and a weak amplitude $A \equiv 0.01$ for the perturbation. 
\item
    (Strong instability.)
    Next to compare our results with Refs.~\cite{Banks.Hittinger.2010.ieee_tps, Rossmanith.Seal.2011.jcp} we set
    $$
    f(t=0,x,v) \equiv \frac{v^2}{\sqrt{2\pi}} e^{-\frac{v^2}{2}} \left( 1 - A \cos(k x) \right)
    $$
    with $k \equiv \frac 12$ and a strong amplitude $A \equiv 0.5$ for the perturbation. 
\end{enumerate}
For the simulations we use a cutoff velocity $v_{\rm max} \equiv 5$ and periodic boundary conditions at $x=0$ and $x=L\equiv 2\pi/k$.

\begin{figure} [ht!]
\begin{center}
\begin{tabular}{ccc}
\includegraphics[height=0.29\textwidth]{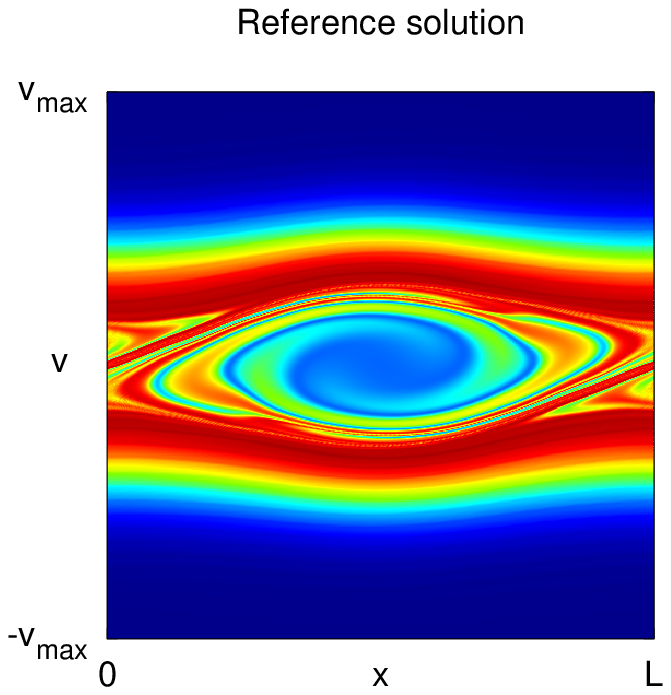} 
& 
\includegraphics[height=0.29\textwidth]{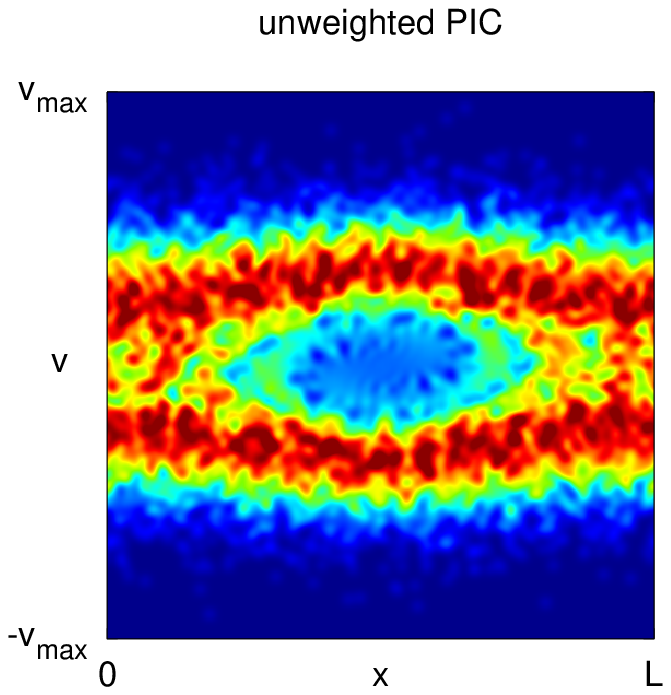} 
& 
\includegraphics[height=0.29\textwidth]{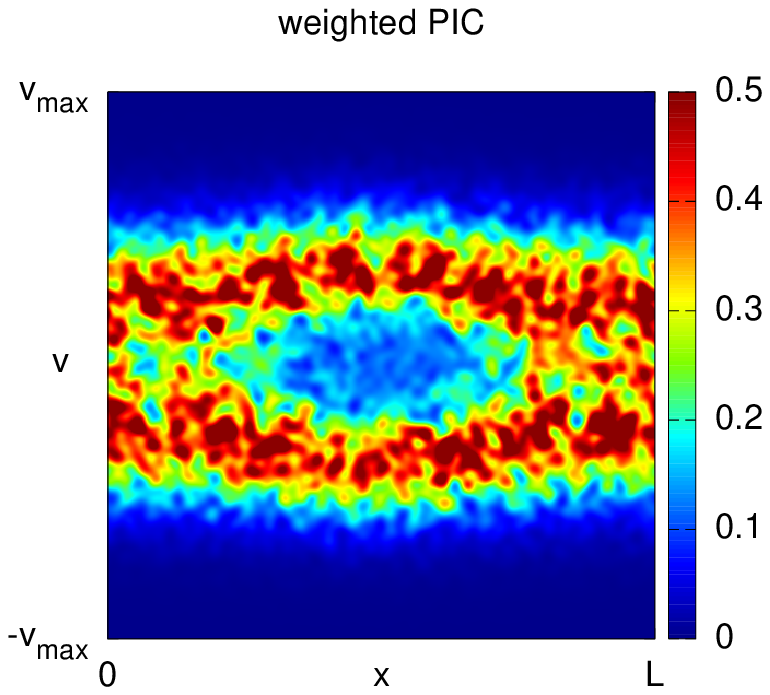} 
\vspace{5pt} 
\\
\includegraphics[height=0.29\textwidth]{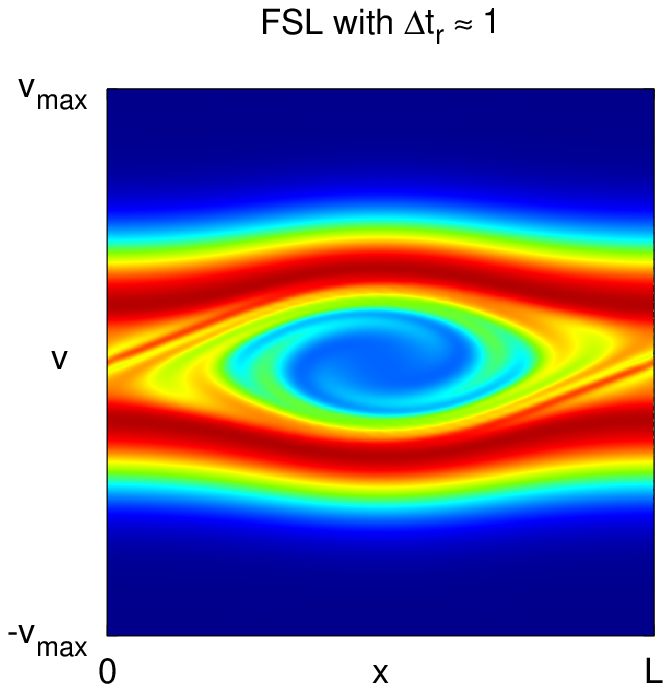} 
& 
\includegraphics[height=0.29\textwidth]{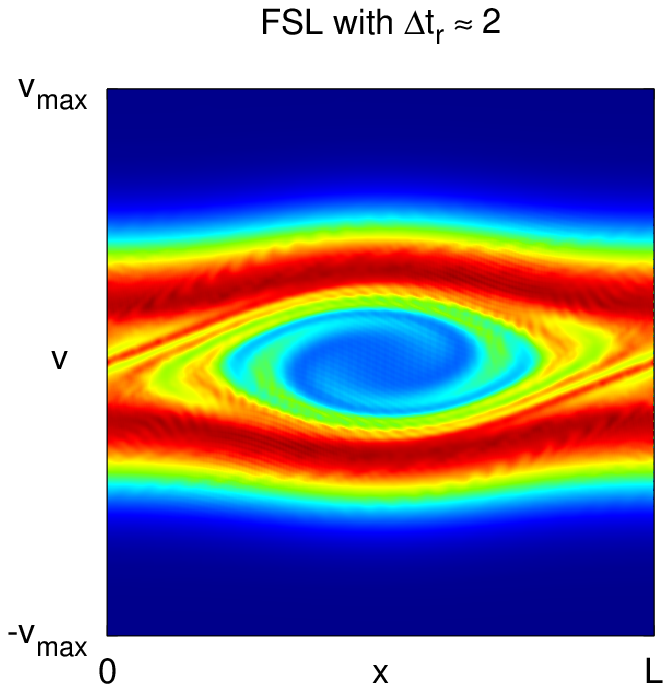} 
& 
\includegraphics[height=0.29\textwidth]{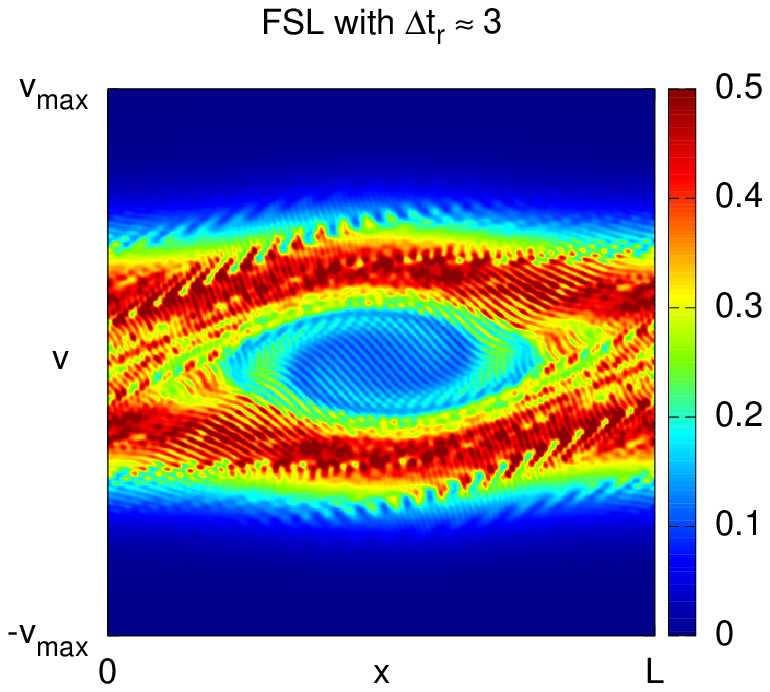} 
\vspace{5pt} 
\\
\includegraphics[height=0.29\textwidth]{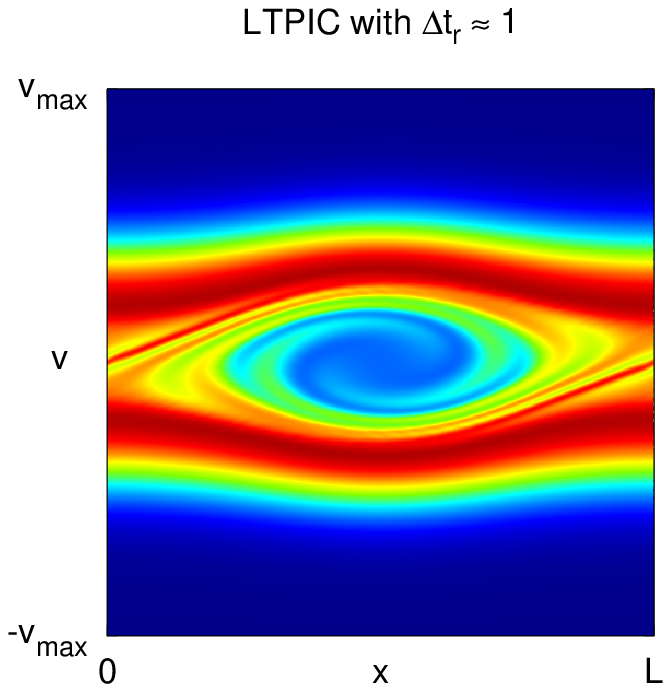} 
& 
\includegraphics[height=0.29\textwidth]{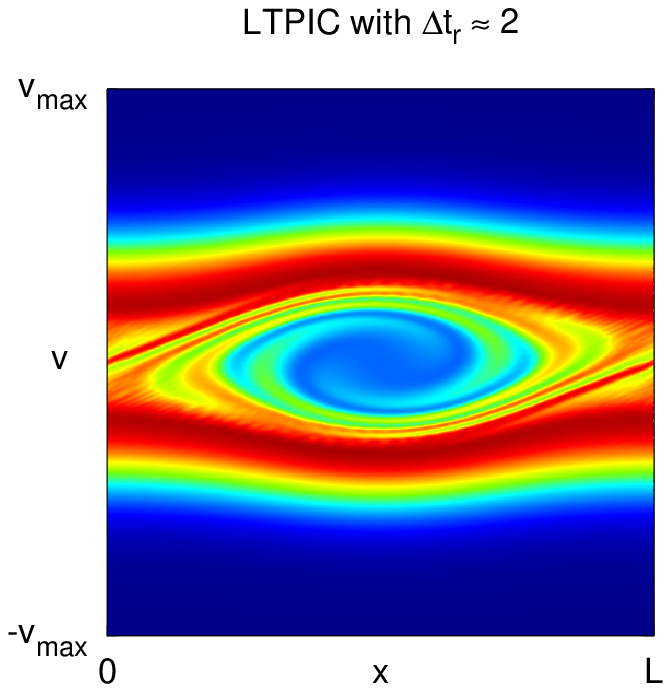} 
& 
\includegraphics[height=0.29\textwidth]{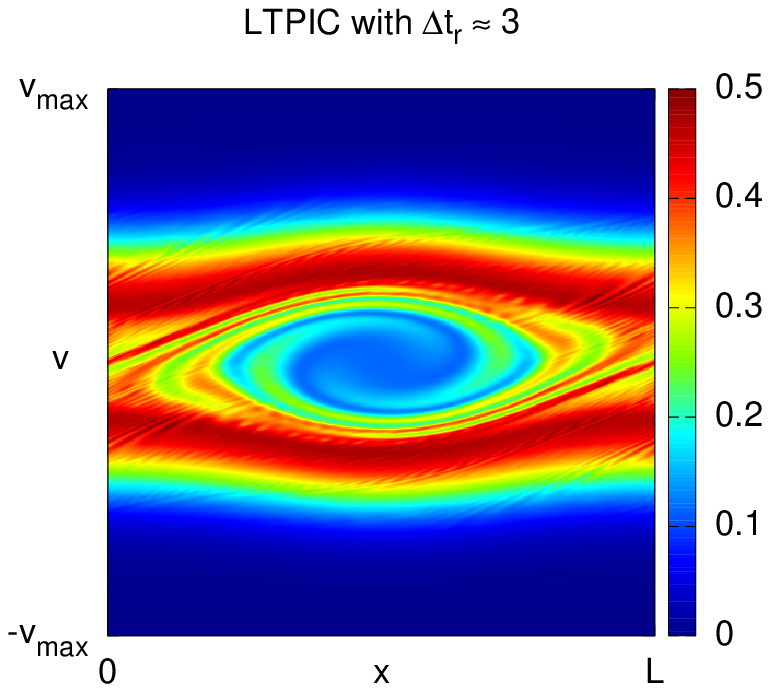} 
\end{tabular}
  \caption{Weak two-stream instability (case 1). 
  Comparisons of phase-space densities obtained at $t_n=53$ with different methods.
  All the runs use a time step $\Dt \approx 1/5$, a Poisson solver with 64 cells and $128\times 128$ particles, except for the reference simulation,
  an LTPIC run with 128 cells and $512\times 512$ particles.
  In the FSL and LTPIC runs the remapping period varies as indicated (in the reference run it is $\Dtrec \approx 2$).
  The approximate cpu times for these runs are 875~s (reference LTPIC), 30~s (unweighted PIC), 26~s (weighted PIC), 31 to 37~s (FSL runs) and 35 to 40~s (LTPIC runs). 
   }
  \label{fig. TSI_FS-t53}
 \end{center}
\end{figure}

In Figure~\ref{fig. TSI_FS-t53} we compare phase-space densities for the weak two-stream instability (case 1) obtained at $t=53$ with PIC, FSL and LTPIC runs
and various numerical parameters indicated in the figure caption. Again, the results lead to several observations. 
\begin{itemize}
\item 
    First, LTPIC and FSL are able to remove the ``noise'' (i.e., the oscillations) for appropriate values of the remapping period 
    $\Dtrec$. In that regard our simulations show again the robustness of LTPIC compared to FSL,
    where strong oscillations appear for $\Dtrec \gtrapprox 2$.
\item 
    For low remapping periods FSL and LTPIC give similar results -- an expected observation since particles have less time
    to deform. However a closer look at the filaments in the $\Dtrec = 1$ case  shows that the latter 
    is less diffusive.
\item
    Again, our measurements indicate that for similar numerical parameters the FSL and LTPIC runs take similar computational time. This signifies that deforming 
    the particles is not an expensive task in our code.
\item 
    Finally we find that the LTPIC scheme is able to achieve the accuracy of some high-resolution state-of-the-art grid-based methods. 
    For instance, the bottom left panel in Figure~\ref{fig. TSI_FS-t53} showing an LTPIC run using $\Dtrec = 1$ and $128\times 128$ particles
    is very similar to the right panel in Figure~11 from Ref.~\cite{Qiu.Christlieb.2010.jcp}, obtained with a conservative third order WENO BSL 
    scheme using a $256 \times 512$ phase-space mesh.
\end{itemize}

\begin{figure} [ht!]
\begin{center}
\begin{tabular}{ccc}
\includegraphics[height=0.29\textwidth]{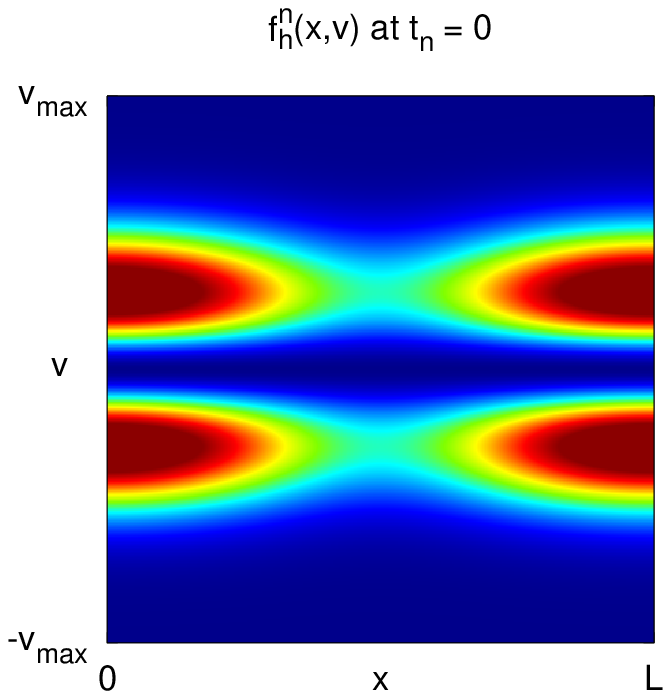}
& \hspace{5pt} 
\includegraphics[height=0.29\textwidth]{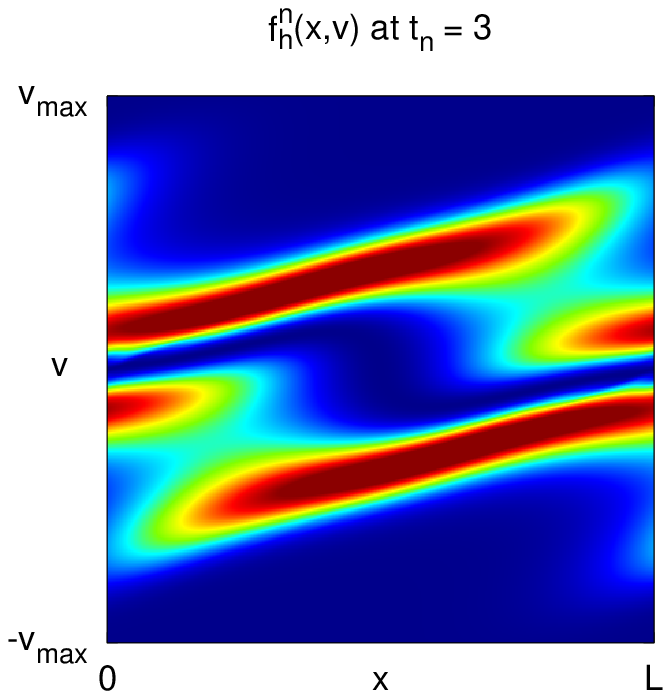}
& \hspace{5pt} 
\includegraphics[height=0.29\textwidth]{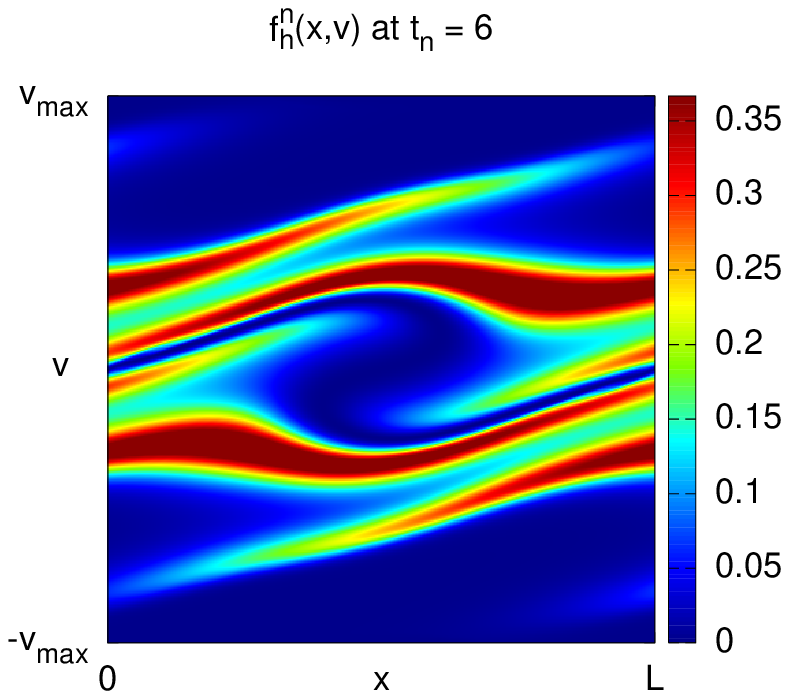}
\vspace{5pt} \\
\includegraphics[height=0.29\textwidth]{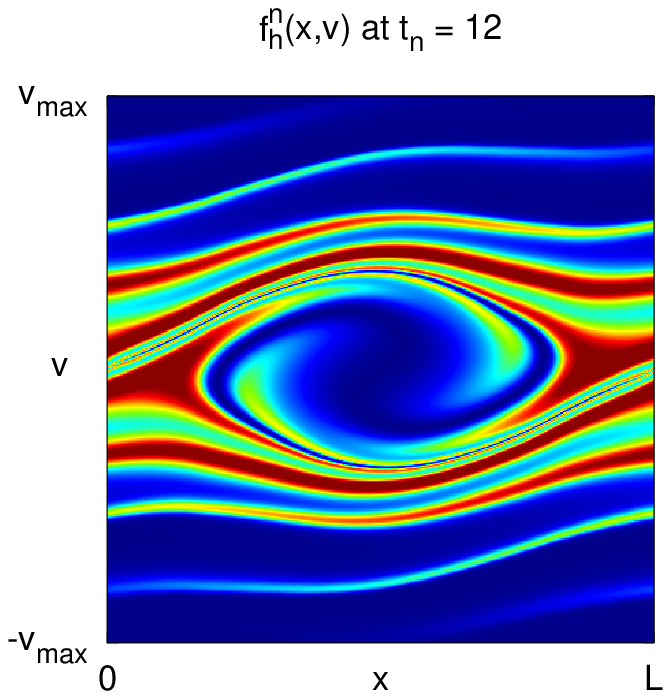}
& \hspace{5pt} 
\includegraphics[height=0.29\textwidth]{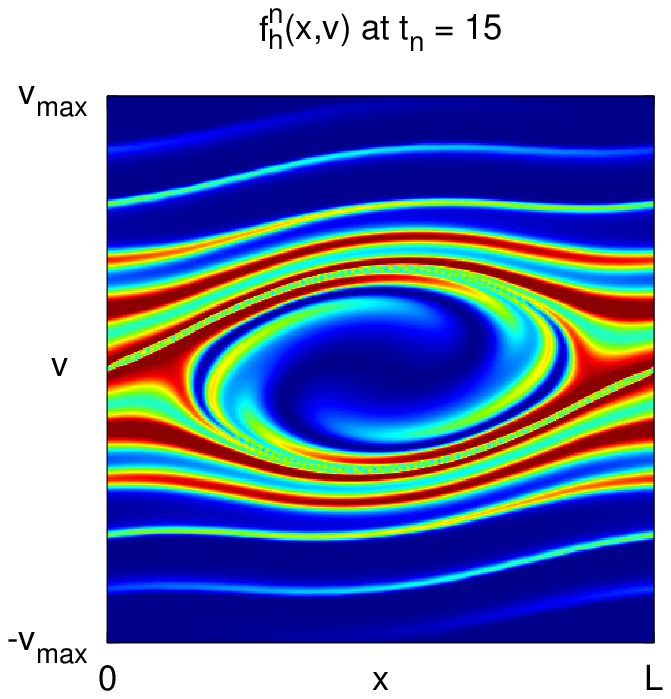}
& \hspace{5pt} 
\includegraphics[height=0.29\textwidth]{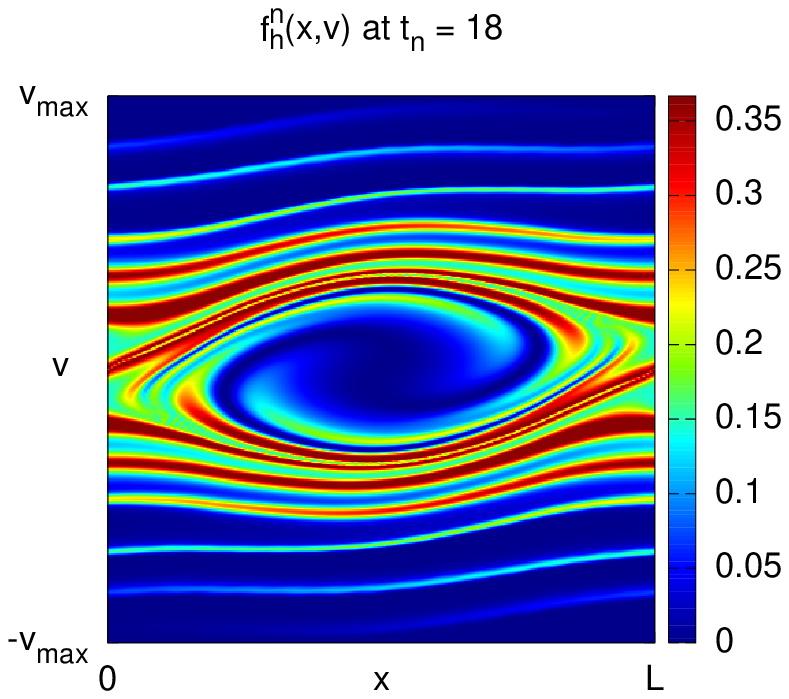}
\vspace{5pt} \\ 
\includegraphics[height=0.29\textwidth]{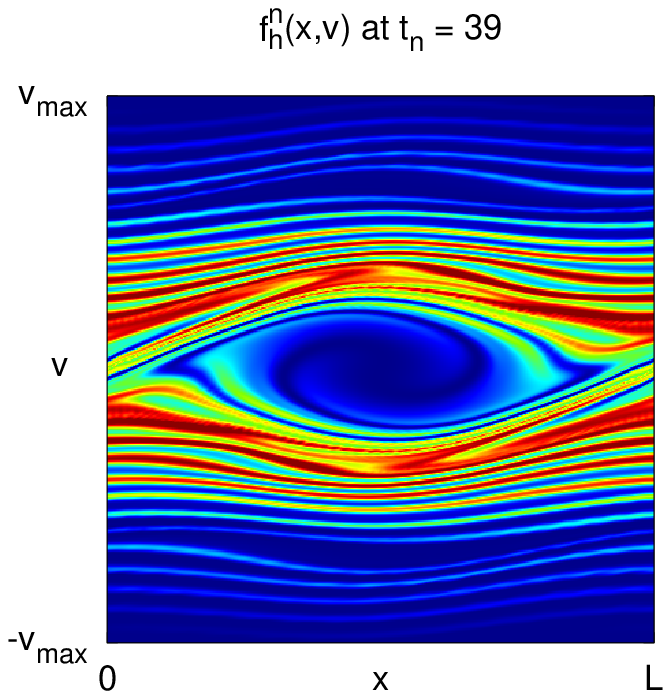}
& \hspace{5pt} 
\includegraphics[height=0.29\textwidth]{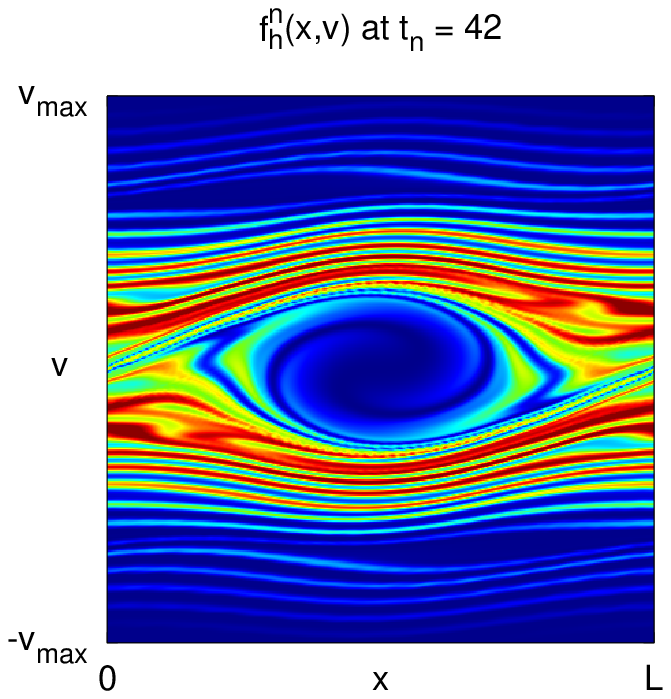}
& \hspace{5pt} 
\includegraphics[height=0.29\textwidth]{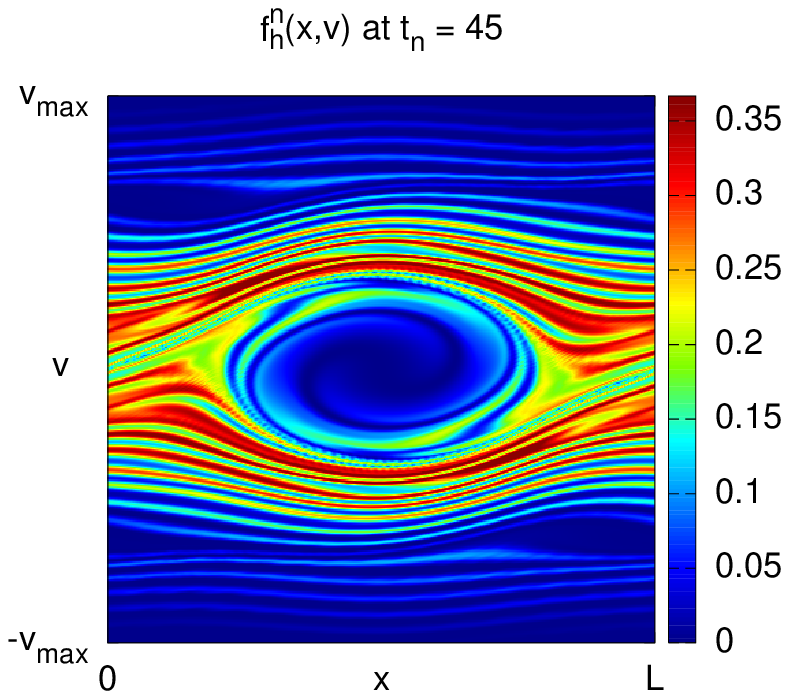}
\end{tabular}
  \caption{Strong two-stream instability (case 2). 
  Time evolution of the phase-space particle density $f^n_h(x,v)$ obtained 
  with an LTPIC simulation. This run uses a time step $\Dt=1/5$, a remapping period $\Dtrec \approx 2$, 
  a Poisson solver with 64 cells and $256 \times 256$ particles.
  The approximate cpu time for this run is 115~s.
  }  
  \label{fig. TSI_NY-evol}
 \end{center}
\end{figure}

Turning next to the strong two-stream instability (case 2), we show in Figure~\ref{fig. TSI_NY-evol} the time evolution of the 
phase-space density obtained with an LTPIC run using $256 \times 256$ particles. Fine phase-space detail is resolved as the strong amplitude 
of the initial perturbation leads to filamentations. Again we can compare our results with high order state-of-the-art grid-based methods. 
For instance we observe that the bottom right panel in Figure~\ref{fig. TSI_NY-evol} showing the LTPIC density at $t=45$ resolves
the filaments with similar accuracy to that of the (center and bottom) panels in Figure~4 from Ref.~\cite{Rossmanith.Seal.2011.jcp}, 
obtained by a fifth-order BSL-DG scheme using $129\times 129$ and $255 \times 255$ phase-space cells, respectively. 
Here some complementary observations are in order. On the one hand indeed, a closer look at the bottom right panel shows that the LTPIC solution 
presents moderate oscillations in the inner filaments. Therefore, it is not strictly as accurate as the mentionned BSL-DG simulations.
(To remove these oscillations one may lower the remapping period $\Dtrec$ but at the cost of more diffusion; 
one should then use remapping operators less diffusive than the cubic spline quasi-interpolation.) 
On the other hand, as a forward particle method LTPIC is simpler to implement, and potentially cheaper to run compared to grid-based or 
BSL methods. 
We also note that the LTPIC simulation shown in Figure~\ref{fig. TSI_NY-evol} involves significantly 
larger time steps (namely $\Dt = 1/5$) than the BSL run where the announced CFL constant corresponds to $\Dt \approx 1/64$.
Moreover, as each fifth-order DG cell contains 15 basis functions in 2d, the $255 \times 255$ BSL-DG run
involves approximatively twice as many degrees of freedom as the plotted LTPIC simulation. 

\begin{figure} [ht!]
\begin{center}
\begin{tabular}{ccc}
\includegraphics[height=0.29\textwidth]{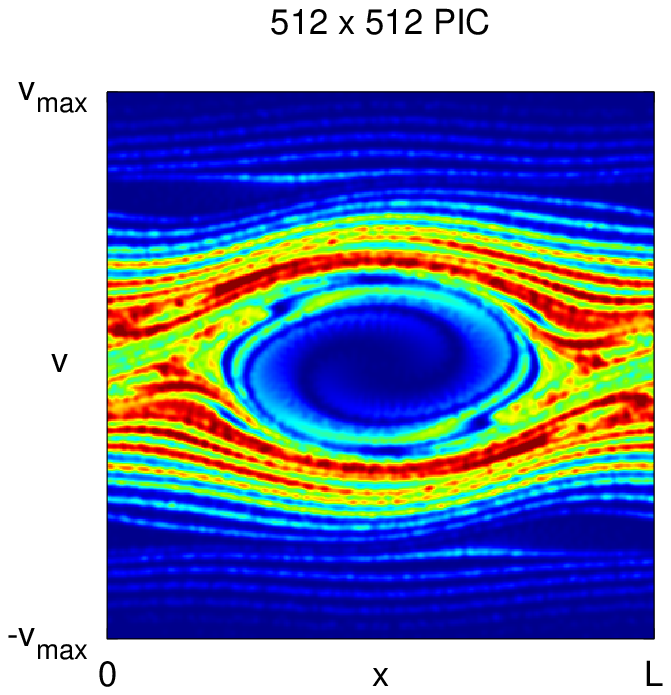} 
& \hspace{5pt}
\includegraphics[height=0.29\textwidth]{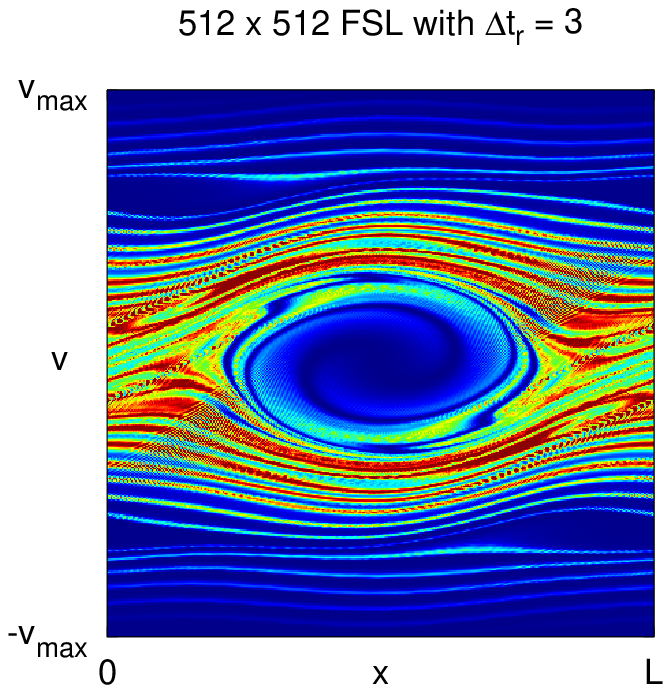} 
& \hspace{5pt}
\includegraphics[height=0.29\textwidth]{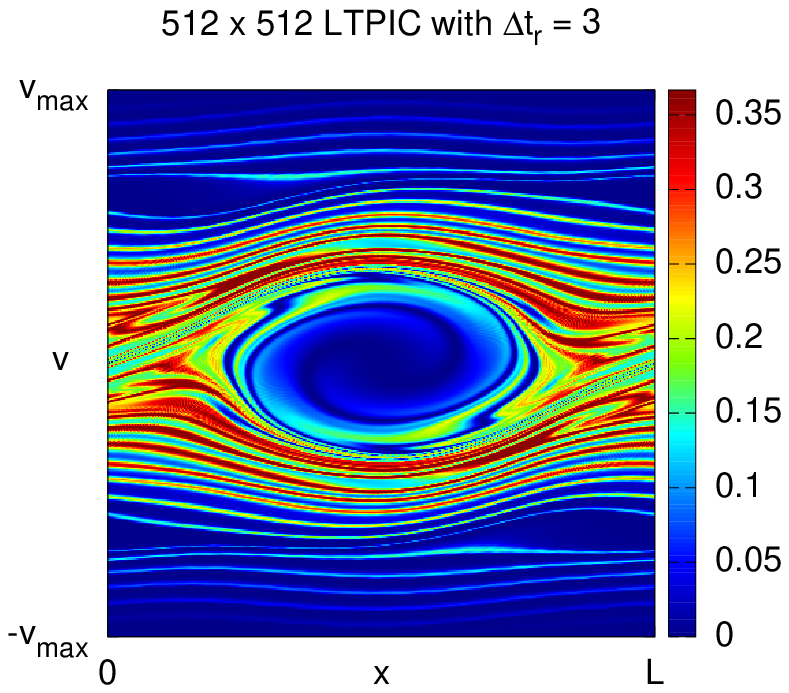} 
\vspace{5pt} \\ 
\includegraphics[height=0.29\textwidth]{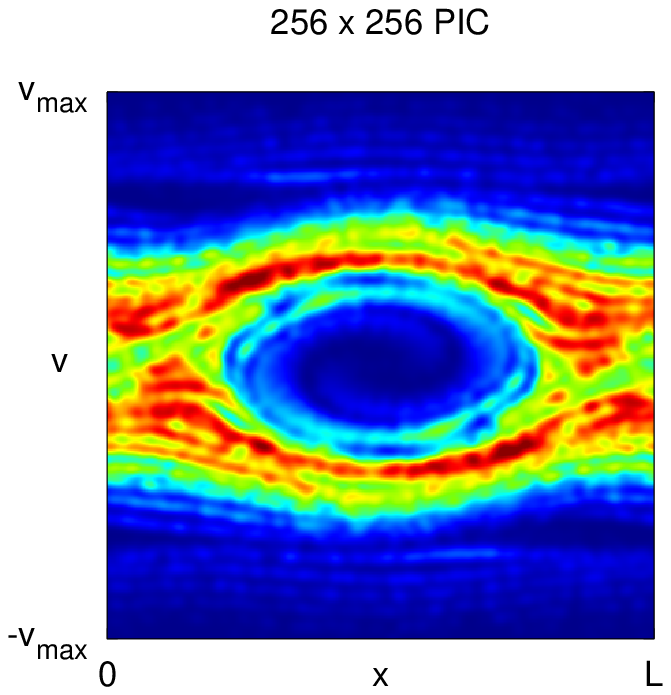} 
& \hspace{5pt} 
\includegraphics[height=0.29\textwidth]{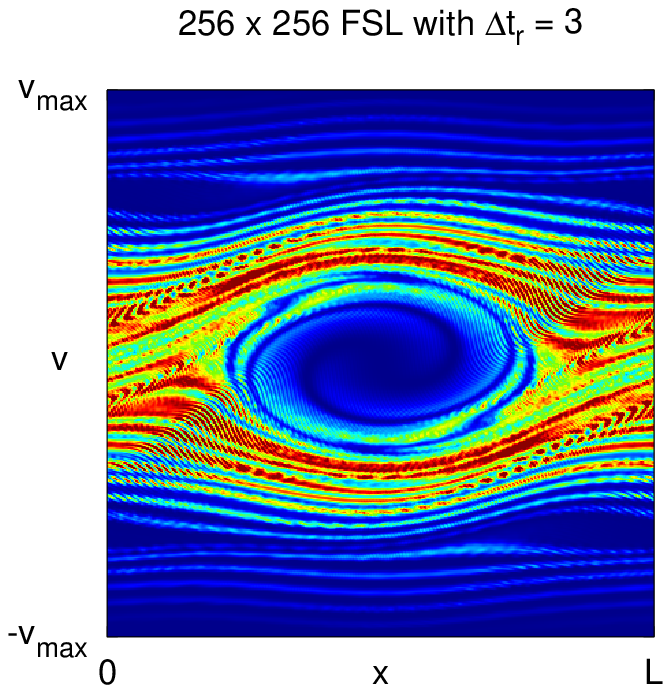} 
& \hspace{5pt} 
\includegraphics[height=0.29\textwidth]{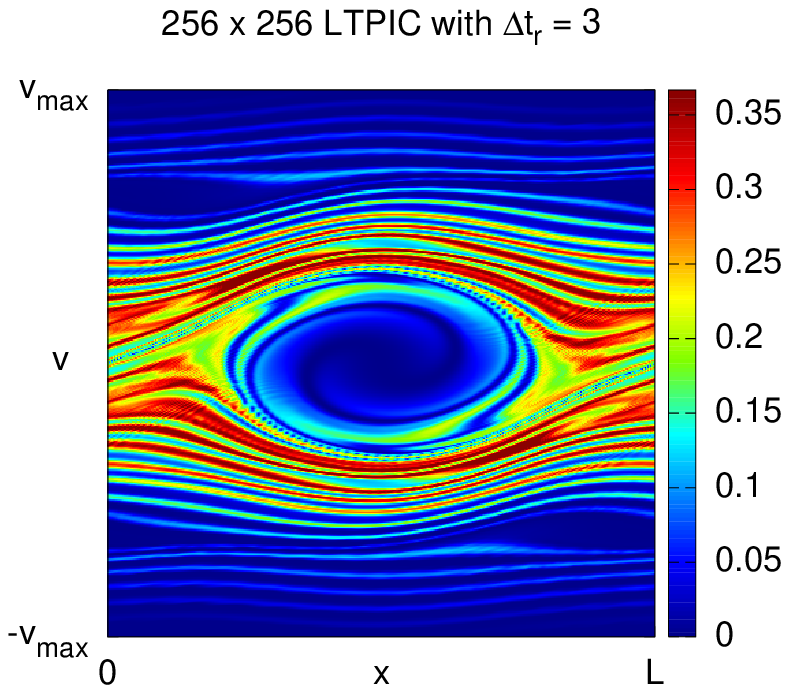} 
\end{tabular}
  
  \caption{Strong two-stream instability (case 2). 
  Comparisons of phase-space densities obtained at $t_n=45$ with high (top) and moderate (bottom) resolution runs. 
  All the runs use a time step $\Dt = 1/5$. The top runs use 128 cells and $512\times 512$ particles, 
  with approximate cpu times 600~s (PIC), 630~s (FSL) and 680~s (LTPIC).
  The bottom runs use 64 cells and $256 \times 256$ particles, with approximate cpu times 
  90~s (PIC), 90~s (FSL) and 110~s (LTPIC).
  Note that the remapping period $\Dtrec = 3$ taken here for the FSL and LTPIC runs is slightly longer than in Figure~\ref{fig. TSI_NY-evol}.
  }
  \label{fig. TSI_NY-t45}
 \end{center}
\end{figure}

Therefore, in the top row of Figure~\ref{fig. TSI_NY-t45} we show PIC, FSL and LTPIC runs obtained with $512 \times 512$ particles. 
Now the high-resolution LTPIC run involves about 1.5 as many degrees of freedom as the $256\times 256$ BSL-DG solution 
shown in Figure~4 from Ref.~\cite{Rossmanith.Seal.2011.jcp}, and it achieves a similar level of details, still with a very large time step.
In the bottom row of Figure~\ref{fig. TSI_NY-t45} we then show PIC, FSL and LTPIC runs using $256 \times 256$ particles,
for comparison. To highlight the robustness of LTPIC compared to FSL the remapping period is taken higher than in Figure~\ref{fig. TSI_NY-evol}.
Results indeed show strong ripples in the FSL solutions, but almost none in the LTPIC ones. 
By running the FSL method with $\Dtrec = 1.8$, we obtain solutions (not shown here) where oscillations are either significantly 
reduced in the $256 \times 256$ case or fully smoothed out in the $512 \times 512$ case.
Finally, our cpu time measurements show that FSL and PIC runs using similar numbers of Poisson cells and particles require 
very similar computational effort. With the same numerical parameters the LTPIC runs are only slightly longer, 
which again indicates that the extra work of deforming the particles is not dominant.

\subsection{Halo formation in a mismatched thermal sheet beam}
\label{sec. MTB}

We now consider the case of a 1D sheet beam in a continuous focusing 
channel with prescribed focusing strength
\begin{equation}
\label{kappa}
\kappa(s) \equiv k_{\beta_0}^2,
\end{equation}
as studied in Ref.~\cite{Lund.2011.prstab}. 
Here no electron cloud is present ($\kappa$ takes the role of a neutralizing species) and the density $f = f(s,x,x')$ 
models an axially thin, transverse slice of a continuous
($\partial/\partial z = \partial/\partial y = 0$)
ion beam composed of single species particles of charge $q$ and rest mass $m$.
The slice propagates with velocity $\beta_b c = {\rm const}$ and relativistic gamma factor 
$\gamma_b \equiv (1-\beta_b^2)^{-1/2}$ along the axial ($z$) direction. Here,
$c$ is the speed of light in vacuum. 
The beam phase space is described by the spatial coordinate $x$ and
the angle $x'$ that the particle trajectories make relative to the longitudinal 
axis, and the independent timelike coordinate $s$ represents
the axial coordinate of a reference particle of the beam (or of the slice being followed), 
measured along the design orbit (nominally the machine axis).
In this model the Vlasov equation reads 
\begin{equation}
\left\{ \frac{\partial}{\partial s} + \frac{\partial H}{\partial x'}\frac{\partial}{\partial x} 
- \frac{\partial H}{\partial x} \frac{\partial}{\partial x'} \right\} f(s,x,x') = 0
\label{Vlasov-eqn}
\end{equation}
with Hamiltonian
\begin{equation}
H \equiv \frac{1}{2}x^{\prime 2} + \frac{1}{2}\kappa x^2 + 
  \frac{q\phi(s,x)}{m\gamma_b^3 \beta_b^2 c^2}.
\label{Hamiltonian}
\end{equation}
Here, the electrostatic potential $\phi$ is the solution to the Poisson equation
\begin{equation}
\label{Poisson}
\frac{\partial^2\phi}{\partial x^2}(s,x) = - \frac{q}{\epsilon_0}n_i(s,x)
\end{equation}
solved with free space boundary conditions 
$- \frac{\partial \phi}{\partial x}(\pm \infty) = \pm \frac{q}{2\epsilon_0} N_i$ to obtain 
\begin{equation}
\label{Poisson-free}
E(s,x) \equiv - \frac{\partial \phi}{\partial x}(s,x) = \frac{q}{\epsilon_0}\Big(\int_{-\infty}^x n_i(s,\tilde x) \rmd \tilde x - \frac 12 N_i\Big)
\end{equation}
where $n_i(s,x) \equiv \int_\RR  f(s,x,x') \rmd x'$ is the ion density in configuration space 
and $N_i \equiv \int_\RR n_i(s,x)\rmd x$ is the integrated ion density, or total number of ions -- a constant, as particles are neither
created nor destroyed.
Following the procedure described therein, we shall first review how thermal equilibrium solutions
can be obtained with physical scales roughly consistent with a recent experiment for beam driven Warm 
Dense Matter called the NDCX-I at Lawrence Berkeley National Laboratory \cite{Bieniosek.al.2009.nimp}.
Specifically, we shall consider a 100 KeV kinetic energy potassium $K^+$ ion beam
(the axial velocity of which can be set nonrelativistically by $m \gamma_b \beta_b^2 c^2 / 2 \approx 10^5q$) with the sheet beam perveance 
\begin{equation}
\label{P}
P \equiv \frac{q^2 N_i}{2\epsilon_0 m \gamma_b^3\beta_b^2 c^2}
\end{equation}
to be specified below.
Following the analysis carried out in \cite{Lund.2011.prstab}, thermal equilibrium distributions can then be obtained as follows.
Given a specific value for the positive, dimensionless parameter 
\begin{equation}
\label{Delta}
\Delta \equiv \frac{\gamma_b^3\beta_b^2c^2 k_{\beta_0}^2}{\hat \omega_p^2} - 1
\end{equation}
where $\hat \omega_p \equiv [q^2\hat n/(\epsilon_0 m)]^{1/2}$ is the plasma frequency formed from the peak density scale $\hat n$,
a normalized effective potential $\psi_\Delta$ is defined as the solution of the transformed Poisson equation
\begin{equation}
\label{Poisson-trans}
\psi_\Delta''(\hat x)  = 1+\Delta - e^{-\psi_\Delta(\hat x)}
\quad \text{ with boundary conditions } \quad 
\psi_\Delta'(0) = \psi_\Delta(0) = 0.
\end{equation}
Then, the thermal distribution given by
\begin{equation}
\label{f0-thermeq}
f^{\rm eq}(x,x') \equiv 
    \frac {\hat n}{\sqrt{2\pi T^*}} \exp\left(-\frac{x^{\prime 2}}{2T^*}\right)\exp\left(-\psi_\Delta\bigg(\frac{x}{\gamma_b \lambda_D}\bigg)\right)
\end{equation}
yields a stationary equilibrium solution $f(s,x,x') = f^{\rm eq}(x,x')$ of the Vlasov-Poisson
system \eqref{Vlasov-eqn}, \eqref{Hamiltonian}-\eqref{Poisson-free}.
Here, $T^* \equiv T/(m \gamma_b \beta_b^2 c^2)$ is the dimensionless temperature associated with the thermodynamic
temperature $T$ (expressed in energy units), and $\lambda_D \equiv [T/(m\hat \omega_p^2)]^{1/2}$ is the corresponding Debye length.
We observe that the parameters in \eqref{f0-thermeq} can be derived by first inverting \eqref{Delta}, i.e.,
$$
\hat n = \frac{\epsilon_0 m \gamma_b^3\beta_b^2c^2 k_{\beta_0}^2}{q^2(1+\Delta)},
$$
and next infering from \eqref{P} and $N_i = \int_{-\infty}^{\infty}  n_i(x) \rmd x = 2 \hat n \,\gamma_b \lambda_D \int_0^\infty e^{-\psi_\Delta(\hat x)} \rmd \hat x$ that
\begin{equation}
\label{gamma_lambda}
\gamma_b\lambda_D = \frac{N_i}{2 \hat n \int_0^\infty e^{-\psi_\Delta(\hat x)}\rmd \hat x} 
= \frac{P}{k_{\beta_0}^2} \frac{(1+\Delta)}{\int_0^\infty e^{-\psi_\Delta(\hat x)} \rmd \hat x}.
\end{equation}
The resulting temperature is then 
$$
T^* = 
\left(\frac{P}{k_{\beta_0}}\right)^2 \frac{1+\Delta}{\Big(\int_0^\infty e^{-\psi_\Delta(\hat x)}\rmd \hat x \Big)^2}.
$$

The tune depression $\sigma/\sigma_0$ -- defined as the ratio between the phase advance of the particles in the presence
and absence of beam charge -- can be calculated \cite{Lund.2011.prstab} as 
\begin{equation}
\label{tunedep}
\sigma/\sigma_0 = \Bigg[1-\frac{1}{\sqrt{3}(1+\Delta)} 
\frac{\big(\int_0^\infty  e^{-\psi_\Delta(\hat x)} \rmd \hat x\big)^{3/2}}{\big(\int_0^\infty \hat x^2 e^{-\psi_\Delta(\hat x)}\rmd \hat x\big)^{1/2}} \Bigg]^{\frac 12}.
\end{equation}
By solving numerically \eqref{Poisson-trans}, \eqref{tunedep}, it is then possible to prescribe a specific tune depression 
and derive the corresponding value of $\Delta$ to specify the equilibrium distribution: the resulting parameters are given in 
\cite[Table II]{Lund.2011.prstab} for regularly spaced values of $\sigma/\sigma_0 \in \{ 0.1, 0.2, \ldots, 0.9\}$, and in Table~\ref{tab. beam-params} below for 
a strong tune depression, i.e. $\sigma/\sigma_0 = 0.1$. 
For the purpose of comparing our results to typical NDCX-I experiments, we set the focusing strength to
$k_{\beta_0} \equiv \sigma_0 / L_p$ in such a way that free particles have a phase advance of $\sigma_0 \sim \pi/3 $ 
per lattice period $L_p = 0.5$ m, and we set the perveance by taking $P/k_{\beta_0} = 0.01$. We note that for a sheet beam, the perveance has 
dimension 1/length and $P/k_{\beta_0}$ is dimensionless. 
The distribution corresponding to $\sigma/\sigma_0 = 0.1$ corresponds to a highly nonlinear form in $x$ due to the radial beam extent and the nonlinear solution
for the effective potential $\psi_\Delta$.

\begin{table}[htbp]
\begin{center} 
\begin{tabular}{cccccc}
\hline\hline
depression  & parameter & temperature    & Debye length & rms radius & peak density
\\ 
$\sigma/\sigma_0$  & $\Delta$ & $T^*$  & $\lambda_D$ & $x^{\rm eq}_b$ & $\hat n$  \vphantom{$\frac{R^t}{L_p}$}
\\
\hline
$0.1$  & $5.522 \times 10^{-8}$ & $3.463 \times 10^{-7}$   & $2.810\times 10^{-4}$m & $4.822\times 10^{-3}$m & $4.848\times 10^{13}$
\\
\hline 
\hline
\end{tabular}
\end{center}
\caption{Physical parameters for the matched thermal sheet beams with 100 KeV $K^+$ ions, corresponding to an axial beam velocity 
of $\beta_b c$ with $\beta_b = 2.343 \times 10^{-3}$ and relativistic factor $\gamma_b \approx 1$.
We take $k_{\beta_0} \approx 2.094 $ corresponding to a $60^\circ$ phase advance per lattice period $L_p = 0.5$ m, 
and the perveance is set to  $P = 0.01 k_{\beta_0}$. }
\label{tab. beam-params}
\end{table}

Finally, from \eqref{f0-thermeq} we derive an initially ``mismatched'' beam through a canonical transformation that dilates the distribution 
in the spatial dimension while preserving its perveance and initial effective phase-space area (emittance) by taking
\begin{equation}
\label{f0-mismatch}
f(s=0,x,x') \equiv f^{\rm eq}\left(\frac {x}{\mu},\mu x'\right), \qquad \mu > 0.
\end{equation}
Here $\mu$ corresponds to the mismatch parameter, defined as the ratio of the initial (rms) 
beam radius to the radius of the matched beam, see e.g. Ref.~\cite{Wangler.P.1996}.

\begin{figure} [hbtp]
\begin{center}
\begin{tabular}{ccc}
\includegraphics[height=0.29\textwidth]{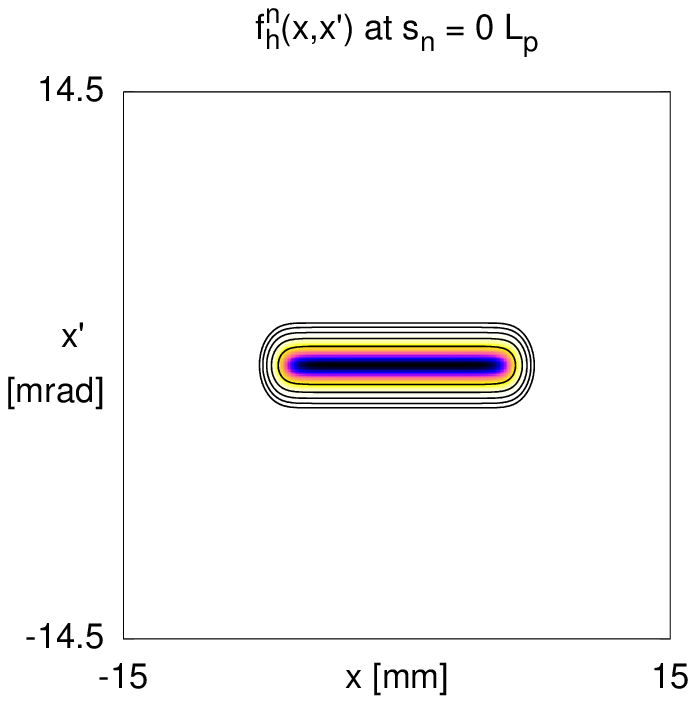}
& 
\includegraphics[height=0.29\textwidth]{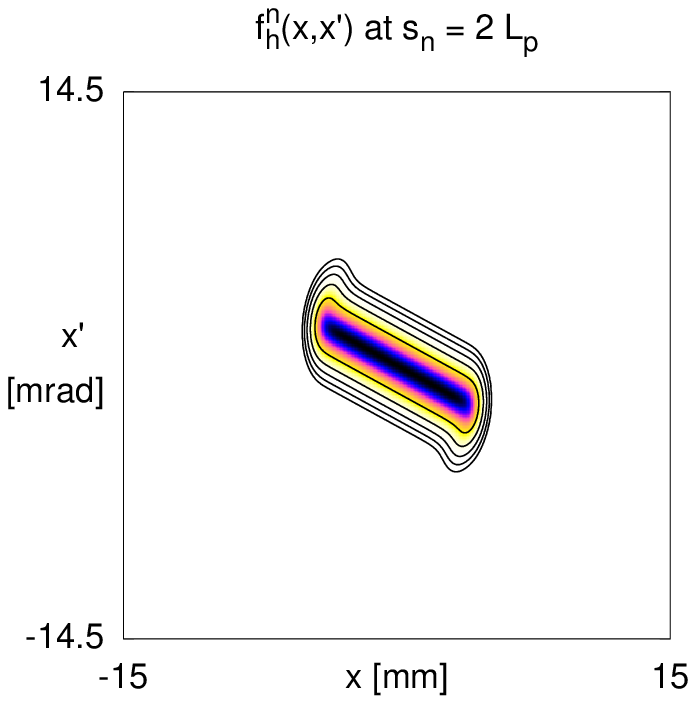}
& 
\includegraphics[height=0.29\textwidth]{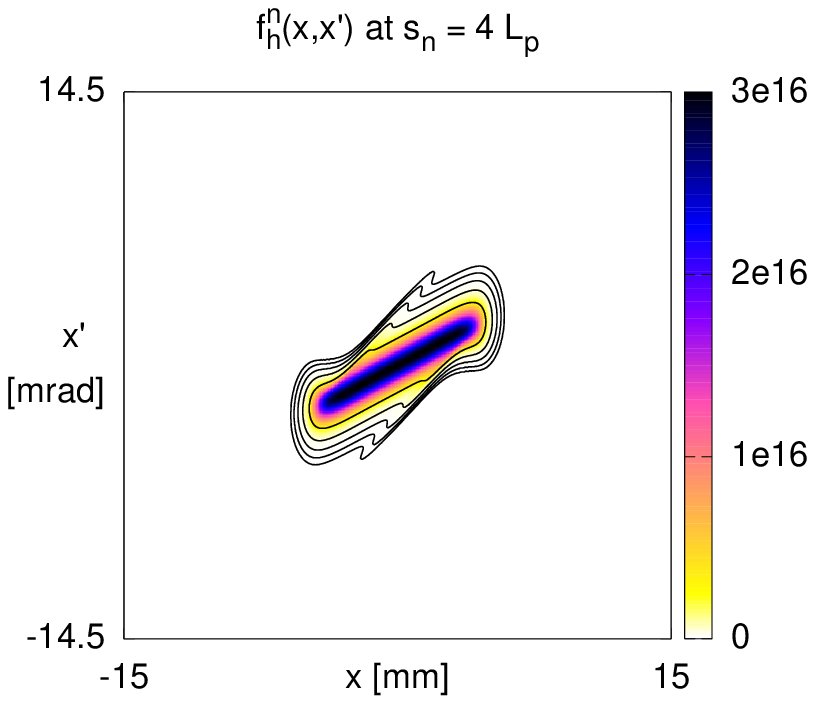}
\vspace{5pt} 
\\
\includegraphics[height=0.29\textwidth]{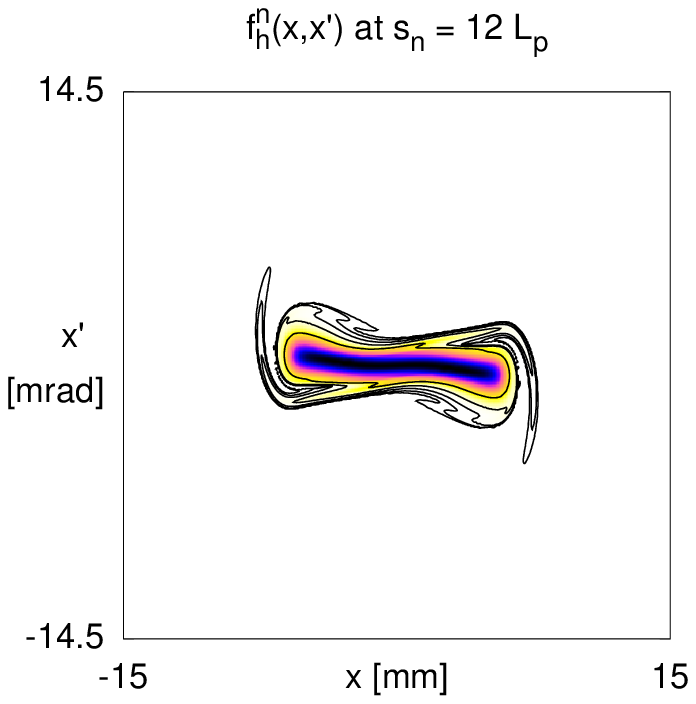}
& 
\includegraphics[height=0.29\textwidth]{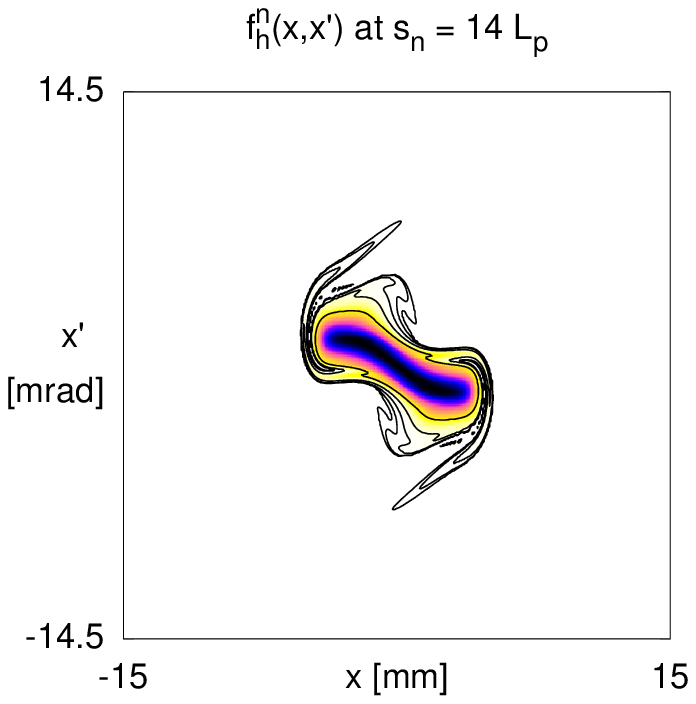}
& 
\includegraphics[height=0.29\textwidth]{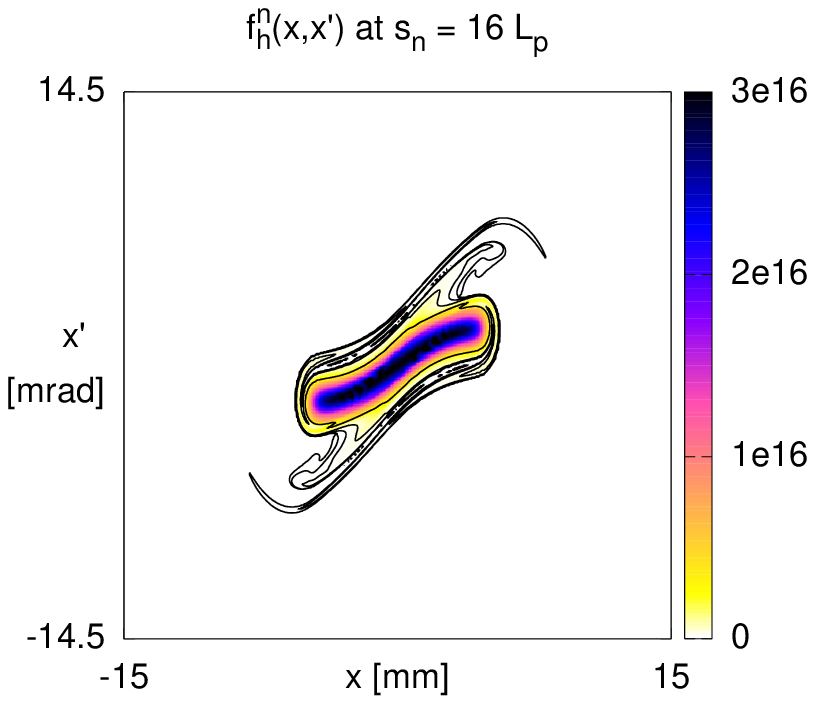}
\vspace{5pt} 
\\ 
\includegraphics[height=0.29\textwidth]{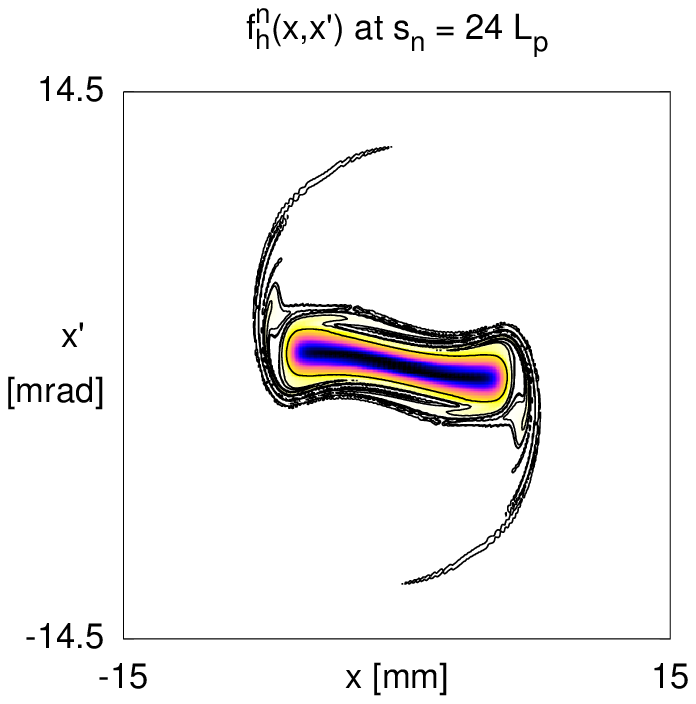}
& 
\includegraphics[height=0.29\textwidth]{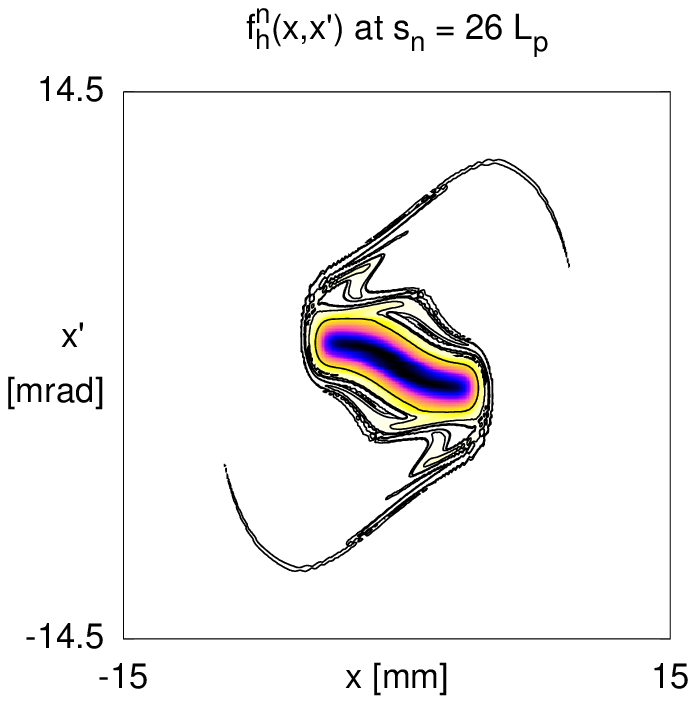}
& 
\includegraphics[height=0.29\textwidth]{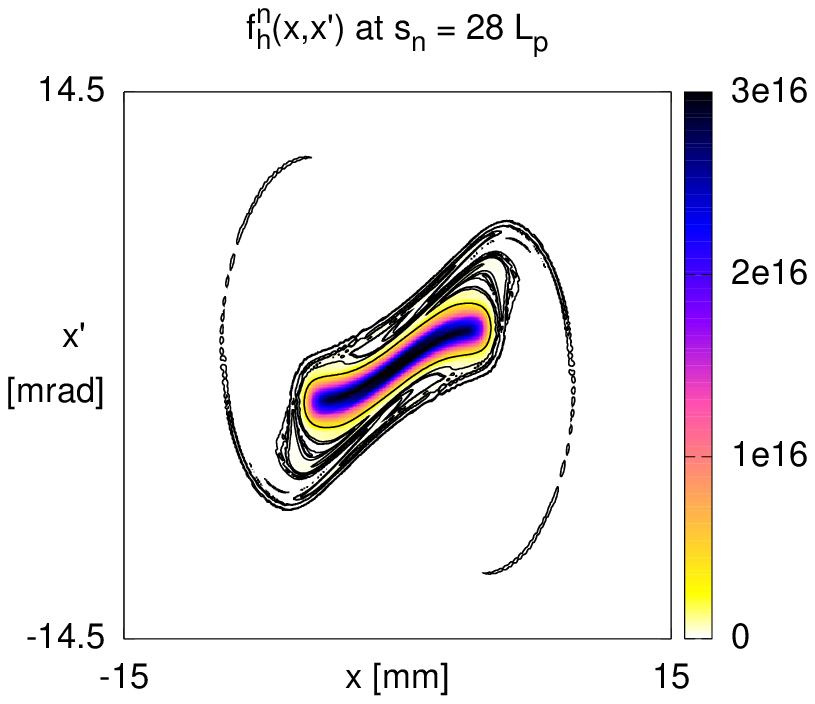}
\vspace{5pt} 
\\ 
\includegraphics[height=0.29\textwidth]{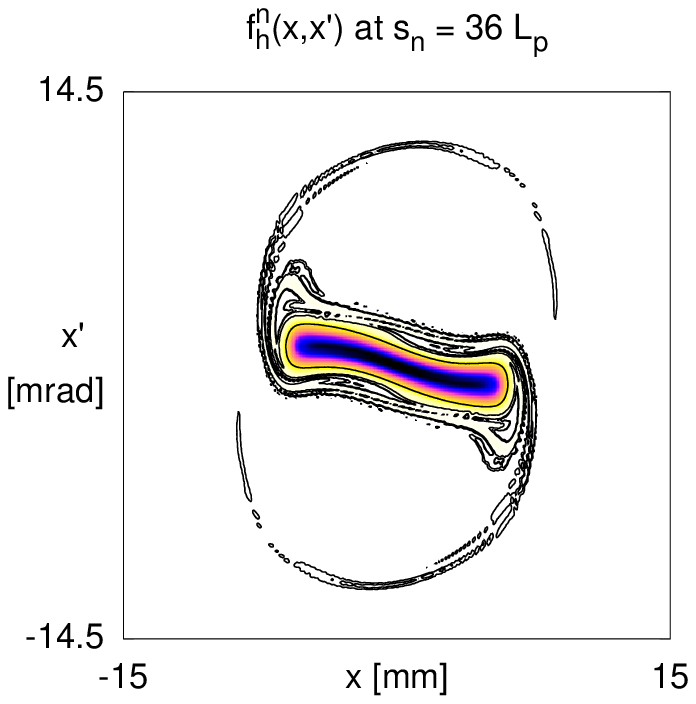}
& 
\includegraphics[height=0.29\textwidth]{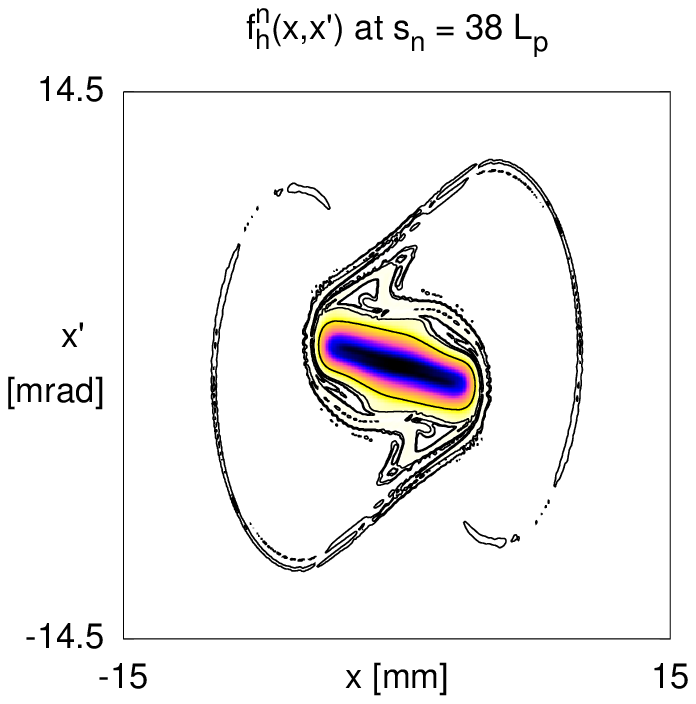}
& 
\includegraphics[height=0.29\textwidth]{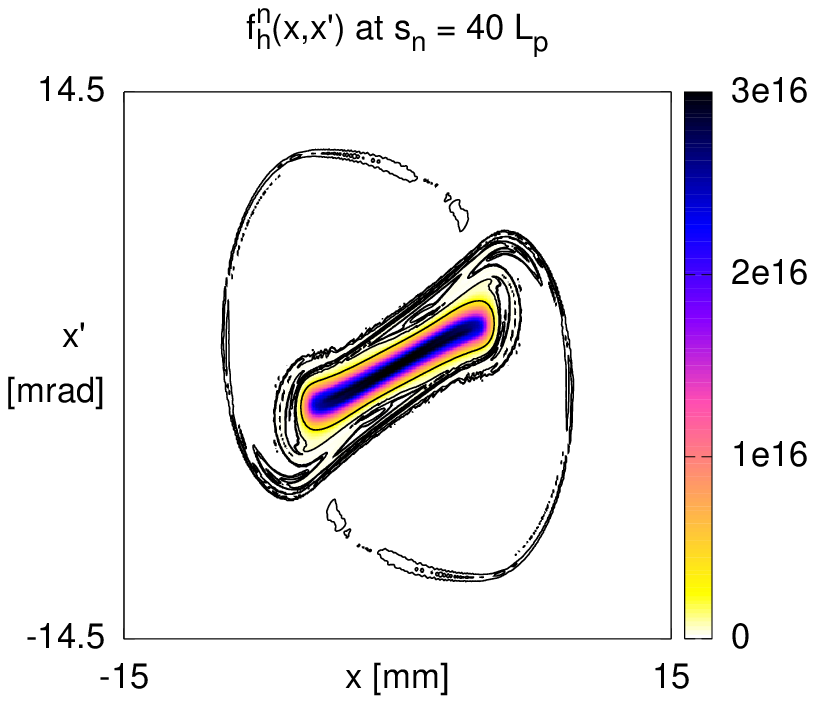}
\end{tabular}
  \caption{Mismatched thermal sheet beam in a continuous focusing lattice with phase advance of $60^\circ$ per period $L_p = 0.5 {\rm m}$,
  and tune depression $\sigma/\sigma_0 = 0.1$. The plots show the evolution of the phase-space density (with respect of the timelike, 
  longitudinal coordinate $s$) obtained with an LTPIC simulation. 
  This run uses a time step $\Ds=L_p/16$, a remapping period $\Dsrec = 2.5 L_p$, a Poisson solver with 128 cells and $256 \times 256$ particles.
  The halo is shown through isolines corresponding to values of $10^{-1}$, $10^{-2}$, \ldots, $10^{-5}$ of the peak initial density.
  The approximate cpu time for this run is 220~s.
  }
  \label{fig. mtb-evol}
 \end{center}
\end{figure}

In Figure~\ref{fig. mtb-evol} we show the evolution of a mismatched beam with a thermal equilibrium form 
specified by \eqref{f0-thermeq} using the procedure outlined above with $\mu = 1.25$ and a tune depression of $\sigma/\sigma_0 = 0.1$.
Here the numerical solution is computed with an LTPIC scheme 
on a computational domain corresponding to $\abs{x} \le 15 {\rm mm}$ and $\abs{x'} \le 14.5 {\rm mrad}$.
In the phase-space plots we vizualise the tenuous halo that evolves from the initial distribution by taking 
contours of the numerical density using exponential increments. Filled color contours illustrate the core of the phase-space density.

\begin{figure} [hbtp]
\begin{center}
\begin{tabular}{ccc}
\includegraphics[height=0.29\textwidth]{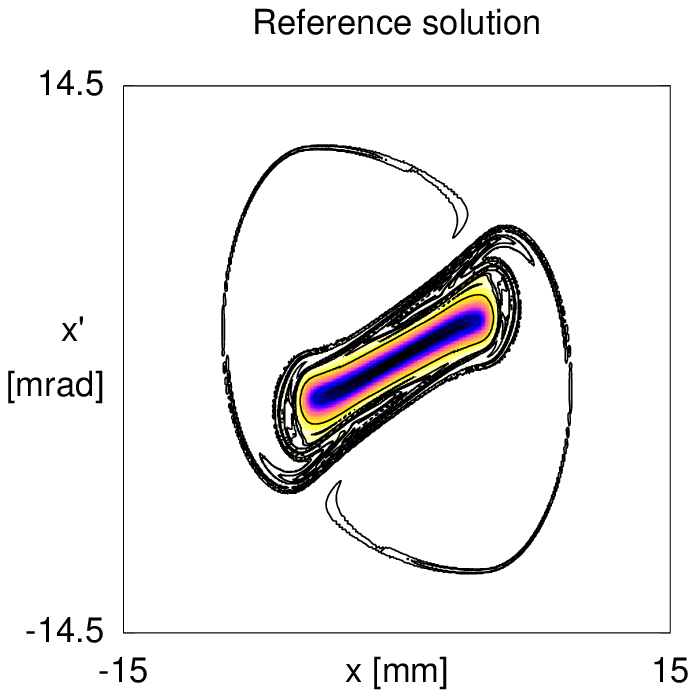} 
&
\includegraphics[height=0.29\textwidth]{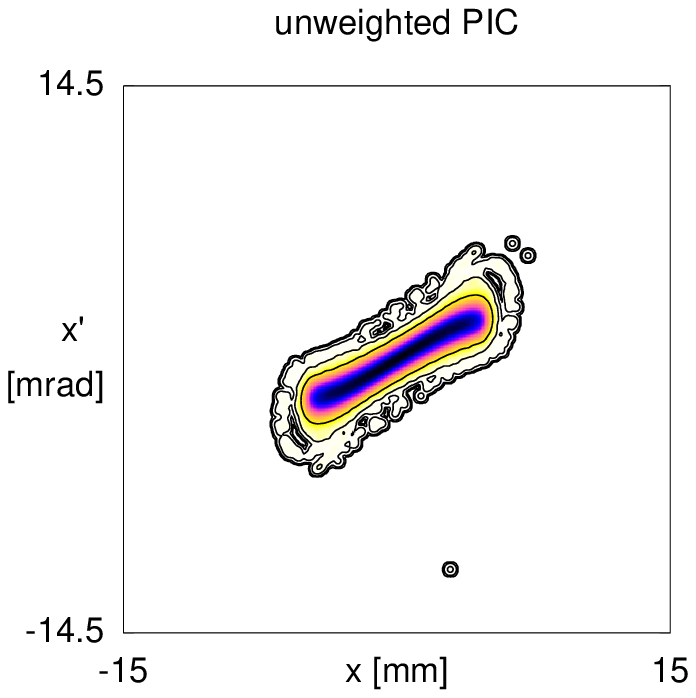} 
& 
\includegraphics[height=0.29\textwidth]{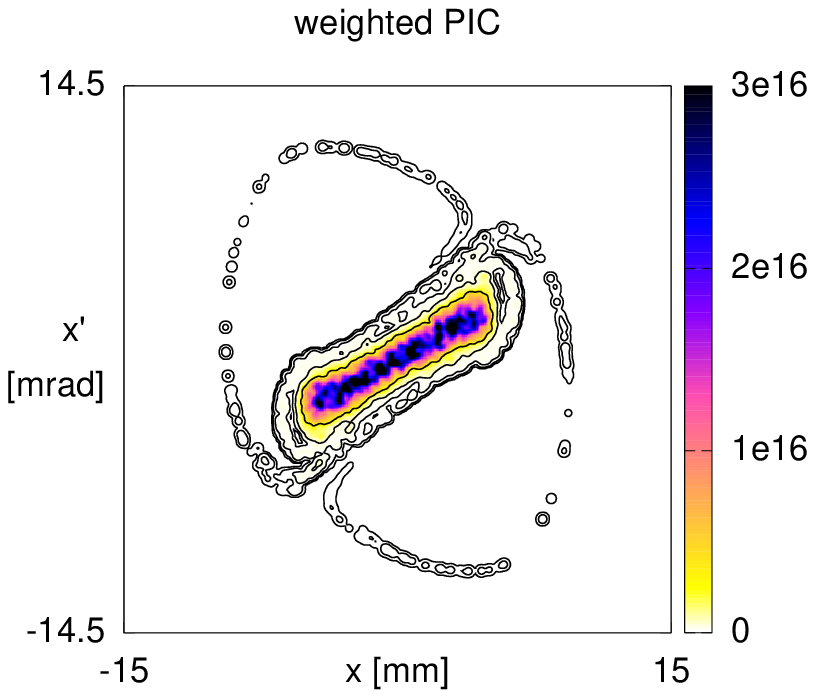} 
\vspace{5pt} 
\\ 
\includegraphics[height=0.29\textwidth]{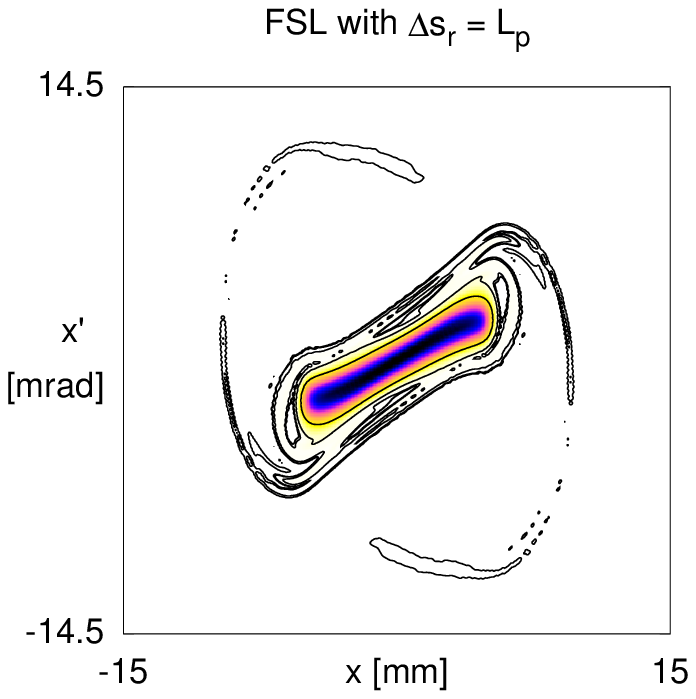} 
& 
\includegraphics[height=0.29\textwidth]{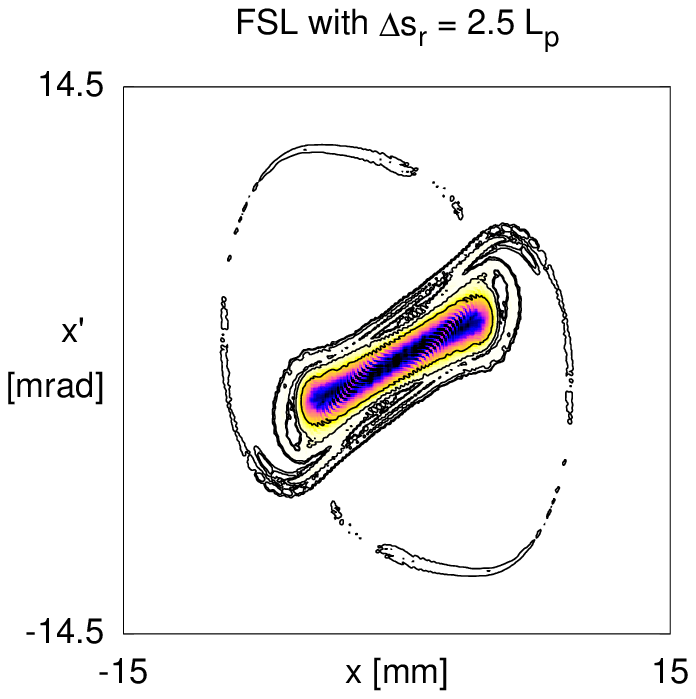} 
& 
\includegraphics[height=0.29\textwidth]{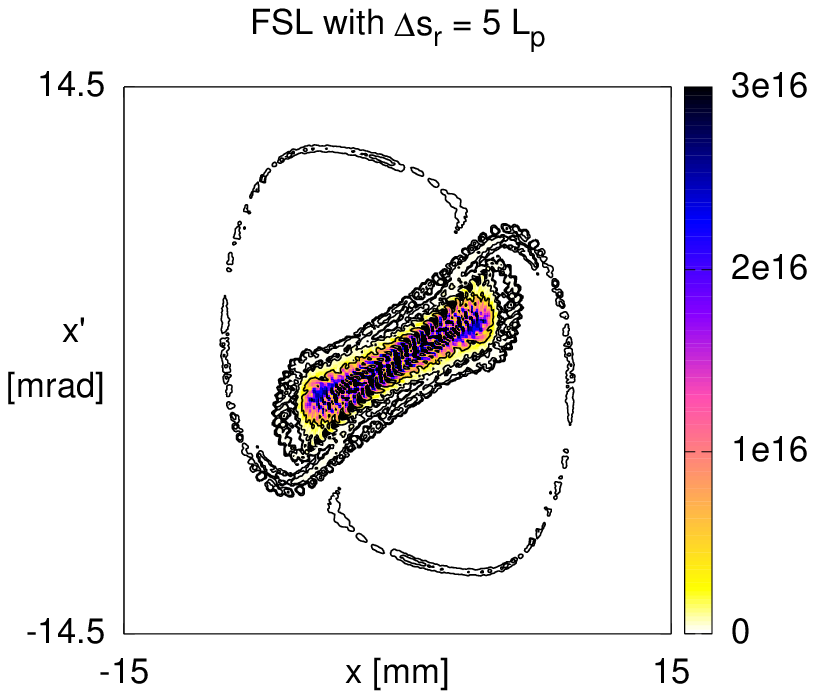} 
\vspace{5pt} 
\\
\includegraphics[height=0.29\textwidth]{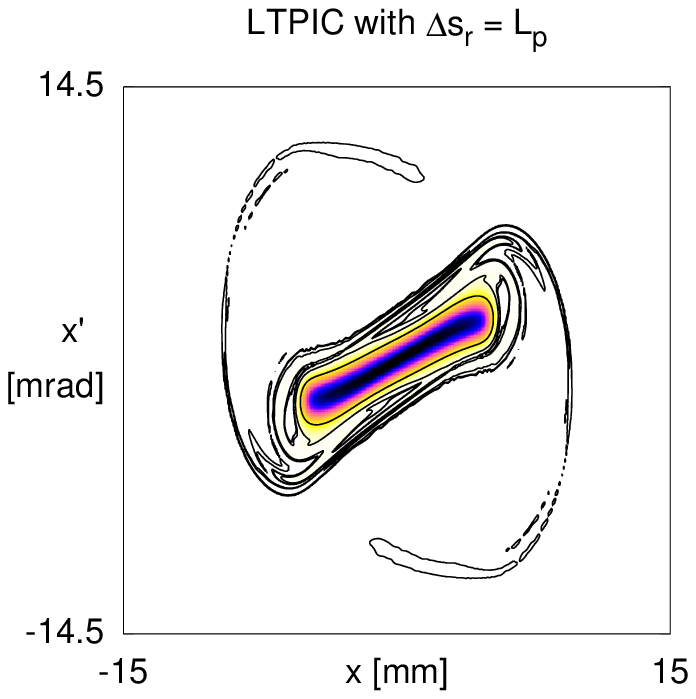} 
& 
\includegraphics[height=0.29\textwidth]{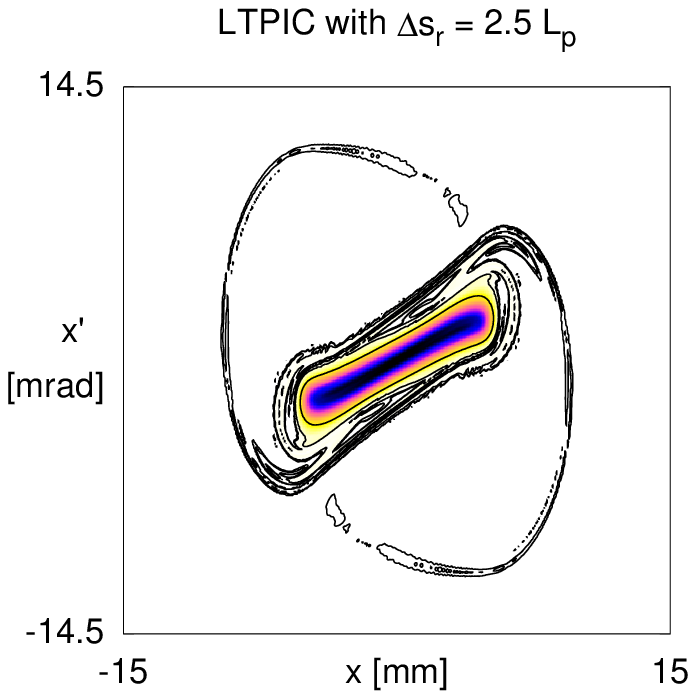} 
& 
\includegraphics[height=0.29\textwidth]{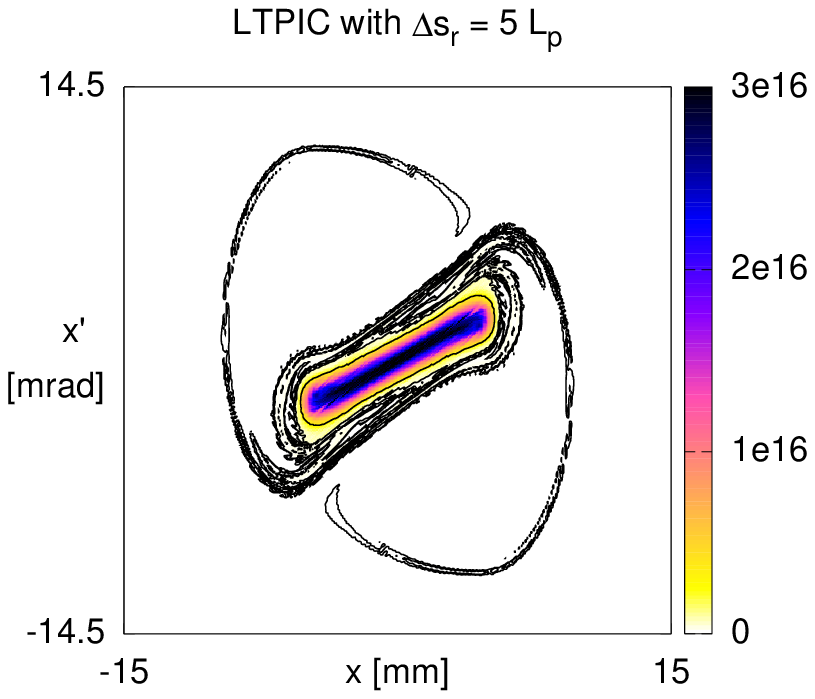} 
\end{tabular}
  \caption{Mismatched thermal beam. 
  Comparisons of phase-space densities obtained at $s_n=40 L_p$ with different methods.
  All the runs use a time step $\Ds = L_p/16$, a Poisson solver with 128 cells and $256\times 256$ particles, except for the reference simulation,
  an LTPIC run with 256 cells and $512\times 512$ particles.
  In the FSL and LTPIC runs the remapping period varies as indicated (in the reference run it is $\Dsrec = 2.5 L_p$).
  The approximate cpu times for these runs are 1450~s (reference LTPIC), 225~s (unweighted PIC), 145~s (weighted PIC), 220 to 265~s (FSL runs) and 200 to 245~s (LTPIC runs). 
  }
  \label{fig. mtb-compar}
 \end{center}
\end{figure}

In Figure~\ref{fig. mtb-compar} we next compare the phase-space densities at $s=20 {\rm m}$ with halo contours obtained 
with different schemes using $256 \times 256$ particles. Here we observe that the unweighted PIC has a low level of noise 
in the core but misses almost all of the halo. The weighted PIC simulation catches a fair proportion of the halo but in a very fragmented way,
and in addition it has a high level of noise in the core.
In contrast, the FSL simulation with short remapping periods (shown on the center left panel) does a reasonable job, although it still
misses some part of the halo arms. For longer remapping periods (center and center right panels) it is severly hampered by phase-space 
oscillations. In the LTPIC simulations (bottom panels) these numerical artifacts are significantly reduced, which shows once more the
ability of this new approach to remove the noise at reasonable computational cost, and with similar or improved accuracy.

\section{Conclusion}

We have presented a new deterministic PIC method for electrostatic plasma simulations, wherein finite-sized particles have a shape function that is linearly transformed in time to approximately follow the flow and thereby reduce the oscillations traditionally observed in standard PIC simulations. Although this method may be seen as an extension of the remapped-particle (FSL) scheme \cite{Denavit.1972.jcp} (due to the practical need to remap the particles after the flow has evolved significantly), its relative robustness to low remapping frequencies makes it actually closer to a proper particle scheme. By testing our method on benchmarked test cases we have demonstrated its ability to effectively reduce the noise and reach accuracy levels similar to those of expensive high order state-of-the-art Vlasov schemes. 

Simulations of real-world systems often need to capture the production of phase space structures by collective interactions. For example, in accelerator physics, beam halo must be minimized in order to limit the unwanted particle loss on machine surfaces. The simulated halo location and density must be quantitatively correct, in a calculation wherein the far denser core of the distribution must also be evolved self-consistently (since the core fields influence the halo dynamics).  On test problems of this type the LTPIC method performs well, and we believe that it has considerable promise to augment the standard PIC methods predominately employed today.


\section*{Acknowledgments}

Most of this work was carried out in 2010-2011 while Martin Campos Pinto was a Fulbright visiting scholar at the Lawrence Berkeley National Laboratory
with partial support from the France-Berkeley Fund. Fruitful interactions with John Verboncoeur are gratefully acknowledged. 
The work of the final three authors was performed under the auspices of the USDOE by LLNL under Contract DE-AC52-07NA27344.







\section*{References}
\bibliographystyle{model1-num-names}
\bibliography{jcp_martin.bib}







\end{document}

%% file: gauss-dep.pstex_t
\begin{picture}(0,0)%
\epsfig{file=gauss-dep.pstex}%
\end{picture}%
\setlength{\unitlength}{3552sp}%
\begingroup\makeatletter\ifx\SetFigFont\undefined%
\gdef\SetFigFont#1#2#3#4#5{%
  \reset@font\fontsize{#1}{#2pt}%
  \fontfamily{#3}\fontseries{#4}\fontshape{#5}%
  \selectfont}%
\fi\endgroup%
\begin{picture}(6212,2972)(726,-5336)
\put(3103,-3151){\makebox(0,0)[lb]{\smash{\SetFigFont{10}{12.0}{\familydefault}{\mddefault}{\updefault}$\hat \rho^n_{h,k}(x_{i-1}) \approx \tilde \rho^n_{h,k}(x_{i-1})$}}}
\put(6512,-4792){\makebox(0,0)[rb]{\smash{\SetFigFont{10}{12.0}{\familydefault}{\mddefault}{\updefault}$x$}}}
\put(2351,-5209){\makebox(0,0)[rb]{\smash{\SetFigFont{10}{12.0}{\familydefault}{\mddefault}{\updefault}$h'$}}}
\put(3711,-4673){\makebox(0,0)[lb]{\smash{\SetFigFont{10}{12.0}{\familydefault}{\mddefault}{\updefault}$\varphi_{h'}(x-x_i)$}}}
\put(3323,-5209){\makebox(0,0)[b]{\smash{\SetFigFont{10}{12.0}{\familydefault}{\mddefault}{\updefault}$x_i = ih'$}}}
\put(2764,-2729){\makebox(0,0)[lb]{\smash{\SetFigFont{10}{12.0}{\familydefault}{\mddefault}{\updefault}$\hat \rho^n_{h,k}(x_{i-2}) \approx \tilde \rho^n_{h,k}(x_{i-2})$}}}
\put(1431,-2558){\makebox(0,0)[lb]{\smash{\SetFigFont{10}{12.0}{\familydefault}{\mddefault}{\updefault}$v$}}}
\put(2212,-3163){\makebox(0,0)[rb]{\smash{\SetFigFont{10}{12.0}{\familydefault}{\mddefault}{\updefault}$\varphi^n_{h,k}(x,v)$}}}
\put(1229,-2804){\makebox(0,0)[rb]{\smash{\SetFigFont{10}{12.0}{\familydefault}{\mddefault}{\updefault}$v^+(i-2)$}}}
\put(1229,-4095){\makebox(0,0)[rb]{\smash{\SetFigFont{10}{12.0}{\familydefault}{\mddefault}{\updefault}$v^-(i-2)$}}}
\put(5074,-3796){\makebox(0,0)[b]{\smash{\SetFigFont{10}{12.0}{\familydefault}{\mddefault}{\updefault}$\boxed{\begin{array}{c} \text{charge deposited on } ~ x_i: \\ \rho^n_i = h'\sum_{|l|\le m_p} a_l \, \hat \rho^n_{h,k}(x_{i+l}) \end{array}}$}}}
\end{picture}

%% file: moment-dep.pstex_t
\begin{picture}(0,0)%
\epsfig{file=moment-dep.pstex}%
\end{picture}%
\setlength{\unitlength}{3552sp}%
\begingroup\makeatletter\ifx\SetFigFont\undefined%
\gdef\SetFigFont#1#2#3#4#5{%
  \reset@font\fontsize{#1}{#2pt}%
  \fontfamily{#3}\fontseries{#4}\fontshape{#5}%
  \selectfont}%
\fi\endgroup%
\begin{picture}(6212,2687)(726,-5336)
\put(2622,-4200){\makebox(0,0)[b]{\smash{\SetFigFont{10}{12.0}{\familydefault}{\mddefault}{\updefault}$\approx \int \rmd v$}}}
\put(1126,-2911){\makebox(0,0)[rb]{\smash{\SetFigFont{10}{12.0}{\familydefault}{\mddefault}{\updefault}$v$}}}
\put(5135,-3459){\makebox(0,0)[b]{\smash{\SetFigFont{10}{12.0}{\familydefault}{\mddefault}{\updefault}$\boxed{\begin{array}{c} \text{charge deposited on } ~ x_i: \\ \rho^n_i = h' \sum_{|l|\le m_p} a_l \, \hat \rho^n_{h,k}(x_{i+l}) \end{array}}$}}}
\put(6512,-4792){\makebox(0,0)[rb]{\smash{\SetFigFont{10}{12.0}{\familydefault}{\mddefault}{\updefault}$x$}}}
\put(2351,-5209){\makebox(0,0)[rb]{\smash{\SetFigFont{10}{12.0}{\familydefault}{\mddefault}{\updefault}$h'$}}}
\put(1802,-4607){\makebox(0,0)[b]{\smash{\SetFigFont{10}{12.0}{\familydefault}{\mddefault}{\updefault}$\hat \rho^n_{h,k}(x)$}}}
\put(4193,-4456){\makebox(0,0)[lb]{\smash{\SetFigFont{10}{12.0}{\familydefault}{\mddefault}{\updefault}$\varphi_{h'}(x-x_i)$}}}
\put(2335,-3092){\makebox(0,0)[rb]{\smash{\SetFigFont{10}{12.0}{\familydefault}{\mddefault}{\updefault}$\varphi^n_{h,k}(x,v)$}}}
\put(3666,-5209){\makebox(0,0)[b]{\smash{\SetFigFont{10}{12.0}{\familydefault}{\mddefault}{\updefault}$x_i =i h'$}}}
\end{picture}

%% file: jcp_ltpic_arxiv-v2.bbl
\begin{thebibliography}{34}
\expandafter\ifx\csname natexlab\endcsname\relax\def\natexlab#1{#1}\fi
\providecommand{\bibinfo}[2]{#2}
\ifx\xfnm\relax \def\xfnm[#1]{\unskip,\space#1}\fi
\bibitem[{Beale and Majda(1982)}]{Beale.Majda.1982b.mcomp}
\bibinfo{author}{J.~T. Beale}, \bibinfo{author}{A.~Majda},
\newblock \bibinfo{title}{{Vortex methods. II. Higher order accuracy in two and
  three dimensions}},
\newblock \bibinfo{journal}{Mathematics of Computation} \bibinfo{volume}{39}
  (\bibinfo{year}{1982}) \bibinfo{pages}{29--52}.
\bibitem[{Raviart(1985)}]{Raviart.1985.lnm}
\bibinfo{author}{P.-A. Raviart},
\newblock \bibinfo{title}{{An analysis of particle methods}},
\newblock in: \bibinfo{booktitle}{Numerical methods in fluid dynamics (Como,
  1983)}, \bibinfo{publisher}{Lecture Notes in Mathematics},
  \bibinfo{address}{Berlin}, \bibinfo{year}{1985}, pp.
  \bibinfo{pages}{243--324}.
\bibitem[{Ghizzo et~al.(1992)Ghizzo, Bertrand, Shoucri, Johnston, Fijalkow,
  Feix, and Demchenko}]{Ghizzo.al.1992.nfus}
\bibinfo{author}{A.~Ghizzo}, \bibinfo{author}{P.~Bertrand},
  \bibinfo{author}{M.~Shoucri}, \bibinfo{author}{T.~Johnston},
  \bibinfo{author}{E.~Fijalkow}, \bibinfo{author}{M.~Feix},
  \bibinfo{author}{V.~Demchenko},
\newblock \bibinfo{title}{{Study of laser-plasma beat wave current drive with
  an Eulerian Vlasov code}},
\newblock \bibinfo{journal}{Nuclear Fusion} \bibinfo{volume}{32}
  (\bibinfo{year}{1992}) \bibinfo{pages}{45--65}.
\bibitem[{Shoucri et~al.(2004)Shoucri, Cardinali, Matte, and
  Spigler}]{Shoucri.al.2004.epjd}
\bibinfo{author}{M.~Shoucri}, \bibinfo{author}{A.~Cardinali},
  \bibinfo{author}{J.~Matte}, \bibinfo{author}{R.~Spigler},
\newblock \bibinfo{title}{{Numerical study of plasma-wall transition using an
  Eulerian Vlasov code}},
\newblock \bibinfo{journal}{The European Physical Journal D}
  \bibinfo{volume}{30} (\bibinfo{year}{2004}) \bibinfo{pages}{81--92}.
\bibitem[{Valsaque and Manfredi(2001)}]{Valsaque.Manfredi.2001.jnucmat}
\bibinfo{author}{F.~Valsaque}, \bibinfo{author}{G.~Manfredi},
\newblock \bibinfo{title}{{Numerical study of plasma--wall transition in an
  oblique magnetic field}},
\newblock \bibinfo{journal}{Journal of nuclear materials}
  \bibinfo{volume}{290-293} (\bibinfo{year}{2001}) \bibinfo{pages}{763--767}.
\bibitem[{Sonnendr{\"u}cker et~al.(2001)Sonnendr{\"u}cker, Barnard, Friedman,
  Grote, and Lund}]{Sonnendrucker.al.2001.nima}
\bibinfo{author}{E.~Sonnendr{\"u}cker}, \bibinfo{author}{J.~J. Barnard},
  \bibinfo{author}{A.~Friedman}, \bibinfo{author}{D.~P. Grote},
  \bibinfo{author}{S.~M. Lund},
\newblock \bibinfo{title}{{Simulation of heavy ion beams with a semi-Lagrangian
  Vlasov solver}},
\newblock \bibinfo{journal}{Nuclear Instruments and Methods in Physics Research
  A} \bibinfo{volume}{464} (\bibinfo{year}{2001}) \bibinfo{pages}{470--476}.
\bibitem[{Goldman et~al.(2008)Goldman, Newman, and
  Pritchett}]{Goldman.Newman.Pritchett.2008.grl}
\bibinfo{author}{M.~V. Goldman}, \bibinfo{author}{D.~L. Newman},
  \bibinfo{author}{P.~Pritchett},
\newblock \bibinfo{title}{{Vlasov simulations of electron holes driven by
  particle distributions from PIC reconnection simulations with a guide
  field}},
\newblock \bibinfo{journal}{Geophysical Research Letters} \bibinfo{volume}{35}
  (\bibinfo{year}{2008}) \bibinfo{pages}{L22109}.
\bibitem[{Denavit(1972)}]{Denavit.1972.jcp}
\bibinfo{author}{J.~Denavit},
\newblock \bibinfo{title}{{Numerical simulation of plasmas with periodic
  smoothing in phase space}},
\newblock \bibinfo{journal}{Journal of Computational Physics}
  \bibinfo{volume}{9} (\bibinfo{year}{1972}) \bibinfo{pages}{75--98}.
\bibitem[{Nair et~al.(2003)Nair, Scroggs, and
  Semazzi}]{Nair.Scroggs.Semazzi.2003.jcp}
\bibinfo{author}{R.~D. Nair}, \bibinfo{author}{J.~S. Scroggs},
  \bibinfo{author}{F.~H.~M. Semazzi},
\newblock \bibinfo{title}{{A forward-trajectory global semi-Lagrangian
  transport scheme}},
\newblock \bibinfo{journal}{Journal of Computational Physics}
  \bibinfo{volume}{190} (\bibinfo{year}{2003}) \bibinfo{pages}{275--294}.
\bibitem[{Crouseilles et~al.(2009)Crouseilles, Respaud, and
  Sonnendr{\"u}cker}]{Crouseilles.Respaud.Sonnendrucker.2009.cpc}
\bibinfo{author}{N.~Crouseilles}, \bibinfo{author}{T.~Respaud},
  \bibinfo{author}{E.~Sonnendr{\"u}cker},
\newblock \bibinfo{title}{{A forward semi-Lagrangian method for the numerical
  solution of the Vlasov equation}},
\newblock \bibinfo{journal}{Computer Physics Communications}
  \bibinfo{volume}{180} (\bibinfo{year}{2009}) \bibinfo{pages}{1730--1745}.
\bibitem[{Bergdorf and Koumoutsakos(2006)}]{Bergdorf.Koumoutsakos.2006.MMS}
\bibinfo{author}{M.~Bergdorf}, \bibinfo{author}{P.~Koumoutsakos},
\newblock \bibinfo{title}{{A lagrangian particle-wavelet method}},
\newblock \bibinfo{journal}{Multiscale Model. Simul.} \bibinfo{volume}{5}
  (\bibinfo{year}{2006}) \bibinfo{pages}{980--995}.
\bibitem[{Wang et~al.(2011)Wang, Miller, and
  Colella}]{Wang.Miller.Colella.2011.jsc}
\bibinfo{author}{B.~Wang}, \bibinfo{author}{G.~Miller},
  \bibinfo{author}{P.~Colella},
\newblock \bibinfo{title}{{A Particle-In-Cell method with adaptive phase-space
  remapping for kinetic plasmas}},
\newblock \bibinfo{journal}{SIAM Journal on Scientific Computing}
  \bibinfo{volume}{33} (\bibinfo{year}{2011}) \bibinfo{pages}{3509--3537}.
\bibitem[{Chehab et~al.(2005)Chehab, Cohen, Jennequin, Nieto, Roland, and
  Roche}]{Chehab.al.2005.irma}
\bibinfo{author}{J.-P. Chehab}, \bibinfo{author}{A.~Cohen},
  \bibinfo{author}{D.~Jennequin}, \bibinfo{author}{J.~Nieto},
  \bibinfo{author}{C.~Roland}, \bibinfo{author}{J.~Roche},
\newblock \bibinfo{title}{An adaptive particle-in-cell method using
  multi-resolution analysis},
\newblock in: \bibinfo{booktitle}{Numerical methods for hyperbolic and kinetic
  problems}, volume~\bibinfo{volume}{7} of \textit{\bibinfo{series}{IRMA Lect.
  Math. Theor. Phys.}}, \bibinfo{publisher}{Eur. Math. Soc., Z\"urich},
  \bibinfo{year}{2005}, pp. \bibinfo{pages}{29--42}.
\bibitem[{Terzic and Pogorelov(2005)}]{Terzic.Pogorelov.2005.nyac}
\bibinfo{author}{B.~Terzic}, \bibinfo{author}{I.~Pogorelov},
\newblock \bibinfo{title}{{Wavelet-Based Poisson Solver for Use in
  Particle-in-Cell Simulations}},
\newblock \bibinfo{journal}{Annals of the New York Academy of Sciences}
  \bibinfo{volume}{1045} (\bibinfo{year}{2005}) \bibinfo{pages}{55--67}.
\bibitem[{Gassama et~al.(2007)Gassama, Sonnendr{\"u}cker, Schneider, Farge, and
  Domingues}]{Gassama.al.2007.esaim}
\bibinfo{author}{S.~Gassama}, \bibinfo{author}{E.~Sonnendr{\"u}cker},
  \bibinfo{author}{K.~Schneider}, \bibinfo{author}{M.~Farge},
  \bibinfo{author}{M.~Domingues},
\newblock \bibinfo{title}{{Wavelet denoising for postprocessing of a 2D
  Particle-In-Cell code}},
\newblock \bibinfo{journal}{ESAIM: Proceedings} \bibinfo{volume}{16}
  (\bibinfo{year}{2007}) \bibinfo{pages}{195--210}.
\bibitem[{Campos~Pinto(2012)}]{Campos-Pinto.2012.sub}
\bibinfo{author}{M.~Campos~Pinto},
\newblock \bibinfo{title}{{Smooth particle methods without smoothing}},
\newblock \bibinfo{journal}{arXiv:1112.1859 (to be submitted for publication)}
  (\bibinfo{year}{2012}) \bibinfo{pages}{1--32}.
\bibitem[{Hou(1990)}]{Hou.1990.sinum}
\bibinfo{author}{T.~Hou},
\newblock \bibinfo{title}{{Convergence of a Variable Blob Vortex Method for the
  Euler and Navier-Stokes Equations}},
\newblock \bibinfo{journal}{SIAM Journal on Numerical Analysis}
  \bibinfo{volume}{27} (\bibinfo{year}{1990}) \bibinfo{pages}{1387--1404}.
\bibitem[{Cohen and Perthame(2000)}]{Cohen.Perthame.2000.sima}
\bibinfo{author}{A.~Cohen}, \bibinfo{author}{B.~Perthame},
\newblock \bibinfo{title}{{Optimal Approximations of Transport Equations by
  Particle and Pseudoparticle Methods}},
\newblock \bibinfo{journal}{SIAM Journal on Mathematical Analysis}
  \bibinfo{volume}{32} (\bibinfo{year}{2000}) \bibinfo{pages}{616--636}.
\bibitem[{Bateson and Hewett(1998)}]{Bateson.Hewett.1998.jcp}
\bibinfo{author}{W.~Bateson}, \bibinfo{author}{D.~Hewett},
\newblock \bibinfo{title}{{Grid and Particle Hydrodynamics}},
\newblock \bibinfo{journal}{Journal of Computational Physics}
  \bibinfo{volume}{144} (\bibinfo{year}{1998}) \bibinfo{pages}{358--378}.
\bibitem[{Hewett(2003)}]{Hewett.2003.jcp}
\bibinfo{author}{D.~Hewett},
\newblock \bibinfo{title}{{Fragmentation, merging, and internal dynamics for
  PIC simulation with finite size particles}},
\newblock \bibinfo{journal}{Journal of Computational Physics}
  \bibinfo{volume}{189} (\bibinfo{year}{2003}) \bibinfo{pages}{390--426}.
\bibitem[{Alard and Colombi(2005)}]{Alard.Colombi.2005.ras}
\bibinfo{author}{C.~Alard}, \bibinfo{author}{S.~Colombi},
\newblock \bibinfo{title}{{A cloudy Vlasov solution}},
\newblock \bibinfo{journal}{Monthly Notices of the Royal Astronomical Society}
  \bibinfo{volume}{359} (\bibinfo{year}{2005}) \bibinfo{pages}{123--163}.
\bibitem[{Lund et~al.(2011)Lund, Friedman, and Bazouin}]{Lund.2011.prstab}
\bibinfo{author}{S.~M. Lund}, \bibinfo{author}{A.~Friedman},
  \bibinfo{author}{G.~Bazouin},
\newblock \bibinfo{title}{{Sheet beam model for intense space charge:
  Application to Debye screening and the distribution of particle oscillation
  frequencies in a thermal equilibrium beam}},
\newblock \bibinfo{journal}{Phys.\ Rev.\ Special Topics -- Accelerators and
  Beams} \bibinfo{volume}{14} (\bibinfo{year}{2011}) \bibinfo{pages}{054201}.
\bibitem[{{de Boor}(1978)}]{deBoor.B.1978}
\bibinfo{editor}{C.~{de Boor}} (Ed.), \bibinfo{title}{A Practical Guide to
  Splines}, \bibinfo{publisher}{Springer-Verlag}, \bibinfo{address}{New York},
  \bibinfo{year}{1978}.
\bibitem[{Chui and Diamond(1990)}]{Chui.Diamond.1990.nm}
\bibinfo{author}{C.~Chui}, \bibinfo{author}{H.~Diamond},
\newblock \bibinfo{title}{{A Characterization of Multivariate
  Quasi-interpolation Formulas and its Applications}},
\newblock \bibinfo{journal}{Numerische Mathematik} \bibinfo{volume}{57}
  (\bibinfo{year}{1990}) \bibinfo{pages}{105--121}.
\bibitem[{Unser and Daubechies(1997)}]{Unser.Daubechies.1997.ieee_tsp}
\bibinfo{author}{M.~Unser}, \bibinfo{author}{I.~Daubechies},
\newblock \bibinfo{title}{{On the approximation power of convolution-based
  least squares versus interpolation}},
\newblock \bibinfo{journal}{IEEE Transactions on Signal Processing}
  \bibinfo{volume}{45} (\bibinfo{year}{1997}) \bibinfo{pages}{1697--1711}.
\bibitem[{Hockney and Eastwood(1988)}]{Hockney.Eastwood.1988.tf}
\bibinfo{author}{R.~Hockney}, \bibinfo{author}{J.~Eastwood},
  \bibinfo{title}{{Computer simulation using particles}},
  \bibinfo{publisher}{Taylor {\&} Francis, Inc}, \bibinfo{address}{Bristol, PA,
  USA}, \bibinfo{year}{1988}.
\bibitem[{Nakamura and Yabe(1999)}]{Nakamura.Yabe.1999.cpc}
\bibinfo{author}{T.~Nakamura}, \bibinfo{author}{T.~Yabe},
\newblock \bibinfo{title}{{Cubic interpolated propagation scheme for solving
  the hyper-dimensional Vlasov-Poisson equation in phase space}},
\newblock \bibinfo{journal}{Computer Physics Communications}
  \bibinfo{volume}{120} (\bibinfo{year}{1999}) \bibinfo{pages}{122--154}.
\bibitem[{Filbet and Sonnendr{\"u}cker(2003)}]{Filbet.Sonnendrucker.2003.cpc}
\bibinfo{author}{F.~Filbet}, \bibinfo{author}{E.~Sonnendr{\"u}cker},
\newblock \bibinfo{title}{{Comparison of Eulerian Vlasov solvers}},
\newblock \bibinfo{journal}{Computer Physics Communications}
  \bibinfo{volume}{150} (\bibinfo{year}{2003}) \bibinfo{pages}{247--266}.
\bibitem[{Qiu and Christlieb(2010)}]{Qiu.Christlieb.2010.jcp}
\bibinfo{author}{J.-M. Qiu}, \bibinfo{author}{A.~Christlieb},
\newblock \bibinfo{title}{{A conservative high order semi-Lagrangian WENO
  method for the Vlasov equation}},
\newblock \bibinfo{journal}{Journal of Computational Physics}
  \bibinfo{volume}{229} (\bibinfo{year}{2010}) \bibinfo{pages}{1130--1149}.
\bibitem[{Banks and Hittinger(2010)}]{Banks.Hittinger.2010.ieee_tps}
\bibinfo{author}{J.~Banks}, \bibinfo{author}{J.~Hittinger},
\newblock \bibinfo{title}{{A New Class of Nonlinear Finite-Volume Methods for
  Vlasov Simulation}},
\newblock \bibinfo{journal}{IEEE Transactions on Plasma Science}
  \bibinfo{volume}{38} (\bibinfo{year}{2010}) \bibinfo{pages}{2198--2207}.
\bibitem[{Crouseilles et~al.(2011)Crouseilles, Mehrenberger, and
  Vecil}]{Crouseilles.Mehrenberger.Vecil.2011.esaim}
\bibinfo{author}{N.~Crouseilles}, \bibinfo{author}{M.~Mehrenberger},
  \bibinfo{author}{F.~Vecil},
\newblock \bibinfo{title}{{Discontinuous Galerkin semi-Lagrangian method for
  Vlasov-Poisson}},
\newblock \bibinfo{journal}{ESAIM: Proceedings} \bibinfo{volume}{32}
  (\bibinfo{year}{2011}) \bibinfo{pages}{211--230}.
\bibitem[{Rossmanith and Seal(2011)}]{Rossmanith.Seal.2011.jcp}
\bibinfo{author}{J.~Rossmanith}, \bibinfo{author}{D.~Seal},
\newblock \bibinfo{title}{{A positivity-preserving high-order semi-Lagrangian
  discontinuous Galerkin scheme for the Vlasov-Poisson equations}},
\newblock \bibinfo{journal}{Journal of Computational Physics}
  \bibinfo{volume}{230} (\bibinfo{year}{2011}) \bibinfo{pages}{6203--6232}.
\bibitem[{Bieniosek et~al.(2009)Bieniosek, Henestroza, Leitner, Logan, More,
  Roy, Ni, Seidl, Waldron, and Barnard}]{Bieniosek.al.2009.nimp}
\bibinfo{author}{F.~Bieniosek}, \bibinfo{author}{E.~Henestroza},
  \bibinfo{author}{M.~Leitner}, \bibinfo{author}{B.~Logan},
  \bibinfo{author}{R.~More}, \bibinfo{author}{P.~Roy}, \bibinfo{author}{P.~Ni},
  \bibinfo{author}{P.~Seidl}, \bibinfo{author}{W.~Waldron},
  \bibinfo{author}{J.~Barnard},
\newblock \bibinfo{title}{High-energy density physics experiments with intense
  heavy ion beams},
\newblock \bibinfo{journal}{Nuclear Instruments and Methods in Physics Research
  Section A: Accelerators, Spectrometers, Detectors and Associated Equipment}
  \bibinfo{volume}{606} (\bibinfo{year}{2009}) \bibinfo{pages}{146--151}.
  \bibinfo{note}{Heavy Ion Inertial Fusion -- Proceedings of the 17th
  International Symposium on Heavy Ion Inertial Fusion}.
\bibitem[{Wangler et~al.(1996)Wangler, Garnett, Gray, Ryne, and
  Wang}]{Wangler.P.1996}
\bibinfo{author}{T.~Wangler}, \bibinfo{author}{R.~Garnett},
  \bibinfo{author}{E.~Gray}, \bibinfo{author}{R.~Ryne},
  \bibinfo{author}{T.~Wang},
\newblock \bibinfo{title}{{Dynamics of beam halo in mismatched beams}},
\newblock in: \bibinfo{booktitle}{XVIII International Linac Conference,
  Gen{\`e}ve, Suisse}, \bibinfo{publisher}{IEEE \#01CH37268C, Piscataway, NJ
  08855}, \bibinfo{year}{1996}, p. \bibinfo{pages}{TPAT061}.

\end{thebibliography}
